\documentclass[final]{elsart}

\date{}
\mathchardef\Omega="700A

\usepackage{verbatim}
\usepackage{graphicx}
\usepackage[noabbrv]{unitsdef}
\usepackage{url}
\usepackage{amsmath}
\usepackage{amssymb}
\usepackage{latexsym}
\usepackage{stmaryrd}
\usepackage{rotating}
\usepackage{pstricks,pst-node,pst-plot,pst-circ}
\usepackage{subfigure}
\usepackage{graphicx}

\newboolean{TR}
\setboolean{TR}{true}
\newcmykcolor{dblue}{1 1 0 .3}

\newcommand*{\coanote}[1]{}
\newcommand*{\TRfootnote}[1]{\ifthenelse{\boolean{TR}}{\footnote{#1}}{}}

\newtheorem{theorem}{Theorem}[section]

\theoremstyle{definition}
\newtheorem{definition}[theorem]{Definition}
\newtheorem{example}[theorem]{Example}

\newcommand{\summary}[1]{\textrm{\textbf{\textup{#1}}}}

\theoremstyle{remark}

\numberwithin{equation}{section}

\newdimen\proofrulebreadth \proofrulebreadth=.05em
\newdimen\proofdotseparation \proofdotseparation=1.25ex
\newdimen\proofrulebaseline \proofrulebaseline=2ex
\newcount\proofdotnumber \proofdotnumber=3
\let\then\relax
\def\hfi{\hskip0pt plus.0001fil}
\mathchardef\squigto="3A3B
\newif\ifinsideprooftree\insideprooftreefalse
\newif\ifonleftofproofrule\onleftofproofrulefalse
\newif\ifproofdots\proofdotsfalse
\newif\ifdoubleproof\doubleprooffalse
\let\wereinproofbit\relax
\newdimen\shortenproofleft
\newdimen\shortenproofright
\newdimen\proofbelowshift
\newbox\proofabove
\newbox\proofbelow
\newbox\proofrulename
\def\shiftproofbelow{\let\next\relax\afterassignment\setshiftproofbelow\dimen0 }
\def\shiftproofbelowneg{\def\next{\multiply\dimen0 by-1 }%
\afterassignment\setshiftproofbelow\dimen0 }
\def\setshiftproofbelow{\next\proofbelowshift=\dimen0 }
\def\setproofrulebreadth{\proofrulebreadth}

\def\prooftree{% NESTED ZERO (\ifonleftofproofrule)
\ifnum  \lastpenalty=1
\then   \unpenalty
\else   \onleftofproofrulefalse
\fi
\ifonleftofproofrule
\else   \ifinsideprooftree
        \then   \hskip.5em plus1fil
        \fi
\fi
\bgroup% NESTED ONE (\proofbelow, \proofrulename, \proofabove,
\setbox\proofbelow=\hbox{}\setbox\proofrulename=\hbox{}%
\let\justifies\proofover\let\leadsto\proofoverdots\let\Justifies\proofoverdbl
\let\using\proofusing\let\[\prooftree
\ifinsideprooftree\let\]\endprooftree\fi
\proofdotsfalse\doubleprooffalse
\let\thickness\setproofrulebreadth
\let\shiftright\shiftproofbelow \let\shift\shiftproofbelow
\let\shiftleft\shiftproofbelowneg
\let\ifwasinsideprooftree\ifinsideprooftree
\insideprooftreetrue
\setbox\proofabove=\hbox\bgroup$\displaystyle % NESTED TWO
\let\wereinproofbit\prooftree
\shortenproofleft=0pt \shortenproofright=0pt \proofbelowshift=0pt
\onleftofproofruletrue\penalty1
}

\def\eproofbit{% NESTED TWO
\ifx    \wereinproofbit\prooftree
\then   \ifcase \lastpenalty
        \then   \shortenproofright=0pt  % 0: some other object, no indentation
        \or     \unpenalty\hfil         % 1: empty hypotheses, just glue
        \or     \unpenalty\unskip       % 2: just had a tree, remove glue
        \else   \shortenproofright=0pt  % eh?
        \fi
\fi
\global\dimen0=\shortenproofleft
\global\dimen1=\shortenproofright
\global\dimen2=\proofrulebreadth
\global\dimen3=\proofbelowshift
\global\dimen4=\proofdotseparation
\global\count255=\proofdotnumber
$\egroup  % NESTED ONE
\shortenproofleft=\dimen0
\shortenproofright=\dimen1
\proofrulebreadth=\dimen2
\proofbelowshift=\dimen3
\proofdotseparation=\dimen4
\proofdotnumber=\count255
}

\def\proofover{% NESTED TWO
\eproofbit % NESTED ONE
\setbox\proofbelow=\hbox\bgroup % NESTED TWO
\let\wereinproofbit\proofover
$\displaystyle
}%
\def\proofoverdbl{% NESTED TWO
\eproofbit % NESTED ONE
\doubleprooftrue
\setbox\proofbelow=\hbox\bgroup % NESTED TWO
\let\wereinproofbit\proofoverdbl
$\displaystyle
}%
\def\proofoverdots{% NESTED TWO
\eproofbit % NESTED ONE
\proofdotstrue
\setbox\proofbelow=\hbox\bgroup % NESTED TWO
\let\wereinproofbit\proofoverdots
$\displaystyle
}%
\def\proofusing{% NESTED TWO
\eproofbit % NESTED ONE
\setbox\proofrulename=\hbox\bgroup % NESTED TWO
\let\wereinproofbit\proofusing
\kern0.3em$
}

\def\endprooftree{% NESTED TWO
\eproofbit % NESTED ONE
  \dimen5 =0pt% spread of hypotheses
\dimen0=\wd\proofabove \advance\dimen0-\shortenproofleft
\advance\dimen0-\shortenproofright
\dimen1=.5\dimen0 \advance\dimen1-.5\wd\proofbelow
\dimen4=\dimen1
\advance\dimen1\proofbelowshift \advance\dimen4-\proofbelowshift
\ifdim  \dimen1<0pt
\then   \advance\shortenproofleft\dimen1
        \advance\dimen0-\dimen1
        \dimen1=0pt
        \ifdim  \shortenproofleft<0pt
        \then   \setbox\proofabove=\hbox{%
                        \kern-\shortenproofleft\unhbox\proofabove}%
                \shortenproofleft=0pt
        \fi
\fi
\ifdim  \dimen4<0pt
\then   \advance\shortenproofright\dimen4
        \advance\dimen0-\dimen4
        \dimen4=0pt
\fi
\ifdim  \shortenproofright<\wd\proofrulename
\then   \shortenproofright=\wd\proofrulename
\fi
\dimen2=\shortenproofleft \advance\dimen2 by\dimen1
\dimen3=\shortenproofright\advance\dimen3 by\dimen4
\ifproofdots
\then
        \dimen6=\shortenproofleft \advance\dimen6 .5\dimen0
        \setbox1=\vbox to\proofdotseparation{\vss\hbox{$\cdot$}\vss}%
        \setbox0=\hbox{%
                \advance\dimen6-.5\wd1
                \kern\dimen6
                $\vcenter to\proofdotnumber\proofdotseparation
                        {\leaders\box1\vfill}$%
                \unhbox\proofrulename}%
\else   \dimen6=\fontdimen22\the\textfont2 % height of maths axis
        \dimen7=\dimen6
        \advance\dimen6by.5\proofrulebreadth
        \advance\dimen7by-.5\proofrulebreadth
        \setbox0=\hbox{%
                \kern\shortenproofleft
                \ifdoubleproof
                \then   \hbox to\dimen0{%
                        $\mathsurround0pt\mathord=\mkern-6mu%
                        \cleaders\hbox{$\mkern-2mu=\mkern-2mu$}\hfill
                        \mkern-6mu\mathord=$}%
                \else   \vrule height\dimen6 depth-\dimen7 width\dimen0
                \fi
                \unhbox\proofrulename}%
        \ht0=\dimen6 \dp0=-\dimen7
\fi
\let\doll\relax
\ifwasinsideprooftree
\then   \let\VBOX\vbox
\else   \ifmmode\else$\let\doll=$\fi
        \let\VBOX\vcenter
\fi
\VBOX   {\baselineskip\proofrulebaseline \lineskip.2ex
        \expandafter\lineskiplimit\ifproofdots0ex\else-0.6ex\fi
        \hbox   spread\dimen5   {\hfi\unhbox\proofabove\hfi}%
        \hbox{\box0}%
        \hbox   {\kern\dimen2 \box\proofbelow}}\doll%
\global\dimen2=\dimen2
\global\dimen3=\dimen3
\egroup % NESTED ZERO
\ifonleftofproofrule
\then   \shortenproofleft=\dimen2
\fi
\shortenproofright=\dimen3
\onleftofproofrulefalse
\ifinsideprooftree
\then   \hskip.5em plus 1fil \penalty2
\fi
}

\newcommand*{\ttv}{\mathrm{tt}}
\newcommand*{\ffv}{\mathrm{ff}}

\newcommand*{\vbar}{\mathrel{\mid}}

\newcommand*{\kw}[1]{\mathop{\textbf{#1}}}

\newcommand*{\Int}{\mathrm{Int}}
\newcommand*{\Bool}{\mathrm{Bool}}
\newcommand*{\Var}{\mathrm{Var}}
\newcommand*{\Aexp}{\mathrm{Aexp}}
\newcommand*{\Bexp}{\mathrm{Bexp}}
\newcommand*{\Stmt}{\mathrm{Stmt}}
\newcommand*{\Store}{\mathord{\mathrm{Store}}}

\newcommand*{\eval}[2]{\mathrel{\buildrel \mathrm{#2} \over #1}}
\newcommand*{\ceval}[1]{\eval{\rightarrow}{#1}}
\newcommand*{\aceval}{\ceval{a}}
\newcommand*{\bceval}{\ceval{b}}
\newcommand*{\sceval}{\ceval{s}}
\newcommand*{\diverges}{\ceval{\infty}\mathord{}}
\newcommand*{\aeval}[1]{\eval{\rightsquigarrow}{#1}}
\newcommand*{\aaeval}{\aeval{a}}
\newcommand*{\baeval}{\aeval{b}}
\newcommand*{\saeval}{\aeval{s}}

\newcommand*{\absplus}{\mathbin{\varoplus}}
\newcommand*{\absminus}{\mathbin{\varominus}}
\newcommand*{\absprod}{\mathbin{\varoast}}
\newcommand*{\absneg}{\mathop{\varobslash}\nolimits}
\newcommand*{\absor}{\mathbin{\varovee}}
\newcommand*{\absand}{\mathbin{\varowedge}}
\newcommand*{\abseq}{\mathrel{\varocircle}}
\newcommand*{\abslt}{\mathrel{\varolessthan}}

\newcommand*{\nohyp}{\phantom{\langle a_0, \sigma \rangle \aceval m_0}}
\newcommand*{\omissis}{\left[\,\mathord{\cdots}\,\right]}

\newcommand*{\con}{\mathrm{con}}

\newcommand{\polyhull}{\mathbin{\uplus}}
\newcommand*{\widen}{\mathbin{\nabla}}

\newcommand{\defrel}[1]{\mathrel{\buildrel \mathrm{def} \over {#1}}}
\newcommand{\defeq}{\defrel{=}}

\providecommand*{\Nset}{\mathbb{N}}             % Naturals
\providecommand*{\Zset}{\mathbb{Z}}             % Integers
\providecommand*{\Rset}{\mathbb{R}}             % Reals
\providecommand*{\nonnegRset}{\mathbb{R}_{\scriptscriptstyle{+}}}
\newcommand{\sset}[2]{{\renewcommand{\arraystretch}{1.2}
                      \left\{\,#1 \,\left|\,
                               \begin{array}{@{}l@{}}#2\end{array}
                      \right.   \,\right\}}}

\newcommand*{\sseq}{\subseteq}
\newcommand*{\union}{\cup}
\newcommand*{\bigunion}{\bigcup}

\newcommand*{\inters}{\cap}
\newcommand*{\setdiff}{\setminus}

\newcommand*{\sqsseq}{\sqsubseteq}

\newcommand*{\lfp}{\mathop{\mathrm{lfp}}\nolimits}

\newcommand*{\vect}[1]{\mathbf{#1}}

\newcommand{\st}{\mathrel{.}}
\newcommand{\itc}{\mathrel{:}}

\newcommand*{\bigland}{\mathop{\bigwedge}}

\newcommand{\lambdadot}{\mathbin{.}}

\newcommand{\relop}{\mathrel{\bowtie}}

\newcommand*{\cA}{\ensuremath{\mathcal{A}}}

\newcommand*{\cC}{\ensuremath{\mathcal{C}}}

\newcommand*{\cF}{\ensuremath{\mathcal{F}}}

\newcommand*{\cH}{\ensuremath{\mathcal{H}}}

\newcommand*{\cK}{\ensuremath{\mathcal{K}}}

\newcommand*{\cP}{\ensuremath{\mathcal{P}}}
\newcommand*{\cQ}{\ensuremath{\mathcal{Q}}}
\newcommand*{\cR}{\ensuremath{\mathcal{R}}}

\newcommand*{\cW}{\ensuremath{\mathcal{W}}}

\newcommand*{\reld}[3]{\mathord{#1}\subseteq#2\times#3}
\newcommand*{\fund}[3]{\mathord{#1}\colon#2\rightarrow#3}
\newcommand*{\pard}[3]{\mathord{#1}\colon#2\rightarrowtail#3}

\newcommand*{\compose}{\mathbin{\circ}}

\providecommand*{\CPset}{\mathbb{CP}}          % Closed polyhedra
\providecommand*{\Pset}{\mathbb{P}}            % (NNC) polyhedra
\providecommand*{\Gset}{\mathbb{G}}            % Rational Lattice

\newcommand*{\Loc}{\mathrm{Loc}}
\newcommand*{\Lab}{\mathrm{Lab}}
\newcommand*{\Init}{\mathrm{Init}}
\newcommand*{\Trans}{\mathrm{Trans}}
\newcommand*{\Inv}{\mathrm{Inv}}
\newcommand*{\Act}{\mathrm{Act}}

\newcommand*{\timeelapse}{\nearrow}

\newcommand*{\plusplus}{\mathord{\scriptstyle ++}}
\newcommand*{\minusminus}{\mathord{\scriptstyle --}}

\newcommand*{\timedstep}[2]{\mathrel{\rightarrow^{#1}_{#2}}}

\newcommand*{\dom}{\mathop{\mathrm{dom}}\nolimits}

\begin{document}

\begin{frontmatter}

\title%
%[Polyhedral Computations and HW/SW Analysis and Verification]%
{Applications of Polyhedral Computations \\
to the Analysis and Verification \\
of Hardware and Software Systems\thanksref{th}}

\thanks[th]{This work has been partly supported by PRIN project
``AIDA: Abstract Interpretation Design and Applications.''}

\author[Parma]{Roberto Bagnara},
\ead{bagnara@cs.unipr.it}
\author[Leeds]{Patricia M. Hill},
\ead{hill@comp.leeds.ac.uk}
\author[Parma]{Enea Zaffanella}
\ead{zaffanella@cs.unipr.it}

\address[Parma]{Department of Mathematics, University of Parma, Italy}
\address[Leeds]{School of Computing, University of Leeds, UK}

\begin{abstract}
Convex polyhedra are the basis for several abstractions used in static
analysis and computer-aided verification of complex and sometimes
mission critical systems. For such applications, the identification of
an appropriate complexity-precision trade-off is a particularly acute
problem, so that the availability of a wide spectrum of alternative
solutions is mandatory.  We survey the range of applications of
polyhedral computations in this area; give an overview of the
different classes of polyhedra that may be adopted;
outline the main polyhedral operations required by
automatic analyzers and verifiers; and look at some possible
combinations of polyhedra with other numerical abstractions that have
the potential to improve the precision of the analysis.
Areas where
further theoretical investigations can result in important
contributions are highlighted.
\end{abstract}


\end{frontmatter}

\section{Introduction}
\label{sec:introduction}

The application of polyhedral computations to the analysis and
verification of computer programs has its origin in a groundbreaking
paper by Cousot and Halbwachs \cite{CousotH78}.
There, the authors applied the theory of abstract interpretation
\ifthenelse{\boolean{TR}}{%%
\cite{CousotC77,CousotC92fr}
}{%%
\cite{CousotC79,CousotC92fr}
}%%\ifthenelse{\boolean{TR}}
to the static determination of linear equality and inequality
relations among program variables.  In essence, the idea consists
in interpreting a program (as will be explained in more detail in
Sections~\ref{sec:abstract-interpretation}
and~\ref{sec:analysis-of-computer-programs})
on a domain of convex polyhedra instead of the concrete domain
of (sets of vectors of) machine numbers.  Each program operation is
correctly approximated by a corresponding operation on polyhedra
and measures are taken to ensure that the approximate computation
always terminates.  At the end of this process, the obtained
polyhedra encode provably correct \emph{linear invariants} of the
analyzed program (i.e., linear equalities and inequalities that are
guaranteed to hold for each program execution and for each program input).

As we show in this paper, relational information
concerning the data objects manipulated by programs or other devices
is crucial for a broad range of applications in the field
of automatic or semi-automatic program manipulation: it can be used to prove
the absence of certain kinds of errors; it can verify that certain processes
always terminate or stabilize; it can pinpoint the position of
errors in the system;
and it can enable the application of optimizations.
Despite this, due to the lack of efficient, robust and publicly available
implementations of convex polyhedra and of the required operations,
the line of work begun by Cousot and Halbwachs did not see much
development until the beginning of the 1990s.  Since then, this
approach has been increasingly adopted and today convex polyhedra
are the basis for several abstractions used in static analysis and
computer-aided verification of complex and sometimes mission critical
systems.  For such applications, the identification of an appropriate
complexity-precision trade-off is a particularly acute problem:
on the one hand, relational information provided by general polyhedra
is extremely valuable; on the other hand, its high computational cost
makes it a fairly scarce resource that must be managed with care.
This implies, among other things,  that general polyhedra must be
combined with simpler polyhedra in order to achieve scalability.
As the complexity-precision trade-off varies considerably between
different applications, the availability of a wide spectrum of alternative
solutions is mandatory.

In this paper, we survey the range of applications of polyhedral
computations in the area of the analysis and verification of hardware
and software systems:
we describe in detail one important ---and historically, first--- application
of polyhedral computations in the field of formal methods,
the linear invariant analysis for imperative programs;
we provide an account of linear hybrid systems that is based directly on
polyhedra;
and we explain with an example how polyhedral approximations
can be applied to analog systems.
The paper also provides
an overview of the main polyhedral operations required by these applications,
brief descriptions of some of the different classes of
polyhedra that may be adopted, depending on the particular context,
and a look at some possible combinations of
polyhedra with other numerical abstractions that have the potential to
improve the precision of the analysis.
Areas where further theoretical investigations can result
in important contributions are highlighted.
\ifthenelse{\boolean{TR}}{}{%%
Some bibliographic references and a few examples have been omitted
from this paper for space reasons; the interested reader can find them
in the technical report version~\cite{BagnaraHZ07TRa}.
}%%\ifthenelse{\boolean{TR}}{}{%%

The plan of the paper is as follows.
Section~\ref{sec:preliminaries} introduces the required notions and
\ifthenelse{\boolean{TR}}{%%
notations, including a minimal exposition of the main concepts of
abstract interpretation theory.
}{%%
notations.
}%%\ifthenelse{\boolean{TR}}{%%
Section~\ref{sec:analysis-of-computer-programs}
demonstrates the use of polyhedral computations in the specification
of a linear invariant analysis for
\ifthenelse{\boolean{TR}}{%%
a simple imperative language;
a few of the many applications for the analysis of computer programs
are briefly recalled.
}{%%
a simple imperative language.
}%%\ifthenelse{\boolean{TR}}{%%
Section~\ref{sec:analysis-of-hybrid-systems}
is devoted to polyhedral approximation techniques for hybrid systems,
which, as shown in Section~\ref{sec:analysis-of-analog-systems} can also
be applied to purely analog systems.
Section~\ref{sec:families-of-polyhedral-approximations}
presents several families of
\ifthenelse{\boolean{TR}}{%%
polyhedral approximations that provide a range of different
solutions to the complexity/precision trade-off.
}{%%
polyhedral approximations.
}%%\ifthenelse{\boolean{TR}}{%%
The most important operations that such approximations must provide
\ifthenelse{\boolean{TR}}{%%
in order to support analysis and verification methods
}{}%%\ifthenelse{\boolean{TR}}{%%
are illustrated in Section~\ref{sec:peculiar-polyhedral-computations}.
Section~\ref{sec:conclusion} concludes.

\section{Preliminaries}
\label{sec:preliminaries}

We assume some basic knowledge about lattice theory~\cite{Birkhoff67}.
Let $(S, \sqsseq)$ and $(T, \preceq)$ be two partially ordered sets;
the function $\fund{f}{S}{T}$ is \emph{monotonic} if,
for all $x_0, x_1 \in S$, $x_0 \sqsseq x_1$ implies $f(x_0) \preceq f(x_1)$.
If $(S, \sqsseq) \equiv (T, \preceq)$, so that $\fund{f}{S}{S}$,
an element $x \in S$ such that $x = f(x)$ is a \emph{fixpoint} of $f$.
If $(S, \sqsseq, \bot, \top, \sqcup, \sqcap)$ is a complete lattice,
then $f$ is \emph{continuous} if it preserves
the least upper bound of all increasing chains, i.e., for all
$x_0 \sqsseq x_1 \sqsseq \cdots$ in $S$, it satisfies
$f\bigl( \bigsqcup x_i \bigr) = \bigsqcup f(x_i)$;
in such a case, the least fixpoint of $f$ with respect to
the partial order `$\mathord{\sqsseq}$', denoted $\lfp f$, can be obtained
by iterating the application of $f$
starting from the bottom element $\bot$,
thereby computing the upward iteration sequence
\ifthenelse{\boolean{TR}}{%%
\[
  \bot = f^0(\bot)
    \sqsseq f^1(\bot)
      \sqsseq f^2(\bot)
        \sqsseq \cdots
          \sqsseq f^i(\bot)
            \sqsseq \cdots,
\]
up to the first non-zero limit ordinal $\omega$;
namely,
\[
  \lfp f
    = f^\omega(\bot)
    \defeq
      \bigsqcup_{i < \omega} f^i(\bot).
\]
}{%%
\(
  \bot = f^0(\bot)
    \sqsseq f^1(\bot)
      \sqsseq f^2(\bot)
        \sqsseq \cdots
          \sqsseq f^i(\bot)
            \sqsseq \cdots
\),
up to the first non-zero limit ordinal $\omega$;
namely,
\(
  \lfp f
    = f^\omega(\bot)
    \defeq
      \bigsqcup_{i < \omega} f^i(\bot)
\).
}%%\ifthenelse{\boolean{TR}}{%%

For each $\fund{f_0}{S_0}{T_0}$ and $\fund{f_1}{S_1}{T_1}$,
the function
$\fund{f_0[f_1]}{(S_0 \union S_1)}{(T_0 \union T_1)}$
is defined, for each $x \in S_0 \union S_1$,
\ifthenelse{\boolean{TR}}{%%
by
\[
  \bigl(f_0[f_1]\bigr)(x)
    \defeq
      \begin{cases}
        f_1(x), &\text{if $x \in S_1$;} \\
        f_0(x), &\text{if $x \in S_0 \setminus S_1$.}
      \end{cases}
\]
}{%%
so that
$f_0[f_1](x) = f_1(x)$, if $x \in S_1$, and
$f_0[f_1](x) = f_0(x)$, otherwise.
}%%\ifthenelse{\boolean{TR}}{%%

For $n > 0$, we denote by $\vect{v} = (v_0, \ldots, v_{n-1}) \in \Rset^n$
an $n$-tuple (vector) of real numbers;
$\nonnegRset$ is the set of non-negative real numbers;
\(
  \langle \vect{v}, \vect{w} \rangle
\)
denotes the scalar product of vectors $\vect{v},\vect{w} \in \Rset^n$;
the vector $\vect{0} \in \Rset^n$ has all components equal to zero.
We write $\vect{v} \mathop{::} \vect{w}$
to denote the \emph{tuple concatenation}
of $\vect{v} \in \Rset^n$ and $\vect{w} \in \Rset^m$,
so that $\vect{v} \mathop{::} \vect{w} \in \Rset^{n+m}$.

Let $\vect{x}$ be an $n$-tuple of distinct variables. Then
$\beta = \bigl( \langle \vect{a}, \vect{x} \rangle \relop b \bigr)$
denotes a linear inequality constraint,
for each vector $\vect{a} \in \Rset^n$, where $\vect{a} \neq \vect{0}$,
each scalar $b \in \Rset$, and
$\mathord{\relop} \in \{ \mathord{\geq}, \mathord{>} \}$.
A linear inequality constraint $\beta$ defines
a (topologically closed or open) affine half-space of $\Rset^n$,
denoted by $\con\bigl(\{\beta\}\bigr)$.

A set $\cP \sseq \Rset^n$ is a \emph{(convex) polyhedron}
if and only if $\cP$ can be expressed as the intersection of
a finite number of affine half-spaces of $\Rset^n$, i.e.,
as the solution $\cP = \con(\cC)$
of a finite set of linear inequality constraints $\cC$
(called a \emph{constraint system}).
The set of all polyhedra on the vector space $\Rset^n$ is
denoted as $\Pset_n$. When partially ordered by set-inclusion,
convex polyhedra form a lattice
\(
  (\Pset_n, \sseq, \emptyset, \Rset^n, \polyhull, \inters)
\)
having the empty set and $\Rset^n$ as the bottom
and top elements, respectively;
the binary meet operation, returning the greatest polyhedron
smaller than or equal to the two arguments, is easily seen
to correspond to set-intersection;
the binary join operation, returning the least polyhedron
greater than or equal to the two arguments,
is denoted `$\mathord{\polyhull}$' and called
\emph{convex polyhedral hull} (poly-hull, for short).
In general, the poly-hull of two polyhedra is different
from their convex hull~\cite{StoerW70}.

A relation $\reld{\psi}{\Rset^n}{\Rset^n}$ (of dimension $n$)
is said to be \emph{affine} if there exists $\ell \in \Nset$ and
$\vect{a}_i, \vect{c}_i \in \Rset^n$,
$b_i \in \Rset$ and
$\mathord{\relop}_i \in \{ \geq, > \}$,
for each $i = 1, \ldots, \ell$,
such that
\begin{equation*}
  \forall \vect{v}, \vect{w} \in \Rset^n
    \itc
      (\vect{v}, \vect{w}) \in \psi
        \iff
          \bigland_{i=1}^{\ell}
            \bigl(
              \langle \vect{c}_i, \vect{w} \rangle
                \relop_i
                  \langle \vect{a}_i, \vect{v} \rangle + b_i
            \bigr).
\end{equation*}
Any affine relation of dimension $n$ can thus be encoded by
$\ell$ linear inequalities on a $2n$-tuple of distinct variables
$\vect{x} \mathop{::} \vect{x}'$
(playing the role of $\vect{v}$ and $\vect{w}$, respectively),
therefore defining a polyhedron in $\Pset_{2n}$.
The set of polyhedra $\Pset_n$ is closed under the
(direct or inverse) application of affine relations:
i.e., for each $\cP \in \Pset_n$
and each affine relation $\reld{\psi}{\Rset^n}{\Rset^n}$,
the image $\psi(\cP)$ and the preimage $\psi^{-1}(\cP)$
are in $\Pset_n$.

\subsection{Abstract Interpretation}
\label{sec:abstract-interpretation}

The semantics of a hardware or software system is a mathematical
description of all its possible run-time behaviors. Different
semantics can be defined for the same system, depending on the
details being recorded.  \emph{Abstract interpretation}
\ifthenelse{\boolean{TR}}{%%
\cite{CousotC77,CousotC79,CousotC92fr}
}{%%
\cite{CousotC79,CousotC92fr}
}%%\ifthenelse{\boolean{TR}}
is a formal method for relating these semantics according to their
level of abstraction, so that questions about the behavior of
a system can be provided with sound, possibly approximate answers.

The concrete semantics $c \in C$ of a program is usually formalized as
the least fixpoint of a continuous semantic function $\fund{\cF}{C}{C}$,
where the concrete domain
$(C, \sqsseq, \bot, \top, \sqcup, \sqcap)$
is a complete lattice of semantic properties;
in many interesting cases, the computational order `$\sqsseq$' corresponds
to the approximation relation, so that $c_1 \sqsseq c_2$ holds
if $c_1$ is a stronger property than $c_2$
(i.e., $c_2$ correctly approximates $c_1$).

For instance, the run-time behavior of a program may be defined in
terms of a transition system $\langle \Sigma, t, \iota \rangle$, where
$\Sigma$ is a set of states, $\iota \sseq \Sigma$ is the subset of initial
states, and $t \in \wp(\Sigma \times \Sigma)$ is a binary transition
relation mapping a state to its possible successor states.
Letting $\Sigma^\star$ denote
the set of all finite sequences of elements in $\Sigma$,
the initial history of a forward computation can be recorded%%
\TRfootnote{This is just one of a wide range of possible semantics;
by the same approach, other semantics may be described and related
by abstract interpretation~\cite{CousotC92fr}.}
as a partial execution trace
$\tau = \sigma_0 \cdots \sigma_m \in \Sigma^\star$
starting from an initial state $\sigma_0 \in \iota$
and such that any two consecutive states
$\sigma_i$ and $\sigma_{i+1}$ are related by the
transition relation, i.e., $(\sigma_i, \sigma_{i+1}) \in t$.
In such a context, an element of the concrete domain
\(
  \bigl(
    \wp(\Sigma^\star), \sseq, \emptyset, \Sigma^\star, \union, \inters
  \bigr)
\)
is a set of partial execution traces and
the concrete semantics is $\lfp(\cF)$,
where the semantic function is defined by
\begin{multline*}
  \cF
    = \lambda X \in \wp(\Sigma^\star)
        \lambdadot
          X
           \union
        \{\,
          \tau \in \Sigma^\star
        \mid
          \tau = \sigma_0 \in \iota
        \,\} \\
          \union
        \bigl\{\,
          \tau \sigma_{i+1} \in \Sigma^\star
        \bigm|
          \tau = \sigma_0 \cdots \sigma_i \in X,
          (\sigma_i, \sigma_{i+1}) \in t
        \,\bigr\}.
\end{multline*}
An abstract domain\footnote{To avoid notational burden,
we will freely overload the lattice-theoretic symbols
`$\sqsseq$', `$\bot$', `$\sqcup$', etc.,
exploiting context to disambiguate their meaning.}
$(D^\sharp, \sqsseq, \bot, \sqcup)$
can be often modeled as a bounded join-semilattice,
so that it has a bottom element $\bot$
and the least upper bound
$d^\sharp_1 \sqcup d^\sharp_2$ exists
for all $d^\sharp_1, d^\sharp_2 \in D^\sharp$.
This domain is related to the concrete domain by a monotonic and injective
concretization function $\fund{\gamma}{D^\sharp}{C}$.
Monotonicity and injectivity mean that the abstract partial order
is equivalent to the approximation relation induced on $D^\sharp$
by the concretization function $\gamma$.
Conversely, the concrete domain is related to the abstract one
by a partial abstraction function $\pard{\alpha}{C}{D^\sharp}$ such that,
for each $c \in C$, if $\alpha(c)$ is defined then
$c \sqsubseteq \gamma\bigl(\alpha(c)\bigr)$.
In particular, we assume that $\alpha(\bot) = \bot$ is always defined;
when needed or useful, we will require a few additional properties.

For example, a first abstraction of the semantics above,
typically adopted for the inference of
invariance properties of programs~\cite{CousotC79,CousotC92fr}, approximates
a set of traces by the set of states occurring in any one of the traces.
The \emph{reachable states} are thus characterized by elements of
the complete lattice
$\bigl( \wp(\Sigma), \sseq, \emptyset, \Sigma, \union, \inters \bigr)$,
which plays here the role of the abstract domain.
The concretization function relating $D^\sharp = \wp(\Sigma)$
to $C = \wp(\Sigma^\star)$ is defined, for each $d^\sharp \in \wp(\Sigma)$, by
\ifthenelse{\boolean{TR}}{%%
\[
}{%%
\(
}%%\ifthenelse{\boolean{TR}}{%%
  \gamma(d^\sharp)
    \defeq
      \{\,
        \tau \in \Sigma^\star
      \mid
        \tau = \sigma_0 \cdots \sigma_m,
        \forall i = 0, \ldots, m \itc \sigma_i \in d^\sharp
      \,\}.
\ifthenelse{\boolean{TR}}{%%
\]
}{%%
\)
}%%\ifthenelse{\boolean{TR}}{%%
The concrete semantic function
$\fund{\cF}{\wp(\Sigma^\star)}{\wp(\Sigma^\star)}$
can thus be approximated by the monotonic abstract semantic function
$\fund{\cA}{\wp(\Sigma)}{\wp(\Sigma)}$
defined by
\[
  \cA
    = \lambda d^\sharp \in \wp(\Sigma)
        \lambdadot
          d^\sharp
            \union
          \iota
            \union
          \bigl\{\,
            \sigma' \in \Sigma
          \bigm|
            \exists \sigma \in d^\sharp \st (\sigma, \sigma') \in t
          \,\bigr\}.
\]
This abstract semantic function is \emph{sound} with respect to
the concrete semantic function in that it satisfies the
local correctness requirement
\begin{equation*}
\label{eq:local-correctness-condition}
  \forall c \in C
    \itc
      \forall d^\sharp \in D^\sharp
        \itc
          c \sqsseq \gamma(d^\sharp)
            \implies
              \cF(c) \sqsseq \gamma\bigl( \cA(d^\sharp) \bigr),
\end{equation*}
ensuring that each iteration $\cF^i(\bot)$
in the concrete fixpoint computation
is approximated by computing
the corresponding abstract iteration $\cA^i\bigl( \alpha(\bot) \bigr)$.
In particular, the least fixpoint of $\cF$ is approximated by
any post-fixpoint of~$\cA$
\ifthenelse{\boolean{TR}}{%%
\cite{CousotC77,CousotC92fr},
}{%%
\cite{CousotC92fr},
}%%\ifthenelse{\boolean{TR}}
i.e., any abstract element $d^\sharp \in D^\sharp$
such that $\cA(d^\sharp) \sqsseq d^\sharp$.

Actually, the abstraction defined above satisfies an even
stronger property, in that the abstract semantic function $\cA$ is
the \emph{most precise} of all the sound approximations of $\cF$
that could be defined on the considered abstract domain.
This happens because the two domains are related by a
Galois connection~\cite{CousotC79}, i.e.,
there exists a total abstraction function
$\fund{\alpha}{C}{D^\sharp}$ satisfying
\[
  \forall c \in C
    \itc
      \forall d^\sharp \in D^\sharp
        \itc
          \alpha(c) \sqsseq d^\sharp
            \iff
              c \sqsseq \gamma(d^\sharp).
\]
Namely,
\ifthenelse{\boolean{TR}}{%%
for all $c \in \wp(\Sigma^\star)$, we can define
\[
  \alpha(c)
    \defeq
      \bigl\{\,
        \sigma_i \in \Sigma
      \bigm|
        \tau = \sigma_0 \cdots \sigma_m \in c,
        i \in \{0, \ldots, m\}
      \,\bigr\}.
\]
}{%%
\(
  \alpha(c)
    \defeq
      \bigl\{\,
        \sigma_i \in \Sigma
      \bigm|
        \tau = \sigma_0 \cdots \sigma_m \in c,
        i \in \{0, \ldots, m\}
      \,\bigr\}
\).
}%%\ifthenelse{\boolean{TR}}{%%

For Galois connections it can be shown
that $\alpha(c)$ is the best possible approximation in $D^\sharp$
for the concrete element $c \in C$;
similarly, $\alpha \compose \cF \compose \gamma$
(i.e., the function $\cA$ defined above)
is the best possible approximation for $\cF$
\cite{CousotC79}.
Such a result is provided with a quite intuitive reading;
in order to approximate the concrete function $\cF$
on an abstract element $d^\sharp \in D^\sharp$:
we first apply the concretization function $\gamma$ so as to obtain
the meaning of $d^\sharp$; then we apply the concrete function $\cF$;
finally, we abstract the result so as to obtain back an element of $D^\sharp$.

Abstract interpretation theory can thus be used to specify (semi-)
automatic program analysis tools that are correct by design.
Of course ---due to well-known undecidability results---
any fully automatic tool can only provide partial, though safe answers.

\subsection{Abstract Domains for Numeric and Boolean Values}
\label{sec:basic-abstract-domains}

The reachable state abstraction described above is just one of the
possible semantic approximations that can be adopted when specifying
an abstract semantics. A further, typical approximation concerns the
description of the states of the transition system.  Each state
$\sigma \in \Sigma$ may be decomposed into, e.g., a set of numerical or
Boolean variables that are of interest for the application at hand;
new abstract domains can be defined (and composed~\cite{CousotC79}) so
as to soundly describe the possible values of these variables.

\ifthenelse{\boolean{TR}}{%%
As an expository example that will be also used in the following sections,
}{%%
As an expository example,
}%%\ifthenelse{\boolean{TR}}
assume that part of a state is characterized by
the value of an integer variable.
Then, the domain
$\bigl( \wp(\Sigma), \sseq, \emptyset, \Sigma, \union, \inters \bigr)$
can be abstracted to the concrete domain of integers
\(
  \bigl(\wp(\Int), \sseq, \emptyset, \Int, \union, \inters\bigr)
\).
This domain is further approximated by an abstract domain
\(
  \bigl(\Int^\sharp, \sqsubseteq, \bot, \sqcup\bigr)
\),
via the concretization function
$\fund{\gamma_\mathrm{I}}{\Int^\sharp}{\wp(\Int)}$.
Elements of $\Int^\sharp$ are denoted by $m^\sharp$,
possibly subscripted.
We assume that the partial abstraction function
$\pard{\alpha_\mathrm{I}}{\wp(\Int)}{\Int^\sharp}$ is defined
on all singletons $\{m\} \in \wp(\Int)$ and on the whole set $\Int$.
We also assume that there are abstract binary operations
`$\absplus$', `$\absminus$' and `$\absprod$'
on $\Int^\sharp$ that are sound with respect to the corresponding
operations on $\wp(\Int)$ which, in turn, are the obvious pointwise
extensions of addition, subtraction and multiplication over the integers.
More formally, for `$\absplus$'
we require
\(
  \gamma_\mathrm{I}(m^\sharp_0 \absplus m^\sharp_1)
    \supseteq
      \bigl\{\,
        m_0 + m_1
      \bigm|
        m_0 \in \gamma_\mathrm{I}(m^\sharp_0),
        m_1 \in \gamma_\mathrm{I}(m^\sharp_1)
      \,\bigr\}
\)
for each $m^\sharp_0, m^\sharp_1 \in \Int^\sharp$,
i.e., soundness with respect to addition.
Similar requirements are imposed on `$\absminus$' and `$\absprod$'.
Even though the definition of $\Int^\sharp$ is completely general,
families of integer intervals come naturally to mind for this role.

Suppose now that some other part of the state is characterized by
the value of a Boolean expression.
Then, the domain
$\bigl( \wp(\Sigma), \sseq, \emptyset, \Sigma, \union, \inters \bigr)$
can be abstracted to the finite domain
\(
  \bigl(
    \wp(\Bool), \sseq, \emptyset, \Bool, \union, \inters
  \bigr)
\),
where $\Bool = \{ \ffv, \ttv \}$ is the set of Boolean values.
In general, such a finite domain may be further approximated by
an abstract domain
\(
  (\Bool^\sharp, \sqsseq, \bot, \top, \sqcup, \sqcap)
\),
related to the concrete domain by a Galois connection.
Elements of $\Bool^\sharp$ are denoted by $t^\sharp$, possibly subscripted,
and we can define abstract operations
`$\absneg$', `$\absor$' and `$\absand$' on $\Bool^\sharp$
that are sound with respect to the pointwise extensions
of Boolean negation, disjunction and conjunction over $\wp(\Bool)$.
For instance, for the operation `$\absor$' to be sound with respect
to disjunction on $\wp(\Bool)$, it is required that,
\(
  \gamma_\mathrm{B}(t^\sharp_0 \absor t^\sharp_1)
    \supseteq
      \bigl\{\,
        t_0 \lor t_1
      \bigm|
        t_0 \in \gamma_\mathrm{B}(t^\sharp_0),
        t_1 \in \gamma_\mathrm{B}(t^\sharp_1)
      \,\bigr\}
\)
for each $t^\sharp_0$ and $t^\sharp_1$ in $\Bool^\sharp$.
Likewise for `$\absand$'.
For `$\absneg$' the correctness requirement is that,
for each $t^\sharp$ in $\Bool^\sharp$,
\(
  \gamma_\mathrm{B}(\absneg t^\sharp)
    \supseteq
      \bigl\{\,
        \neg t
      \bigm|
        t \in \gamma_\mathrm{B}(t^\sharp)
      \,\bigr\}
\).
Abstract comparison operations
$\fund{\abseq,\abslt}{\Int^\sharp\times\Int^\sharp}{\Bool^\sharp}$
can then be defined to correctly approximate the equal-to and less-than
tests:
for each $m^\sharp_0, m^\sharp_1 \in \Int^\sharp$,
\(
  \gamma_\mathrm{B}(m^\sharp_0 \abseq m^\sharp_1)
    \supseteq
      \bigl\{\,
        m_0 = m_1
      \bigm|
        m_0 \in \gamma_\mathrm{I}(m^\sharp_0),
        m_1 \in \gamma_\mathrm{I}(m^\sharp_1)
      \,\bigr\}
\);
likewise for `$\abslt$'.

Simple abstract domains such as the ones above can be combined in different
ways so as to obtain quite accurate approximations~\cite{CousotC79}.
In some cases, however, the required precision level may only be obtained
by a suitable initial choice of the abstract domain.
As a notable example, suppose that some part of the state $\sigma \in \Sigma$
is characterized by $n$ (integer or real valued) numeric variables and
the application at hand needs some \emph{relational} information about
these variables. In such a context, an approximation based on a simple
conjunctive combination of $n$ copies of the domain $\Int^\sharp$
described above will be almost useless. Rather, a new approximation
scheme can be devised by modeling states using the domain
\(
  \bigl(
    \wp(\Rset^n), \sseq, \emptyset, \Rset^n, \union, \inters
  \bigr)
\),
where each vector $\vect{v} \in \Rset^n$ is meant to describe a possible
valuation for the $n$ variables. A further abstraction should map this
domain so as to retain some of the relations holding between the values
of the $n$ variables.
If a finite set of linear inequalities provides a good enough approximation,
then the natural choice is to abstract this domain
into the abstract domain of convex polyhedra
\(
  (\Pset_n, \sseq, \emptyset, \Rset^n, \polyhull, \inters)
\)
\cite{CousotH78}.
In this case, the concrete and abstract domains are not related
by a Galois connection and, hence, a best approximation might not exist.%%
\footnote{This happens, for instance, when approximating an $n$-dimensional
ball with a convex polyhedron.}
Nonetheless, the convex polyhedral hull (partial) abstraction function
$\pard{\polyhull}{\wp(\Rset^n)}{\Pset_n}$
is defined in most of the cases of interest
and provides the best possible approximation.
Most of the arithmetic operations seen before can be encoded
(or approximated) by computing images of affine relations.

\subsection{Widening Operators}

It should be stressed that, in general, the abstract semantics
just described is not finitely computable. For instance,
both the domain of convex polyhedra and the domain of integer intervals
have infinite ascending chains, so that the limit of a converging
fixpoint computation cannot generally be reached in a finite number
of iterations.

A finite computation can be enforced by further approximations
resulting in a \emph{Noetherian} abstract domain,
i.e., a domain where all ascending chains are finite.
Alternatively, and more generally, it is possible
to keep an abstract domain with infinite chains, while enforcing
that these chains are traversed in a finite number of
\ifthenelse{\boolean{TR}}{%%
iteration steps \cite{CousotC92plilp}.
}{%%
iteration steps.
}%%\ifthenelse{\boolean{TR}}{%%
In both cases, termination is usually achieved to the detriment
of precision, so that an appropriate trade-off should be pursued.
\emph{Widening operators}
\ifthenelse{\boolean{TR}}{%%
\cite{CousotC76,CousotC77,CousotC92fr,CousotC92plilp}
}{%%
\cite{CousotC76,CousotC92fr}
}%\ifthenelse{\boolean{TR}}
provide a simple and general characterization for the second option.

\begin{definition}
\label{def:widening-operator}
The partial operator $\pard{\widen}{D^\sharp \times D^\sharp}{D^\sharp}$
is a \emph{widening} if:
\begin{enumerate}
\item
for all $d^\sharp, e^\sharp \in D^\sharp$,
$d^\sharp \sqsubseteq e^\sharp$ implies that
$d^\sharp \widen e^\sharp$ is defined and
$e^\sharp \sqsubseteq d^\sharp \widen e^\sharp$;
\item
for all increasing chains
$e^\sharp_0 \sqsubseteq e^\sharp_1 \sqsubseteq \cdots$,
the increasing chain defined by
$d^\sharp_0 \defeq e^\sharp_0$ and
\(
  d^\sharp_{i+1}
    \defeq
      d^\sharp_i \widen (d^\sharp_i \sqcup e^\sharp_{i+1})
\),
for $i \in \Nset$, is not strictly increasing.
\end{enumerate}
\end{definition}

It can be proved that,
for any monotonic operator $\fund{\cA}{D^\sharp}{D^\sharp}$,
the upward iteration sequence with widenings
starting at the bottom element $d^\sharp_0 \defeq \bot$
and defined by
\begin{equation*}
  d^\sharp_{i+1}
    \defeq
      \begin{cases}
        d^\sharp_i,
          &\text{if $\cA(d^\sharp_i) \sqsubseteq d^\sharp_i$,} \\
	d^\sharp_i
          \widen
            \bigl(d^\sharp_i \sqcup \cA(d^\sharp_i)\bigr),
          &\text{otherwise,}
      \end{cases}
\end{equation*}
converges to a post-fixpoint of $\mathord{\cA}$
\ifthenelse{\boolean{TR}}{%%
after a finite number of iterations \cite{CousotC92plilp}.
}{%%
after a finite number of iterations.
}%%\ifthenelse{\boolean{TR}}{%%
Clearly, the choice of the widening has a deep impact on the precision
of the results obtained. Designing a widening which is appropriate
for a given application is therefore a difficult (but possibly rewarding)
activity.

\section{Analysis and Verification of Computer Programs}
\label{sec:analysis-of-computer-programs}

In this section we begin a review of the applications of polyhedral
computations to analysis and verification problems starting with the
\ifthenelse{\boolean{TR}}{%%
the work of Cousot and Halbwachs \cite{CousotH78,Halbwachs79th}.
These seminal papers
}{%%
the work of Cousot and Halbwachs \cite{CousotH78}.
This seminal paper
}%% \ifthenelse{\boolean{TR}}
on the automatic inference of linear invariants for
imperative programs constituted a major leap forward for at least two reasons.
First, the polyhedral domain proposed by Cousot and Halbwachs
was considerably more powerful than all the data-flow analyses known
at that time, including the rather sophisticated one by Karr
\ifthenelse{\boolean{TR}}{%%
which was limited to linear equalities \cite{Karr76,Muller-OlmS04ICALP}.
}{%%
which was limited to linear equalities \cite{Karr76}.
}%% \ifthenelse{\boolean{TR}}
Secondly, the use of convex polyhedra as an abstract domain established
abstract interpretation as the right methodology for the definition
of complex and correct program analyzers.

We illustrate the basic ideas by partially specifying the analysis
of linear invariants for a very simple imperative language.
The simplicity of the language we have chosen for expository purposes
should not mislead the reader: the approach is generalizable to any
imperative (and, for that matter, functional and logic) language
\cite{BagnaraHPZ07TR}.
The abstract syntax of the language is presented in
Figure~\ref{fig:IMP-syntax}.
\begin{figure}
\ifthenelse{\boolean{TR}}{%%
\begin{description}
\item[Integers]
$m \in \Int \defeq \Zset$
\item[Booleans]
$t \in \Bool \defeq \{ \ttv, \ffv \}$
\item[Variables]
$x \in \Var \defeq \{ x_0, x_1, x_2, \ldots \}$
\end{description}
\begin{description}
\item[Arithmetic expressions]
\[
  \Aexp \ni
  a ::= m \vbar x \vbar a_0 + a_1 \vbar a_0 - a_1 \vbar a_0 * a_1
\]
\item[Boolean expressions]
\[
  \Bexp \ni
  b ::= t \vbar a_0 = a_1 \vbar a_0 < a_1
\]
\item[Statements]
\begin{align*}
  \Stmt \ni
  s ::= \kw{skip} \vbar x := a \vbar s_0 ; s_1
     \vbar \kw{if} b \kw{then} s_0 \kw{else} s_1
     \vbar \kw{while} b \kw{do} s
\end{align*}
\end{description}
}{%%
\begin{gather*}
  m \in \Int \defeq \Zset
\qquad
  t \in \Bool \defeq \{ \ttv, \ffv \}
\qquad
  x \in \Var \defeq \{ x_0, x_1, x_2, \ldots \} \\
\begin{aligned}
  \Aexp \ni
  a &::= m \vbar x \vbar a_0 + a_1 \vbar a_0 - a_1 \vbar a_0 * a_1 \\
  \Bexp \ni
  b &::= t \vbar a_0 = a_1 \vbar a_0 < a_1 \\
  \Stmt \ni
  s &::= \kw{skip} \vbar x := a \vbar s_0 ; s_1
     \vbar \kw{if} b \kw{then} s_0 \kw{else} s_1
     \vbar \kw{while} b \kw{do} s
\end{aligned}
\end{gather*}
}%% \ifthenelse{\boolean{TR}}{%%
\caption{Abstract syntax of the simple imperative language}
\label{fig:IMP-syntax}
\end{figure}
The basic syntactic categories, corresponding to the sets $\Int$, $\Bool$
and $\Var$, are defined directly.
From these, the categories of arithmetic and Boolean expressions
and of statements are defined by means of BNF rules.
Notice the use of syntactic meta-variables: for instance,
to save typing we will consistently denote by $s$, possibly subscripted
or superscripted, any element of $\Stmt$.

The concrete semantics of programs is formally defined using the
\emph{natural semantics} approach \cite{Kahn87}.
This, in turn, is a ``big-step'' operational semantics defined by structural
induction on program structures in the style of Plotkin \cite{Plotkin81}.
First we define the notion of \emph{store}, which is any mapping
between a finite set of variables and elements of $\Int$.
Formally, a store is an element of the set
\ifthenelse{\boolean{TR}}{%%
\[
}{%%
\(
}%%\ifthenelse{\boolean{TR}}
  \Store
    \defeq
      \{\,
        \fund{\sigma}{V}{\Int}
      \mid
        V \sseq \Var, \,\text{$V$ finite}
      \,\}
\ifthenelse{\boolean{TR}}{%%
\]
}{%%
\)
}%%\ifthenelse{\boolean{TR}}
and denoted by the letter $\sigma$,
possibly subscripted or superscripted.
The store obtained from $\sigma \in \Store$ by the assignment
of $m \in \Int$ to $x \in \dom(\sigma)$, denoted by $\sigma[m/x]$,
\ifthenelse{\boolean{TR}}{%%
is defined as follows, for each $x' \in \dom(\sigma)$:%
\[
  \sigma[m/x](x')
    \defeq
     \begin{cases}
        m,           &\text{if $x' = x$;} \\
        \sigma(x'),  &\text{if $x' \neq x$.}
     \end{cases}
\]
}{%%
is defined so that,
for each $x' \in \dom(\sigma)$,
$\sigma[m/x](x') = m$, if $x' = x$, and
$\sigma[m/x](x') = \sigma(x')$, otherwise.
}%%\ifthenelse{\boolean{TR}}{%%

The concrete evaluation relations that complete the definition of the
concrete semantics for our simple language are defined by structural
induction from a set of rule schemata.
The evaluation relations for terminating computations are given by
$\reld{\aceval}{(\Aexp \times \Store)}{\Int}$,
for arithmetic expressions,
$\reld{\bceval}{(\Bexp \times \Store)}{\Bool}$,
for Boolean expressions, and
$\reld{\sceval}{(\Stmt \times \Store)}{\Store}$,
for statements.
The judgment $\langle a, \sigma \rangle \aceval m$ means that
when expression $a$ is executed in store $\sigma$ it results in
the integer $m$.
The judgment $\langle b, \sigma \rangle \bceval t$ is
similar.  Note that expressions do not have, in our simple language,
side effects.
The judgment $\langle s, \sigma \rangle \sceval \sigma'$ means that
the statement $s$, executed in store $\sigma$, results in a (possibly
modified) store $\sigma'$.
The rule schemata, in the form $\frac{\text{premise}}{\text{conclusion}}$,
that define these relations are given
in Figure~\ref{fig:IMP-concrete-semantics}.
\begin{figure}
\begin{gather*}
\prooftree
  \nohyp
\justifies
  \langle m, \sigma \rangle
    \aceval
      m
\endprooftree
\quad\hfill
\prooftree
  \nohyp
\justifies
  \langle x, \sigma \rangle
    \aceval
      \sigma(x)
\endprooftree
\quad\hfill
\prooftree
  \langle a_0, \sigma \rangle
    \aceval
      m_0
\quad
  \langle a_1, \sigma \rangle
    \aceval
      m_1
\justifies
  \langle a_0 + a_1, \sigma \rangle
    \aceval
      m_0 + m_1
\endprooftree \\[2ex]
\prooftree
  \langle a_0, \sigma \rangle
    \aceval
      m_0
\quad
  \langle a_1, \sigma \rangle
    \aceval
      m_1
\justifies
  \langle a_0 - a_1, \sigma \rangle
    \aceval
      m_0 - m_1
\endprooftree
\quad
\prooftree
  \langle a_0, \sigma \rangle
    \aceval
      m_0
\quad
  \langle a_1, \sigma \rangle
    \aceval
      m_1
\justifies
  \langle a_0 * a_1, \sigma \rangle
    \aceval
      m_0 \cdot m_1
\endprooftree \\[2ex]
\prooftree
  \langle a_0, \sigma \rangle
    \aceval
      m_0
\quad
  \langle a_1, \sigma \rangle
    \aceval
      m_1
\justifies
  \langle a_0 = a_1, \sigma \rangle
    \bceval
      (m_0 = m_1)
\endprooftree
\quad
\prooftree
  \langle a_0, \sigma \rangle
    \aceval
      m_0
\quad
  \langle a_1, \sigma \rangle
    \aceval
      m_1
\justifies
  \langle a_0 < a_1, \sigma \rangle
    \bceval
      (m_0 < m_1)
\endprooftree \\[2ex]
\prooftree
  \nohyp
\justifies
  \langle \kw{skip}, \sigma \rangle
    \sceval
      \sigma
\endprooftree
\quad
\prooftree
  \langle a, \sigma \rangle
    \aceval
      m
\justifies
  \langle x := a, \sigma \rangle
    \sceval
      \sigma[m/x]
\endprooftree
\quad
\prooftree
  \langle s_0, \sigma \rangle
    \sceval
      \sigma''
\quad
  \langle s_1, \sigma'' \rangle
    \sceval
      \sigma'
\justifies
  \langle s_0 ; s_1, \sigma \rangle
    \sceval
      \sigma'
\endprooftree \\[2ex]
\prooftree
  \langle b, \sigma \rangle
    \bceval
      \ttv
\quad
  \langle s_0, \sigma \rangle
    \sceval
      \sigma'
\justifies
  \langle \kw{if} b \kw{then} s_0 \kw{else} s_1, \sigma \rangle
    \sceval
      \sigma'
\endprooftree
\qquad\qquad\qquad
\prooftree
  \langle b, \sigma \rangle
    \bceval
      \ffv
\quad
  \langle s_1, \sigma \rangle
    \sceval
      \sigma'
\justifies
  \langle \kw{if} b \kw{then} s_0 \kw{else} s_1, \sigma \rangle
    \sceval
      \sigma'
\endprooftree \\[2ex]
\prooftree
  \langle b, \sigma \rangle
    \bceval
      \ffv
\justifies
  \langle \kw{while} b \kw{do} c, \sigma \rangle
    \sceval
      \sigma
\endprooftree
\quad
\prooftree
  \langle b, \sigma \rangle
    \bceval
      \ttv
\quad
  \langle c, \sigma \rangle
    \sceval
      \sigma''
\quad
  \langle \kw{while} b \kw{do} c, \sigma'' \rangle
    \sceval
      \sigma'
\justifies
  \langle \kw{while} b \kw{do} c, \sigma \rangle
    \sceval
      \sigma'
\endprooftree
\end{gather*}
\caption{Concrete semantics rule schemata for the finite computations
of the simple imperative language}
\label{fig:IMP-concrete-semantics}
\end{figure}
Rule instances can be composed in the obvious way to form finite
tree structures, representing finite computations.
\ifthenelse{\boolean{TR}}{%%
Figure~\ref{fig:example-of-concrete-tree} shows one such tree.

\begin{sidewaysfigure}
\[
\prooftree
  \prooftree
    \prooftree
      \nohyp
    \justifies
      \langle 0, \sigma_0 \rangle \aceval 0
    \endprooftree
    %\;
    \prooftree
      \nohyp
    \justifies
      \langle x_0, \sigma_0 \rangle \aceval 1
    \endprooftree
  \justifies
    \langle 0 < x_0, \sigma_0 \rangle
      \bceval
        \ttv
  \endprooftree
%\;
  \prooftree
    \prooftree
      \prooftree
        \prooftree
          \nohyp
        \justifies
          \langle x_1, \sigma_0 \rangle
            \aceval
              1
        \endprooftree
      %\;
        \prooftree
          \nohyp
        \justifies
          \langle 2, \sigma_0 \rangle
            \aceval
              2
        \endprooftree
      \justifies
        \langle x_1+2, \sigma_0 \rangle
          \aceval
            3
      \endprooftree
    \justifies
      \langle x_1 := x_1+2, \sigma_0 \rangle
        \sceval
          \sigma_1
    \endprooftree
  %\;
    \prooftree
      \prooftree
        \prooftree
          \nohyp
        \justifies
          \langle x_0, \sigma_1 \rangle
            \aceval
              1
        \endprooftree
      %\;
        \prooftree
          \nohyp
        \justifies
          \langle x_1, \sigma_1 \rangle
            \aceval
              3
        \endprooftree
      \justifies
        \langle x_0-x_1, \sigma_1 \rangle
          \aceval
            -2
      \endprooftree
    \justifies
      \langle x_0 := x_0-x_1, \sigma_1 \rangle
        \sceval
          \sigma_2
    \endprooftree
  \justifies
    \bigl\langle (x_1 := x_1+2; x_0 := x_0-x_1), \sigma_0 \bigr\rangle
      \sceval
        \sigma_2
  \endprooftree
%\;
  \prooftree
    \prooftree
      \prooftree
        \nohyp
      \justifies
        \langle 0, \sigma_2 \rangle \aceval 0
      \endprooftree
      %\;
      \prooftree
        \nohyp
      \justifies
        \langle x_0, \sigma_2 \rangle \aceval -2
      \endprooftree
    \justifies
      \langle 0 < x_0, \sigma_2 \rangle
        \bceval
          \ffv
    \endprooftree
  \justifies
    \langle w, \sigma_2 \rangle
      \sceval
        \sigma_2
  \endprooftree
\justifies
  \bigl\langle
    \kw{while}\, 0 < x_0\, \kw{do}\, (x_1 := x_1+2; x_0 := x_0-x_1),
    \sigma_0
  \bigr\rangle
    \sceval
      \sigma_2
\endprooftree
\]

\medskip
\noindent
Legend:
\begin{align*}
  \sigma_0 &\defeq \bigl\{ (x_0, 1), (x_1, 1) \bigr\}, \\
  \sigma_1 &\defeq \bigl\{ (x_0, 1), (x_1, 3) \bigr\}, \\
  \sigma_2 &\defeq \bigl\{ (x_0, -2), (x_1, 3) \bigr\}, \\
  w        &\defeq \bigl(\kw{while}\, 0 < x_0\, \kw{do}\, (x_1 := x_1+2; x_0 := x_0-x_1)\bigr).
\end{align*}
\caption{The tree representing a concrete execution of a program}
\label{fig:example-of-concrete-tree}
\end{sidewaysfigure}
}{}%%\ifthenelse{\boolean{TR}}

The possibly infinite set of all finite trees is obtained by means
of a least fixpoint computation, corresponding to the classical
inductive interpretation of the rules
in Figure~\ref{fig:IMP-concrete-semantics}.
The rule schemata in Figure~\ref{fig:IMP-concrete-semantics-divergence}
can be used to directly model non-terminating computations and need to
\ifthenelse{\boolean{TR}}{%%
be interpreted coinductively \cite{CousotC92,Leroy06,Schmidt98}.
}{%%
be interpreted coinductively \cite{CousotC92}.
}%%\ifthenelse{\boolean{TR}}
\begin{figure}
\begin{gather*}
\prooftree
  \langle s_0, \sigma \rangle
    \diverges
\justifies
  \langle s_0 ; s_1, \sigma \rangle
    \diverges
\endprooftree
\qquad\qquad\qquad
\prooftree
  \langle s_0, \sigma \rangle
    \sceval
      \sigma'
\quad
  \langle s_1, \sigma' \rangle
    \diverges
\justifies
  \langle s_0 ; s_1, \sigma \rangle
    \diverges
\endprooftree \\[2ex]
\prooftree
  \langle b, \sigma \rangle
    \bceval
      \ttv
\quad
  \langle s_0, \sigma \rangle
    \diverges
\justifies
  \langle \kw{if} b \kw{then} s_0 \kw{else} s_1, \sigma \rangle
    \diverges
\endprooftree
\qquad\qquad\qquad
\prooftree
  \langle b, \sigma \rangle
    \bceval
      \ffv
\quad
  \langle s_1, \sigma \rangle
    \diverges
\justifies
  \langle \kw{if} b \kw{then} s_0 \kw{else} s_1, \sigma \rangle
    \diverges
\endprooftree \\[2ex]
\prooftree
  \langle b, \sigma \rangle
    \bceval
      \ttv
\quad
  \langle c, \sigma \rangle
    \diverges
\justifies
  \langle \kw{while} b \kw{do} c, \sigma \rangle
    \diverges
\endprooftree
\quad
\prooftree
  \langle b, \sigma \rangle
    \bceval
      \ttv
\quad
  \langle c, \sigma \rangle
    \sceval
      \sigma'
\quad
  \langle \kw{while} b \kw{do} c, \sigma' \rangle
    \diverges
\justifies
  \langle \kw{while} b \kw{do} c, \sigma \rangle
    \diverges
\endprooftree
\end{gather*}
\caption{Additional concrete semantics rule schemata
for the infinite computations of the simple imperative language}
\label{fig:IMP-concrete-semantics-divergence}
\end{figure}
The judgment $\langle s, \sigma \rangle \diverges$ means that
the statement $s$ diverges when executed in store $\sigma$.
By a suitable adaptation of the computational ordering,
both sets of finite and infinite trees can be jointly computed
in a single least fixpoint computation
\ifthenelse{\boolean{TR}}{%%
\cite{CousotC92,Leroy06,Schmidt98}.
}{%%
\cite{CousotC92}.
}%%\ifthenelse{\boolean{TR}}{%%
While these semantics characterizations contain all the information we need
to perform a wide range of program reasoning tasks, they are generally
not computable: we have thus to resort to approximation.

Following the abstract interpretation approach, as instantiated
\ifthenelse{\boolean{TR}}{%%
in \cite{Schmidt95,Schmidt97,Schmidt98},
}{%%
in \cite{Schmidt95},
}%%\ifthenelse{\boolean{TR}}{%%
the concrete rule schemata
are paired with abstract rule schemata that correctly approximate them.
Before doing that, we need to formalize abstract domains for each concrete
domain used by the concrete semantics.

For simple approximations of integers and Boolean expressions,
we consider the abstract domains $\Int^\sharp$ and $\Bool^\sharp$
introduced in Section~\ref{sec:basic-abstract-domains}.
The last (and most interesting) abstraction we need is one that approximates
sets of stores.  We thus require an abstract domain
\(
  \bigl(\Store^\sharp, \sqsubseteq, \bot, \sqcup \bigr)
\)
that is related, by means of a concretization function
$\gamma_\mathrm{S}$ such that $\gamma_\mathrm{S}(\bot) = \emptyset$,
to the concrete domain
\(
  \bigl(\wp(\Store), \sseq, \emptyset, \Store, \union, \inters \bigr)
\).
Elements of $\Store^\sharp$ are denoted by
$\sigma^\sharp$, possibly subscripted.
The abstract store evaluation and update operators
\begin{gather*}
  \fund{\cdot[\cdot]}%
       {(\Store^\sharp \times \Aexp)}%
       {\Int^\sharp}, \\
  \fund{\cdot[\cdot := \cdot]}%
       {\bigl(\Store^\sharp \times \Var \times \Aexp\bigr)}%
       {\Store^\sharp}, \\
  \fund{\cdot[\cdot / \cdot]}%
       {\bigl(\Store^\sharp \times \Var \times \Int^\sharp\bigr)}%
       {\Store^\sharp}
\end{gather*}
are assumed to be sound with respect to their concrete counterparts,
i.e., such that,
for each $\sigma^\sharp \in \Store^\sharp$, $a \in \Aexp$,
$x \in \Var$ and $m^\sharp \in \Int^\sharp$:
\begin{align*}
  \gamma_\mathrm{I}\bigl(\sigma^\sharp[a]\bigr)
    &\supseteq
      \bigl\{\,
        m \in \Int
      \bigm|
        \sigma \in \gamma_\mathrm{S}(\sigma^\sharp),
        \langle a, \sigma \rangle \aceval m
      \,\bigr\}, \\
  \gamma_\mathrm{S}\bigl(\sigma^\sharp\bigl[x := a]\bigr)
    &\supseteq
      \bigl\{\,
        \sigma' \in \Store
      \bigm|
        \sigma \in \gamma_\mathrm{S}(\sigma^\sharp),
        \langle x := a, \sigma \rangle \sceval \sigma'
      \,\bigr\}, \\
  \gamma_\mathrm{S}\bigl(\sigma^\sharp\bigl[m^\sharp/x]\bigr)
    &\supseteq
      \bigl\{\,
        \sigma[m/x] \in \Store
      \bigm|
        \sigma \in \gamma_\mathrm{S}(\sigma^\sharp),
        m \in \gamma_\mathrm{I}(m^\sharp)
      \,\bigr\}.
\end{align*}
We also need computable ``Boolean filters'' to refine the information
contained in abstract stores, i.e., two functions
$\fund{\phi_\ttv,\phi_\ffv}{\Store^\sharp \times \Bexp}{\Store^\sharp}$
such that,
\ifthenelse{\boolean{TR}}{%%
for each $\sigma^\sharp \in \Store^\sharp$ and $b \in \Bexp$,
\begin{align*}
  \gamma_\mathrm{S}\bigl(\phi_\ttv(\sigma^\sharp, b)\bigr)
    &\supseteq
      \bigl\{\,
        \sigma \in \gamma_\mathrm{S}(\sigma^\sharp)
      \bigm|
        \langle b, \sigma \rangle \bceval \ttv
      \,\bigr\}, \\
  \gamma_\mathrm{S}\bigl(\phi_\ffv(\sigma^\sharp, b)\bigr)
    &\supseteq
      \bigl\{\,
        \sigma \in \gamma_\mathrm{S}(\sigma^\sharp)
      \bigm|
        \langle b, \sigma \rangle \bceval \ffv
      \,\bigr\}.
\end{align*}
}{%%
for each $t \in \Bool$, $\sigma^\sharp \in \Store^\sharp$ and $b \in \Bexp$:
\[
  \gamma_\mathrm{S}\bigl(\phi_t(\sigma^\sharp, b)\bigr)
    \supseteq
      \bigl\{\,
        \sigma \in \gamma_\mathrm{S}(\sigma^\sharp)
      \bigm|
        \langle b, \sigma \rangle \bceval t
      \,\bigr\}.
\]
}%%\ifthenelse{\boolean{TR}}{%%
We are now in a position to present,
in Figure~\ref{fig:IMP-abstract-semantics},
a possible set of domain-independent abstract rule schemata.
\begin{figure}
\begin{gather*}
\prooftree
  \nohyp
\justifies
  \langle m, \sigma^\sharp \rangle
    \aaeval
      \alpha_\mathrm{I}\bigl(\{m\}\bigr)
\endprooftree
\quad\hfill
\prooftree
  \nohyp
\justifies
  \langle x, \sigma^\sharp \rangle
    \aaeval
      \sigma^\sharp[x]
\endprooftree
\quad\hfill
\prooftree
  \langle a_0, \sigma^\sharp \rangle
    \aaeval
      m^\sharp_0
\quad
  \langle a_1, \sigma^\sharp \rangle
    \aaeval
      m^\sharp_1
\justifies
  \langle a_0 + a_1, \sigma^\sharp \rangle
    \aaeval
      m^\sharp_0 \absplus m^\sharp_1
\endprooftree \\[2ex]
\prooftree
  \langle a_0, \sigma^\sharp \rangle
    \aaeval
      m^\sharp_0
\quad
  \langle a_1, \sigma^\sharp \rangle
    \aaeval
      m^\sharp_1
\justifies
  \langle a_0 - a_1, \sigma^\sharp \rangle
    \aaeval
      m^\sharp_0 \absminus m^\sharp_1
\endprooftree
\quad\hfill
\prooftree
  \langle a_0, \sigma^\sharp \rangle
    \aaeval
      m^\sharp_0
\quad
  \langle a_1, \sigma^\sharp \rangle
    \aaeval
      m^\sharp_1
\justifies
  \langle a_0 * a_1, \sigma^\sharp \rangle
    \aaeval
      m^\sharp_0 \absprod m^\sharp_1
\endprooftree \\[2ex]
\prooftree
  \nohyp
\justifies
  \langle t, \sigma^\sharp \rangle
    \baeval
      \alpha_\mathrm{B}\bigl(\{t\}\bigr)
\endprooftree
\quad\hfill
\prooftree
  \langle a_0, \sigma^\sharp \rangle
    \aaeval
      m^\sharp_0
\quad
  \langle a_1, \sigma^\sharp \rangle
    \aaeval
      m^\sharp_1
\justifies
  \langle a_0 = a_1, \sigma^\sharp \rangle
    \baeval
      m^\sharp_0 \abseq m^\sharp_1
\endprooftree \\[2ex]
\prooftree
  \langle a_0, \sigma^\sharp \rangle
    \aaeval
      m^\sharp_0
\quad
  \langle a_1, \sigma^\sharp \rangle
    \aaeval
      m^\sharp_1
\justifies
  \langle a_0 < a_1, \sigma^\sharp \rangle
    \baeval
      m^\sharp_0 \abslt m^\sharp_1
\endprooftree
\quad\hfill
\prooftree
  \nohyp
\justifies
  \langle \kw{skip}, \sigma^\sharp \rangle
    \saeval
      \sigma^\sharp
\endprooftree \\[2ex]
\prooftree
  \langle a, \sigma^\sharp \rangle
    \aaeval
      m^\sharp
\using\;\text{(i)}
\justifies
  \langle x := a, \sigma^\sharp \rangle
    \saeval
      \sigma^\sharp[x := a]
\endprooftree
\quad\hfill
\prooftree
  \langle a, \sigma^\sharp \rangle
    \aaeval
      m^\sharp
\using\;\text{(ii)}
\justifies
  \langle x := a, \sigma^\sharp \rangle
    \saeval
      \sigma^\sharp[m^\sharp/x]
\endprooftree \\
\prooftree
  \langle s_0, \sigma^\sharp_0 \rangle
    \saeval
      \sigma^\sharp_1
\quad
  \langle s_1, \sigma^\sharp_1 \rangle
    \saeval
      \sigma^\sharp_2
\justifies
  \langle s_0 ; s_1, \sigma^\sharp_0 \rangle
    \saeval
      \sigma^\sharp_2
\endprooftree \\[2ex]
\prooftree
  \langle b, \sigma^\sharp \rangle
    \baeval
      t^\sharp
\quad
  \bigl\langle s_0, \phi_\ttv(\sigma^\sharp, b) \bigr\rangle
    \saeval
      \sigma^\sharp_0
\quad
  \bigl\langle s_1, \phi_\ffv(\sigma^\sharp, b) \bigr\rangle
    \saeval
      \sigma^\sharp_1
\justifies
  \langle \kw{if} b \kw{then} s_0 \kw{else} s_1, \sigma^\sharp \rangle
    \saeval
      \sigma^\sharp_0 \sqcup \sigma^\sharp_1
\endprooftree \\[2ex]
\prooftree
  \langle b, \sigma^\sharp \rangle
    \baeval
      t^\sharp
\quad
  \bigl\langle c, \phi_\ttv(\sigma^\sharp, b) \bigr\rangle
    \saeval
      \sigma^\sharp_1
\quad
  \langle \kw{while} b \kw{do} c, \sigma^\sharp_1 \rangle
    \saeval
      \sigma^\sharp_2
\justifies
  \langle \kw{while} b \kw{do} c, \sigma^\sharp \rangle
    \saeval
      \phi_\ffv(\sigma^\sharp, b) \sqcup \sigma^\sharp_2
\endprooftree
\end{gather*}

\begin{quote}
{
\footnotesize
\begin{itemize}
\item[Notes:]
\item[(i)]  This rule is used if the domain $\Store^\sharp$ can
            capture the assignment precisely (e.g., when
            $\Store^\sharp$ is a domain of convex polyhedra and $a$ is
            an affine expression). Notice that the premise is intentionally
            not used: its presence is required in order to ensure that the
            abstract tree approximates the concrete tree in its entirety.
\item[(ii)] This rule is used when (i) is not applicable.
\end{itemize}
}
\end{quote}
\caption{Abstract semantics rule schemata for the simple imperative language}
\label{fig:IMP-abstract-semantics}
\end{figure}
These schemata allow for the free approximation of the `$\rightsquigarrow$'
right-hand sides in the conclusions.  This means that if, e.g.,
\ifthenelse{\boolean{TR}}{%%
\[
\prooftree
  \text{premise}
\justifies
  \langle s, \sigma \rangle \saeval \sigma^\sharp_1
\endprooftree
\]
is an instance of some rule, then
\[
\prooftree
  \text{premise}
\justifies
  \langle s, \sigma \rangle \saeval \sigma^\sharp_2
\endprooftree
\]
}{%%
$\frac{\text{premise}}{\langle s, \sigma \rangle \saeval \sigma^\sharp_1}$
is an instance of some rule, then
$\frac{\text{premise}}{\langle s, \sigma \rangle \saeval \sigma^\sharp_2}$
}%%\ifthenelse{\boolean{TR}}{%%
is also an instance of the same rule for each $\sigma^\sharp_2$
such that $\sigma^\sharp_1 \sqsseq \sigma^\sharp_2$.
Hence the schemata in Figure~\ref{fig:IMP-abstract-semantics}
ensure correctness yet leaving complete freedom about precision.
The ability to give up some precision, as we will see, is crucial
in order to ensure the (reasonably quick) termination of the analysis.

It is possible to prove that,
for each (possibly infinite) concrete tree $T$
built using the schemata of Figures~\ref{fig:IMP-concrete-semantics}
and~\ref{fig:IMP-concrete-semantics-divergence},
for each (possibly infinite) abstract tree $T^\sharp$
built using the schemata of Figure~\ref{fig:IMP-abstract-semantics},
if the concrete tree root is of the form
$\langle s, \sigma \rangle \sceval \sigma_1$
(when the tree is finite)
or $\langle s, \sigma \rangle \diverges$
(when the tree is infinite)
and the abstract tree root is of the form
$\langle s, \sigma^\sharp \rangle \saeval \sigma^\sharp_1$ with
$\sigma \in \gamma_\mathrm{S}(\sigma^\sharp)$,
then $T^\sharp$ correctly approximates $T$.
This means not only that $\sigma_1 \in \gamma_\mathrm{S}(\sigma^\sharp_1)$
(when $T$ is finite), but also that \emph{each} node in $T$ is correctly
approximated by at least one node in $T^\sharp$.
In other words, the abstract tree correctly approximates the entire
concrete computation (see \cite{BagnaraHPZ07TR} for the details).

\ifthenelse{\boolean{TR}}{%%
It is worth stressing the observation in~\cite{Schmidt98} that,
}{%%
It is worth stressing the observation by Schmidt that,
}%%\ifthenelse{\boolean{TR}}{%%
even when disregarding the non-terminating concrete computations,
the abstract rules still have to be interpreted coinductively
because most of the finite concrete trees can only be approximated
by infinite abstract trees; for instance, all abstract trees containing
a while loop are infinite.
Since, in general, we cannot effectively compute infinite abstract trees,
we still do not have a viable analysis technique.  The solution is to restrict
ourselves to the class of rational trees, i.e., trees with only finitely many
subtrees and that, consequently, admit a finite representation.

The analysis algorithm is sketched in \cite{Schmidt95}.
For expository purposes, we describe here a simplified version that,
however, is enough to handle the considered programming language
features.
The algorithm works by recursively constructing a finite approximation
for the (possibly infinite) abstract subtree rooted in
the current node (initially, the root of the whole tree).
The current node
$n = \bigl(\langle p, \sigma^\sharp_n \rangle \rightsquigarrow r_n\bigr)$,
where $r_n$ is a placeholder for the ``yet to be computed'' conclusion,
is processed according to the following alternatives:
\begin{enumerate}
\item
\label{enum-case:expand}
If no ancestor of $n$ has $p$ in the label, the node has to be
expanded using an applicable abstract rule instance.
Namely, descendants of the premises of the rule are (recursively)
processed, one at a time and from left to right.
When the expansion of all the premises has been completed, including
the case when the rule has no premise at all, the marker $r_n$
is replaced by an abstract value computed according to the conclusion
of the rule.
\item
\label{enum-case:subsumed}
If there exists an ancestor node
$m = \langle p, \sigma^\sharp_m \rangle \rightsquigarrow r_m$ of $n$
labeled by the same syntax $p$ and such that
$\sigma^\sharp_n \sqsubseteq \sigma^\sharp_m$,
i.e., if node $n$ is subsumed by node $m$,
then the node is not expanded further and
the placeholder $r_n$ is replaced by the least fixpoint of the equation
$r_n = f_m(r_n)$, where $f_m$ is the expression corresponding to
the conclusion of the abstract rule that was used for the expansion
of node $m$.%
\footnote{As explained in
\ifthenelse{\boolean{TR}}{%%
\cite{Schmidt95,Schmidt98},
}{%%
\cite{Schmidt95},
}%%\ifthenelse{\boolean{TR}}{%%
the computation of such a \emph{least} fixpoint
(in the context of a coinductive interpretation of the abstract rules)
is justified by the fact that here we only need to approximate
the conclusions produced by the terminating concrete
\ifthenelse{\boolean{TR}}{%%
computations,
i.e., by the concrete rules of Figure~\ref{fig:IMP-concrete-semantics},
which are interpreted inductively. Also note that the divergence rules of
Figure~\ref{fig:IMP-concrete-semantics-divergence} have no conclusion at all.
}{%%
computations.
}%%\ifthenelse{\boolean{TR}}
}
\item
\label{enum-case:widen}
Otherwise, there must be an ancestor node
$m = \langle p, \sigma^\sharp_m \rangle \rightsquigarrow r_m$ of $n$
labeled by the same syntax $p$, but the subsumption condition
$\sigma^\sharp_n \sqsubseteq \sigma^\sharp_m$ does not hold.
Then there are two options:
\begin{enumerate}
\item
\label{enum-case:widen-finite}
if the abstract domain $\Store^\sharp$ is finite,
we proceed as in case~(\ref{enum-case:expand});
\item
\label{enum-case:widen-infinite}
if the abstract domain $\Store^\sharp$ is infinite,
to ensure convergence,
a widening `$\widen$' over $\Store^\sharp$ can be employed%
\TRfootnote{If $\Store^\sharp$ is infinite but Noetherian,
we can choose $\widen \defeq \sqcup$ as a widening.}
and store $\sigma^\sharp_n$ in node $n$ is replaced by
$\sigma^\sharp_m \widen (\sigma^\sharp_m \sqcup \sigma^\sharp_n)$.
Then, we proceed again as in case~(\ref{enum-case:expand}).
\end{enumerate}
\end{enumerate}

The abstract semantics of Figure~\ref{fig:IMP-abstract-semantics}
and the given algorithm for computing a rational abstract tree are
fully generic in that any choice for the abstract domains $\Int^\sharp$,
$\Bool^\sharp$ and $\Store^\sharp$ will result into a provably
correct analysis algorithm.
Focusing on numerical domains, the role of $\Int^\sharp$ can be played
by any domain of intervals, so that the operations
`$\absplus$', `$\absminus$' and `$\absprod$' are the standard
ones of interval arithmetic~\cite{AlefeldH83};
for instance,
\(
  [m^\mathrm{l}_0, m^\mathrm{u}_0] \absplus [m^\mathrm{l}_1,m^\mathrm{u}_1]
    \defeq
      [m^\mathrm{l}_0+m^\mathrm{l}_1, m^\mathrm{u}_0+m^\mathrm{u}_1]
\).
More sophisticated domains, such as \emph{modulo intervals}
\cite{NakanishiJPF99}, are able to encode more precise information
about the set of \emph{integer} values each variable can take.
For $\Store^\sharp$, a common choice is to abstract from the integrality
of variables and consider a domain of convex polyhedra which, in exchange,
allows the tracking of relational information.
With reference to Figure~\ref{fig:IMP-abstract-semantics},
rule (i) can be applied directly when the arithmetic expression
$a = \langle \vect{a}, \vect{x} \rangle + b$ is affine;
the corresponding polyhedral operation is
the computation of the image of a polyhedron
by a special case of affine relation $\reld{\psi}{\Rset^n}{\Rset^n}$,
called \emph{single-update affine function}:
\[
  (\vect{v}, \vect{w}) \in \psi
    \iff
      w_k = \langle \vect{a}, \vect{v} \rangle + b
      \land
      \bigland_{\genfrac{}{}{0pt}{}{\scriptstyle 0 \leq i < n}{\scriptstyle i \neq k}}
        w_i = v_i.
\]
Another special case, slightly more general than the one above and called
\emph{single-update bounded affine relation}, allows among other things
to approximate nonlinear assignments and to realize rule (ii).
For fixed vectors $\vect{a}, \vect{c} \in \Rset^n$
and scalars $b, d \in \Rset$:
\[
  (\vect{v}, \vect{w}) \in \psi
    \iff
      \langle \vect{a}, \vect{v} \rangle + b
        \leq w_k
          \leq \langle \vect{c}, \vect{v} \rangle + d
      \land
        \bigland_{\genfrac{}{}{0pt}{}{\scriptstyle 0 \leq i < n}{\scriptstyle i \neq k}}
          w_i = v_i.
\]
Both the rules for the if-then-else and the while constructs
require the Boolean filters and least upper bound operations:
these are realized by means of intersections (or the addition
of individual constraints) and poly-hulls, respectively.
These, together with the containment test used to detect the reaching
of post-fixpoints and the widening
(see Section~\ref{sec:peculiar-polyhedral-computations})
used to ensure termination of the analysis algorithm,
are all the operations required for the analysis
of our simple imperative language.  More complex languages
require other operations: for instance, the analysis of languages with
command blocks needs to have the possibility of embedding polyhedra into
a space of higher dimension, reorganizing the dimensions, and projecting
polyhedra on spaces of lower dimension.  Other operations are needed
to accommodate different semantic constructions (e.g., affine preimages
for backward semantics), to allow for the efficient modeling of data objects
(e.g., summarized dimensions to approximate the values of unbounded
collections \cite{GopanRS05}), and to help scalability
(e.g., simplifications of polyhedra \cite{Frehse05}).

\ifthenelse{\boolean{TR}}{%%
Figure~\ref{fig:example-of-abstract-tree} illustrates
an abstract computation that, by following the analysis algorithm above,
approximates the concrete tree in Figure~\ref{fig:example-of-concrete-tree}:
intervals and polyhedra approximate
sets of integers and sets of stores, respectively.
\begin{sidewaysfigure}
\begin{gather*}
\prooftree
  \prooftree
    \prooftree
      \nohyp
    \justifies
      \langle 0, \cQ_0 \rangle \aaeval [0,0]
    \endprooftree
    %\;
    \prooftree
      \nohyp
    \justifies
      \langle x_0, \cQ_0 \rangle \aaeval \top
    \endprooftree
  \justifies
    \langle 0 < x_0, \cQ_0 \rangle
      \baeval
        \top
  \endprooftree
%\;
  \prooftree
    \prooftree
      \omissis
    \justifies
      \langle x_1 := x_1+2, \cQ_0^t \rangle
        \saeval
          \cQ_0'
    \endprooftree \text{(i)}
  \;
    \prooftree
      \omissis
    \justifies
      \langle x_0 := x_0-x_1, \cQ_0' \rangle
        \saeval
          \cQ_1
    \endprooftree \text{(i)}
  \justifies
    \bigl\langle (x_1 := x_1+2;\, x_0 := x_0-x_1), \cQ_0^t \bigr\rangle
      \saeval
        \cQ_1
  \endprooftree
%\;
  \prooftree
    \nohyp
  \justifies
    \langle w, \cQ_1 \rangle
      \saeval
        \cQ  \dblue{\mathord{} = \cQ_0^f}
  \endprooftree \text{(\ref{enum-case:subsumed})}
\justifies
  \langle w, \cQ_0 \rangle
    \saeval
      \cP = (\cQ_0^f \polyhull \cQ) \dblue{\mathord{} = \cQ_0^f}
  \black{\text{ (\ref{enum-case:subsumed})}}
\endprooftree \\[4mm]
\prooftree
  \prooftree
    \prooftree
      \nohyp
    \justifies
      \langle 0, \cP_0 \rangle \aaeval [0,0]
    \endprooftree
    %\;
    \prooftree
      \nohyp
    \justifies
      \langle x_0, \cP_0 \rangle \aaeval [1,\infty]
    \endprooftree
  \justifies
    \langle 0 < x_0, \cP_0 \rangle
      \baeval
        \alpha_\mathrm{B}\bigl(\{\ttv\}\bigr)
  \endprooftree
%\;
  \prooftree
    \prooftree
      \omissis
    \justifies
      \langle x_1 := x_1+2, \cP_0^t \rangle
        \saeval
          \cP_0'
    \endprooftree \text{(i)}
  \;
    \prooftree
      \omissis
    \justifies
      \langle x_0 := x_0-x_1, \cP_0' \rangle
        \saeval
          \cP_1
    \endprooftree \text{(i)}
  \justifies
    \bigl\langle (x_1 := x_1+2;\, x_0 := x_0-x_1), \cP_0^t \bigr\rangle
      \saeval
        \cP_1
  \endprooftree
%\;
    \langle w, \cP_1 \rangle
      \saeval
        \cP
  \black{\text{ (\ref{enum-case:widen-infinite})}}
\justifies
  \langle w, \cP_0 \rangle
    \saeval
      (\cP_0^f \polyhull \cP) = (\emptyset \polyhull \cP) = \cP
\endprooftree
\end{gather*}

\medskip
\noindent
Legend:
\begin{align*}
  w
    &\defeq
      \bigl(
        \kw{while}\, 0 < x_0\, \kw{do}\, (x_1 := x_1+2;\, x_0 := x_0-x_1)
      \bigr), \\
  \cP_0 &\defeq \con\bigl(\{x_0 \geq 1, x_1 = 1\}\bigr), \quad
  &\cQ_0 &\defeq \cP_0 \widen (\cP_0 \polyhull \cP_1)
            =\con\bigl(\{2 x_0 + 3 x_1 \geq 5, x_1 \geq 1\}\bigr), \\
  \cP_0^t &\defeq \phi_\ttv(\cP_0, 0 < x_0)
            = \cP_0, \quad
  &  \cQ_0^t &\defeq \phi_\ttv(\cQ_0, 0 < x_0)
            = \con\bigl(\{x_0 \geq 1, x_1 \geq 1\}\bigr), \\
  \cP_0^f &\defeq \phi_\ffv(\cP_0, 0 < x_0)
            = \emptyset, \quad
  &\cQ_0^f &\defeq \phi_\ffv(\cQ_0, 0 < x_0)
            = \con\bigl(\{2 x_0 + 3 x_1 \geq 5, x_0 \leq 0\}\bigr), \\
  \cP_0' &\defeq \con\bigl(\{x_0 \geq 1, x_1 = 3\}\bigr), \quad
  &\cQ_0' &\defeq \con\bigl(\{x_0 \geq 1, x_1 \geq 3\}\bigr), \\
  \cP_1 &\defeq \con\bigl(\{x_0 \geq -2, x_1 = 3\}\bigr), \quad
  &\cQ_1 &\defeq \con\bigl(\{x_0 + x_1 \geq 1, x_1 \geq 3\}\bigr).
\end{align*}
Notes:
\begin{itemize}
\item[(i)]
Rule (i) of Figure~\ref{fig:IMP-abstract-semantics} is used here.
\item[(\ref{enum-case:subsumed})]
Case (\ref{enum-case:subsumed}) of the algorithm is applied here.
\item[(\ref{enum-case:widen-infinite})]
Case (\ref{enum-case:widen-infinite}) of the algorithm is applied here.
\end{itemize}
\caption{Finite approximation of an infinite abstract computation tree}
\label{fig:example-of-abstract-tree}
\end{sidewaysfigure}
The initial abstract store is given by the polyhedron
$\cP_0 = \con\bigl(\{x_0 \geq 1, x_1 = 1\}\bigr)$,
which approximates all concrete stores $\sigma$ satisfying
$\sigma(x_0) \geq 1$ and $\sigma(x_1) = 1$ including
the concrete store $\sigma_0$ in~Figure~\ref{fig:example-of-concrete-tree}.
Consider first the lower tree in~Figure~\ref{fig:example-of-concrete-tree}.
This corresponds to the stage in the computation when
all possible instances of case~(\ref{enum-case:expand}) of the algorithm
have been applied.
In particular, the two leftmost subtrees are derived according
to the abstract semantics rules in Figure~\ref{fig:IMP-abstract-semantics}
by only using case~(\ref{enum-case:expand}) of the algorithm.
For the rightmost child which has still to be expanded,
$\cP$ is a placeholder for its conclusion.
It is also noted that, in the root of this tree,
since $\cP_0^f = \emptyset$, the final result will be
the same as the value assigned to $\cP$.
Since the rightmost child,
satisfies the conditions of case~(\ref{enum-case:widen-infinite})
of the algorithm, the abstract store $\cP_1$ must undergo a
widening computation, yielding the abstract store $\cQ_0$.
Thus this node has to be replaced by
$\langle w, \cQ_0 \rangle \saeval \cP$.
Consider now the upper tree in~Figure~\ref{fig:example-of-concrete-tree}
which has the root $\langle w, \cQ_0 \rangle \saeval \cP$ as above.
The two left-most immediate subtrees
are derived, as in the lower tree,
by only using case~(\ref{enum-case:expand}) of the algorithm.
The rightmost child is initially given $\cQ$ as a placeholder
for its conclusion.
Since this node
satisfies the conditions for case~(\ref{enum-case:subsumed})
of the algorithm, it is not expanded further;
and the value of $\cQ$ is obtained by finding the least fixpoint solution
for the equation
$\cQ = \cQ_0^f \polyhull \cQ$;
namely,
$\cQ_0^f = \con\bigl(\{2 x_0 + 3 x_1 \geq 5, x_0 \leq 0\}\bigr)$.
Thus in the conclusion of the root of the upper tree we have
$\cP = \cQ_0^f \polyhull \cQ = \cQ_0^f$.
Finally, the completed abstract tree can be obtained by
replacing the rightmost child of the lower tree by the upper tree
and the placeholder $\cP$ in the conclusion of the root
of the lower tree by $\cQ_0^f$.
}{}%%\ifthenelse{\boolean{TR}}

Based on suitable variations of the simple linear invariant analysis
outlined in this section (possibly combined with other analyses),
many different applications have been proposed in the literature.
Examples include the absence of common run-time arithmetic errors,
such as floating-point exceptions, overflows and divisions by zero
\cite{BlanchetCCFMMMR03};
the absence of out-of-bounds array indexing~\cite{CousotH78,VenetB04},
as well as other buffer overruns caused by
incorrect string manipulations~\cite{DorRS01,Ellenbogen04th};
the analysis of programs manipulating (possibly unbounded)
heap-allocated data structures, so as to prove the absence
of several kinds of pointer errors (e.g., memory leaks)
\cite{GopanRS05,ShahamKS00};
the computation of input/output argument size relations
\ifthenelse{\boolean{TR}}{%%
in logic programs~\cite{BenoyK97,GobertLC07,HenriksenG06};
}{%%
in logic programs~\cite{BenoyK97};
}%%\ifthenelse{\boolean{TR}}{%%
the detection of potential security vulnerabilities in x86 binaries
that allow to bypass intrusion detection systems~\cite{KruegelKMRV05};
the inference of temporal schedulability constraints that a partially
specified set of real-time tasks has to satisfy~\cite{DooseM05}.
All of the above are examples of \emph{safety} properties, whereby a
computer program is proved to be free from some undesired behavior.
However, the computation of invariant linear relations is also
an important, often indispensable step when aiming at proving
\emph{progress} properties,
such as termination~\cite{Cousot05,MesnardB05TPLP,SohnVG91}.
It should be also stressed that the same approach, after some minor
adaptations, can be applied to the analysis of alternative computation
paradigms such as, e.g.,
\emph{gated data dependence graphs}~\cite{HymansU04}
(an intermediate representation for compilers)
and \emph{batch workflow networks}~\cite{vanHeeOSV06}
(a form of Petri net used in workflow management).

\section{Analysis and Verification of Hybrid Systems}
\label{sec:analysis-of-hybrid-systems}

Hybrid systems (that is, dynamical systems with both continuous
and discrete components)
are commonly modeled by hybrid automata~\cite{AlurCHH93,Frehse05,Henzinger96}.
These, often highly complex, systems are usually nonlinear (making
them computationally intractable as they are).
However, linear approximations, which allow the use of polyhedral computations
for the model checking operations, have been used successfully for
the verification of useful safety properties
\ifthenelse{\boolean{TR}}{%%
\cite{DoyenHR05,Frehse04,Frehse05,FrehseHK04,SankaranarayananSM06,SongCR06}.
}{%%
\cite{DoyenHR05,Frehse05,SongCR06}.
}%\ifthenelse{\boolean{TR}}

\ifthenelse{\boolean{TR}}{%%
In this section, we illustrate, by means of examples, how polyhedral
computations can be used for verifying simple properties of hybrid automata.
The examples are all instances of \emph{linear hybrid systems}, a
particular class of hybrid systems that can be modeled using polyhedra
where the continuous behavior is specified
by linear constraints over the time-derivatives of the variables.
}{}
\begin{definition}
\summary{(Linear hybrid automaton.)}
\label{def:linear-hybrid-automaton}
A \emph{linear hybrid automaton} (of dimension $n$)
is a tuple
\ifthenelse{\boolean{TR}}{%%
\[
  (\Loc, \Init, \Act, \Inv, \Lab, \Trans)
\]
}{%%
$(\Loc, \Init, \Act, \Inv, \Lab, \Trans)$
}%%\ifthenelse{\boolean{TR}}
where the first component
$\Loc$ is a finite set of \emph{locations}.
The functions
$\fund{\Init}{\Loc}{\Pset_n}$, $\fund{\Act}{\Loc}{\Pset_n}$
and $\fund{\Inv}{\Loc}{\Pset_n}$ define polyhedra. In particular,
for each location $\ell \in \Loc$:
$\Init(\ell)$ specifies the set of possible
initial values the $n$ variables can take if the automaton
starts at $\ell$;
$\Act(\ell)$ specifies the possible derivative values of the $n$ variables,
so that, if the automaton reaches $\ell$ with values
given by the vector $\vect{v}$,
then after staying there for a delay of
$t \in \nonnegRset$, the values will be given by a vector
$\vect{v} + t\vect{w}$, where $\vect{w} \in \Act(\ell)$;
$\Inv(\ell)$ specifies the values that an $n$-vector $\vect{v}$
may have at $\ell$.
The fifth and sixth components
provide a set of \emph{synchronization labels} $\Lab$
and a labeled set of affine transition relations
$\reld{\Trans}{\Loc}{\Lab \times \Pset_{2n} \times \Loc}$,
required to hold when moving from the \emph{source} location
(the first argument)
to the \emph{target} location (the fourth argument).
\end{definition}
Observe that the only differences between this definition of a
linear hybrid automaton and those in, for
\ifthenelse{\boolean{TR}}{%%
example~\cite{AlurCHHHNOSY95,Frehse04,HalbwachsPR97,Henzinger96},
}{%%
example~\cite{HalbwachsPR97,Henzinger96},
}%%\ifthenelse{\boolean{TR}}{%%
are presentational; in particular, as we have used polyhedra to
represent the linear constraints, there is no need to provide, as is
the case in these other definitions, an explicit component of the
system consisting of the set of $n$ variables.

The synchronization labels $\Lab$ are required for specifying large
systems. Each part of the system is specified by a separate automaton,
and then \emph{parallel composition} is employed
to combine the components into an automaton for the complete system.
This ensures that communication between the automata occurs,
via selected input/output variables,
between transitions that have the same label.
Example~\ref{ex:scheduler} provides a very simple
illustration of parallel composition;
formal definitions are available
\ifthenelse{\boolean{TR}}{%%
in~\cite{AlurCHH93,Henzinger96}
and a larger application can be found in~\cite{MullerS00}.
}{%%
in~\cite{AlurCHH93,Henzinger96}.
}%%\ifthenelse{\boolean{TR}}

A linear hybrid automaton can be represented by a directed graph
whose nodes are the locations and edges are the transitions
from the source to the target locations.
Each node $\ell$ is labeled by two sets of constraints defining
the polyhedra $\Inv(\ell)$ and $\Act(\ell)$.
To distinguish these constraints, if, for example $x$ is a variable
used for the constraints defining $\Inv(\ell)$,
$\dot{x}$ will be used in the constraints defining $\Act(\ell)$.%
\footnote{The dot notation reflects the fact
that these variables denote the derivatives of the state variables.}
In the examples, the initial polyhedron $\Init(\ell)$
is assumed to be empty unless there is an arrow
to $\ell$ (with no source node) labeled by the constraint system
defining $\Init(\ell)$.
Each edge $\tau = \bigl(\ell, a, \cP, \ell') \in \Trans$,
is labeled by a constraint system $\cC$ defining $\cP$ and, optionally,
by $a$ which is only included
where it is used for the parallel composition of automata.
Since $\cP \in \Pset_{2n}$, we specify $\cC$ by using two $n$-tuples
of variables $\vect{x}$ and $\vect{x}'$, which are interpreted as usual
to denote the variables in the source $\ell$ and target $\ell'$ locations,
respectively.
We also adopt some helpful shorthand notation:
$x\plusplus$ and $x\minusminus$ denote
$x' = x + 1$ and $x' = x - 1$, respectively;
also, constraints of the form $x' = x$ are omitted.
The following examples, taken (with some minor modifications)
from~\cite{AlurCHH93,HalbwachsPR97}, illustrate the automata.
\begin{example}
\label{ex:water-monitor}
A graphical view of a water-level monitor automaton
is given in \textup{Figure~\ref{fig:water-level-monitor}}.
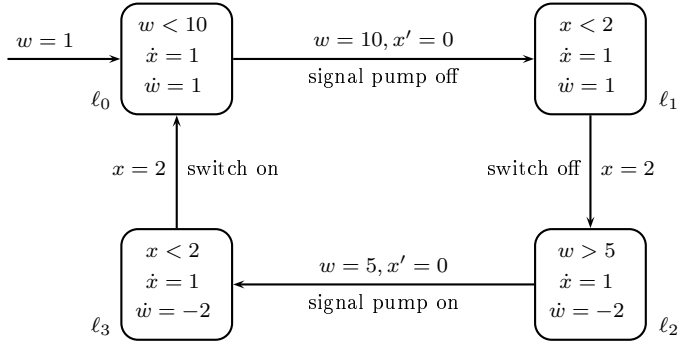
\begin{figure}[p]
\centering
{\scriptsize
\begin{picture}(330,160)(0,-5)
\put(0,0){
\setlength{\unitlength}{0.7pt}%
\psset{xunit=1cm,yunit=1cm,runit=1cm}
\psset{origin={0,0}}
\pspicture*[](-1,0)(10,10)
\psline{->}(-0.5,3.75)(1,3.75)
\rput(0,4){$w = 1$}
\rput(0.75,3.2){$\ell_0$}
\psframe[framearc=.3](1,3)(2.5,4.5)
\rput(1.7,4.2){$w < 10$}
\rput(1.7,3.8){$\dot{x} = 1$}
\rput(1.7,3.4){$\dot{w} = 1$}
\rput(4.5,4){$w = 10, x' = 0$}
\psline{->}(2.5,3.75)(6.5,3.75)
\rput(4.5,3.5){signal pump off}
\rput(8.3,3.2){$\ell_1$}
\psframe[framearc=.3](6.5,3)(8,4.5)
\rput(7.2,4.2){$x < 2$}
\rput(7.2,3.8){$\dot{x} = 1$}
\rput(7.2,3.4){$\dot{w} = 1$}
\rput(6.5,2.3){switch off}
\psline{->}(7.25,3)(7.25,1.5)
\rput(7.75,2.3){$x = 2$}
\rput(8.3,0.2){$\ell_2$}
\psframe[framearc=.3](6.5,0)(8,1.5)
\rput(7.2,1.2){$w > 5$}
\rput(7.2,0.8){$\dot{x} = 1$}
\rput(7.2,0.4){$\dot{w} = -2$}
\rput(4.5,1){$w = 5, x' = 0$}
\psline{<-}(2.5,0.75)(6.5,0.75)
\rput(4.5,0.5){signal pump on}
\rput(0.75,0.2){$\ell_3$}
\psframe[framearc=.3](1,0)(2.5,1.5)
\rput(1.7,1.2){$x < 2$}
\rput(1.7,0.8){$\dot{x} = 1$}
\rput(1.7,0.4){$\dot{w} = -2$}
\rput(1.25,2.3){$x = 2$}
\psline{->}(1.75,1.5)(1.75,3)
\rput(2.5,2.3){switch on}

\endpspicture
}
\end{picture}
}
\caption{Water-level monitor}
\label{fig:water-level-monitor}
\end{figure}
This models a system describing
how the water level in a tank is controlled by a monitor that senses the
water level $w$ and
\ifthenelse{\boolean{TR}}{%%
turns a pump on and off.
}{%%
operates a pump.
}%%\ifthenelse{\boolean{TR}}{%%
When the pump is off, $w$ falls by 2\centimeter per second;
when the pump is on,
$w$ rises by 1\centimeter per second.
However, there is a delay of 2 seconds
from the moment the monitor signals the pump to change
from on to off or vice versa before the switch is actually operated.
Initially the automaton is at $\ell_0$ with
$w = 1$ and it is required that $1 \leq w \leq 12$ at all times.
Thus the monitor must signal the pump to turn on when $w = 5$
and signal it to turn off when $w = 10$.

The automaton illustrated in \textup{Figure~\ref{fig:water-level-monitor}}
has 2 dimensions with variables $w$ and $x$,
where $x$ denotes the time (in seconds) since the previous,
most recent, signal from the monitor.
There are four locations $\ell_i$ where $i = 0$,~$1$,~$2$,~$3$.
At $\ell_0$ and $\ell_1$ the pump is on,
while at $\ell_2$ and $\ell_3$ the pump is off.
At $\ell_1$ and $\ell_3$ the monitor has signaled
a change to the pump switch, but this has not yet been operated.
Thus we have:
\begin{align*}
&\begin{aligned}
  \Init(\ell_0) &= \con\bigl(\{w = 1\}\bigr),
& \Init(\ell_1) &= \Init(\ell_2) = \Init(\ell_3) = \emptyset, \\
  \Inv(\ell_0) &= \con\bigl(\{w < 10\}\bigr),
& \Inv(\ell_1) &= \Inv(\ell_3) =  \con\bigl(\{x < 2\}\bigr), \\
  \Inv(\ell_2) &= \con\bigl(\{w > 5\}\bigr),
& \Act(\ell_0) &= \Act(\ell_1) = \con\bigl(\{\dot{x} = \dot{w} = 1\}\bigr), \\
\end{aligned} \\
& \Act(\ell_2) = \Act(\ell_3) = \con\bigl(\{\dot{x} = 1, \dot{w} = -2\}\bigr).
\end{align*}
There are four transitions
$\tau_{ij} = (\ell_i, a_i, \cP_i, \ell_j) \in \Trans$,
where $i \in \{0,1,2,3\}$ and $j = i+1 \pmod 4$;
the affine relations are
\ifthenelse{\boolean{TR}}{%%
\begin{align*}
  \cP_0 &= \con\bigl(\{w = 10, x' = 0, w' = w\}\bigr), \\
  \cP_1 &= \con\bigl(\{x = 2, x' = x, w' = w\}\bigr), \\
  \cP_2 &= \con\bigl(\{w = 5, x' = 0, w' = w\}\bigr), \\
  \cP_3 &= \cP_1.
\end{align*}
}{%%
\begin{align*}
  \cP_0 &= \con\bigl(\{w = w' = 10, x' = 0\}\bigr),
&\qquad
  \cP_1 &= \con\bigl(\{x = x' = 2, w' = w\}\bigr), \\
  \cP_2 &= \con\bigl(\{w = w' = 5, x' = 0\}\bigr),
&\qquad
  \cP_3 &= \cP_1.
\end{align*}
}%%\ifthenelse{\boolean{TR}}
\end{example}

\ifthenelse{\boolean{TR}}{%%
\begin{example}
\label{ex:fischer-protocol}
A graphical representation of an automaton
for a simplified version of the Fischer protocol is given in
\textup{Figure~\ref{fig:fisher-protocol}}.
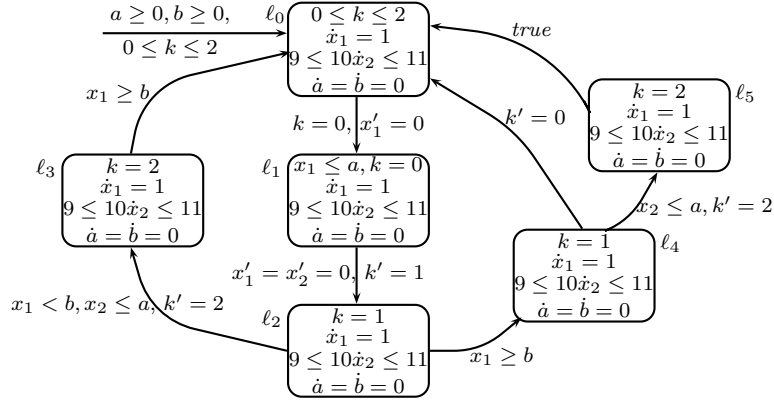
\begin{figure}
\centering
{\scriptsize
\begin{picture}(330,160)(0,0)
\put(0,0){
\setlength{\unitlength}{0.7pt}%
\psset{xunit=1cm,yunit=1cm,runit=1cm}
\psset{origin={0,0}}
\pspicture*[](-1,-1)(12,10)
\psline{->}(0.5,5.1)(2.95,5.1)
\rput(1.4,5.35){$a \geq 0, b \geq 0,$}
\rput(1.4,4.9){$0 \leq k \leq 2$}
\rput(2.8,5.35){$\ell_0$}
\psframe[framearc=.3](2.95,4.25)(4.85,5.52)
\rput(3.9,5.34){$0 \leq k \leq 2$}
\rput(3.9,5.05){$\dot{x}_1 = 1$}
\rput(3.9,4.75){$9 \leq 10\dot{x}_2 \leq 11$}
\rput(3.9,4.45){$\dot{a} = \dot{b} = 0$}

\rput(3.9,3.875){$k = 0,\, x_1' = 0$}
\psline{->}(3.875,4.25)(3.875,3.5)

\rput(2.75,3.3){$\ell_1$}
\psframe[framearc=.3](2.95,2.25)(4.85,3.5)
\rput(3.9,3.34){$x_1 \leq a, k = 0$}
\rput(3.9,3.05){$\dot{x}_1 = 1$}
\rput(3.9,2.75){$9 \leq 10\dot{x}_2 \leq 11$}
\rput(3.9,2.45){$\dot{a} = \dot{b} = 0$}

\rput(3.5,1.9){$x_1' = x_2' = 0,\, k' = 1$}
\psline{->}(3.875,2.25)(3.875,1.5)
\rput(4.8,2){$$}

\rput(2.75,1.3){$\ell_2$}
\psframe[framearc=.3](2.95,0.25)(4.85,1.5)
\rput(3.9,1.34){$k = 1$}
\rput(3.9,1.05){$\dot{x}_1 = 1$}
\rput(3.9,0.75){$9 \leq 10\dot{x}_2 \leq 11$}
\rput(3.9,0.45){$\dot{a} = \dot{b} = 0$}

\psline[linearc=1]{->}(0.875,3.5)(1.05,4.4)(3,4.85)
\rput(0.7,4.3){$x_1 \geq b$}

\rput(-0.25,3.3){$\ell_3$}
\psframe[framearc=.3](-0.05,2.25)(1.85,3.5)
\rput(0.9,3.34){$k = 2$}
\rput(0.9,3.05){$\dot{x}_1 = 1$}
\rput(0.9,2.75){$9 \leq 10\dot{x}_2 \leq 11$}
\rput(0.9,2.45){$\dot{a} = \dot{b} = 0$}

\psline[linearc=1]{->}(2.95,0.875)(1.15,1.3)(0.875,2.25)
\rput(0.7,1.5){$x_1 < b, x_2 \leq a,\, k' = 2$}
\rput(1.5,0.8){$$}

\psline[linearc=1]{->}(4.85,0.875)(5.53,0.875)(6.07,1.32)
\rput(5.8,0.8){$x_1 \geq b$}

\rput(8.05,2.3){$\ell_4$}
\psframe[framearc=.3](5.95,1.25)(7.85,2.5)
\rput(6.9,2.34){$k = 1$}
\rput(6.9,2.05){$\dot{x}_1 = 1$}
\rput(6.9,1.75){$9 \leq 10\dot{x}_2 \leq 11$}
\rput(6.9,1.45){$\dot{a} = \dot{b} = 0$}

\rput(6.25,4){$k' = 0$}
\psline[linearc=2]{->}(6.875,2.5)(6,3.8)(4.85,4.5)

\psline[linearc=1]{->}(7.25,2.5)(7.5,2.6)(7.875,3.25)
\rput(8.5,2.8){$x_2 \leq a, k' = 2$}

\rput(9.05,4.3){$\ell_5$}
\psframe[framearc=.3](6.95,3.25)(8.85,4.5)
\rput(7.9,4.34){$k = 2$}
\rput(7.9,4.05){$\dot{x}_1 = 1$}
\rput(7.9,3.75){$9 \leq 10\dot{x}_2 \leq 11$}
\rput(7.9,3.45){$\dot{a} = \dot{b} = 0$}

\rput(6.2,5.15){$\mathit{true}$}
\psline[linearc=2]{->}(6.9,4.2)(6.5,5)(4.85,5.2)

\endpspicture
}
\end{picture}
}
\caption{Fischer protocol (simplified)}
\label{fig:fisher-protocol}
\end{figure}
This models mutual exclusion for a system with two processors
$P_1$ and $P_2$ with skewed clocks $x_1$ and $x_2$, respectively.
Each processor has a critical
section and, at any one moment in time, at most one may be in its
critical section. This mutual exclusion is ensured by a version of
the Fischer protocol which requires that $P_1$ and $P_2$ share a
variable $k$; a process $P_i$ ($i = 1$,~$2$) is only able to enter its critical
section if $k = i$ and $P_i$ may only write to $k$ if $k=0$.  However,
it takes at most $a$ time units, as measured by $P_i$'s clock for
$P_i$ to set the value of $k$ to $i$ and it could be that the other
process $P_j$ may also have started writing $j$ to $k$. To avoid any
resulting conflict, the protocol requires that $P_i$ must wait for a
further $b$ time units, also measured by $P_i$'s clock, before checking
that $k=i$ still holds. The time $b$ is called the \emph{delay time}.
The protocol ensures mutual exclusion only for certain values of $a$
and $b$ which depend on the relative rates of $x_1$ and $x_2$.
Here it is assumed that the rate of
$x_2$ is between 0.9 and 1.1 times that of $x_1$
and that, for $i = 1$,~$2$, the clock $x_i$ is reset to zero
at the start of both the write process and the delay time
for $P_i$.

The automaton illustrated in \textup{Figure~\ref{fig:fisher-protocol}} has
5 dimensions with variables $a$, $b$, $x_1$, $x_2$, $k$.
Note that here, $a$ and $b$ are constant for all runs of the automaton
and this is indicated in the graph
by  the inclusion of the derivative constraints
$\dot{a} = \dot{b} = 0$ at every location.
There are six locations:
$\ell_0$ where $P_1$ is idle;
at $\ell_1$ where $k=0$ and $P_1$ is in the process of writing to $k$;
at $\ell_2$ where $k=1$ and $P_1$ waits for the delay time of $b$
time units;
at $\ell_3$ where $k=2$ since $P_2$ managed to complete writing
to $k$ before the delay time of $b$ had expired;
at $\ell_4$ where the process $P_1$ is in the critical section;
at $\ell_5$ where $P_2$ has set $k = 2$ and the mutual
exclusion guarantee is violated.
All the functions and transitions for these locations are as given
in \textup{Figure~\ref{fig:fisher-protocol}}.
\end{example}
}{}%% \ifthenelse{\boolean{TR}}

\begin{example}
\label{ex:scheduler}
A representation of an automaton
for a simple task scheduler is given in \textup{Figure~\ref{fig:scheduler}}.
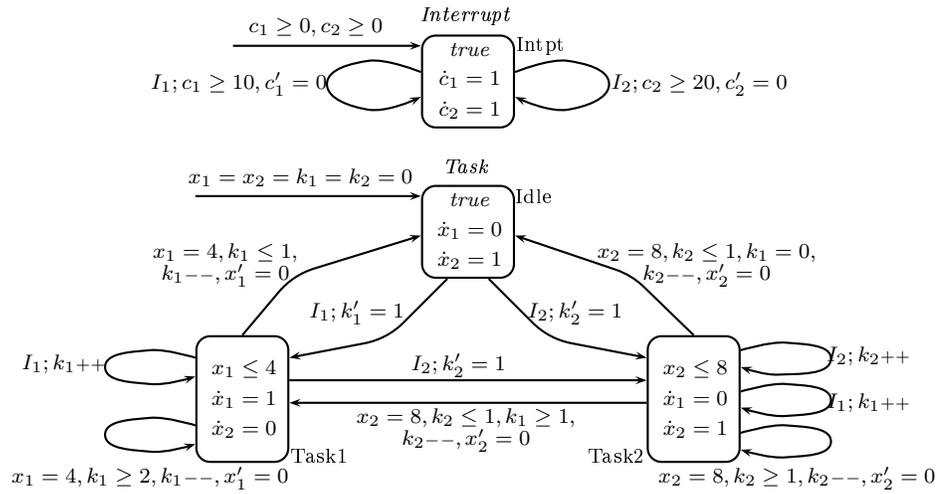
\begin{figure}[p]
\centering
{\scriptsize
\begin{picture}(330,180)(15,-10)
\put(0,0){
\setlength{\unitlength}{0.7pt}%
\psset{xunit=1cm,yunit=1cm,runit=1cm}
\psset{origin={0,0}}
\pspicture*[](-2.5,-1)(10,10)
\rput(3.6,5.75){\emph{Interrupt}}
\psline{->}(0.5,5.35)(3,5.35)
\rput(1.6,5.6){$c_1 \geq 0, c_2 \geq 0$}
\rput(4.6,5.35){Intpt}
\psframe[framearc=.3](3,4.25)(4.25,5.5)
\rput(3.65,5.3){$\mathit{true}$}
\rput(3.65,4.9){$\dot{c}_1 = 1$}
\rput(3.65,4.5){$\dot{c}_2 = 1$}

\rput(0.58,4.845){$I_1; c_1 \geq 10, c_1' = 0$}
\pscurve{->}(3,5)(2.3,5.15)(1.75,4.845)(2.3,4.5)(3,4.67)
\pscurve{->}(4.25,5)(4.95,5.15)(5.50,4.845)(4.95,4.5)(4.25,4.67)
\rput(6.7,4.845){$I_2; c_2 \geq 20, c_2' = 0$}

\rput(3.6,3.75){\emph{Task}}
\psline{->}(0,3.35)(3,3.35)
\rput(1.4,3.6){$x_1 = x_2 = k_1 = k_2 = 0$}
\rput(4.5,3.35){Idle}
\psframe[framearc=.3](3,2.25)(4.25,3.5)
\rput(3.65,3.3){$\mathit{true}$}
\rput(3.65,2.9){$\dot{x}_1 = 0$}
\rput(3.65,2.5){$\dot{x}_2 = 1$}

\rput(0.4,2.6){$x_1 = 4, k_1 \leq 1,$}
\rput(0.4,2.3){$k_1\minusminus, x_1' = 0$}
\psline[linearc=1]{->}(0.625,1.5)(1.3,2.3)(3,2.825)

\rput(2.15,1.8){$I_1; k_1' = 1$}
\rput(2.1,1.7){$$}
\psline[linearc=1]{<-}(1.25,1.2)(2.55,1.45)(3.35,2.25)

\rput(6.8,2.6){$x_2 = 8, k_2 \leq 1, k_1 = 0,$}
\rput(6.8,2.3){$k_2\minusminus, x_2' = 0$}
\psline[linearc=1]{->}(6.625,1.5)(5.8,2.3)(4.25,2.825)

\rput(5.05,1.8){$I_2; k_2' = 1$}
\rput(5.05,1.7){$$}
\psline[linearc=1]{<-}(6,1.2)(4.7,1.45)(3.9,2.25)

\rput(1.65,-0.1){Task1}
\psframe[framearc=.3](0,-0.2)(1.25,1.5)
\rput(0.65,1.05){$x_1 \leq 4$}
\rput(0.65,0.65){$\dot{x}_1 = 1$}
\rput(0.65,0.25){$\dot{x}_2 = 0$}

\rput(5.6,-0.1){Task2}
\psframe[framearc=.3](6,-0.2)(7.25,1.5)
\rput(6.65,1.05){$x_2 \leq 8$}
\rput(6.65,0.65){$\dot{x}_1 = 0$}
\rput(6.65,0.25){$\dot{x}_2 = 1$}

\rput(3.5,1.1){$I_2; k_2' = 1$}
\psline{->}(1.25,0.9)(6,0.9)
\psline{<-}(1.25,0.6)(6,0.6)
\rput(3.6,0.4){$x_2 = 8, k_2 \leq 1, k_1 \geq 1,$}
\rput(3.6,0.1){$k_2\minusminus, x_2' = 0$}

\rput(-1.75,1.1135){$I_1; k_1\plusplus$}
\pscurve{->}(0,1.2)(-0.6,1.3)(-1.2,1.075)(-0.6,0.85)(0,0.95)
\rput(-0.6,-0.4){$x_1 = 4, k_1 \geq 2, k_1\minusminus, x_1' = 0$}
\pscurve{->}(0,0.3)(-0.6,0.4)(-1.2,0.175)(-0.6,-0.05)(0,0.05)

\rput(8.95,1.225){$I_2; k_2\plusplus$}
\pscurve{->}(7.25,1.3)(7.85,1.4)(8.45,1.1865)(7.75,0.975)(7.25,1.075)
\rput(8.95,0.6){$I_1; k_1\plusplus$}
\pscurve{->}(7.25,0.75)(7.85,0.85)(8.45,0.6365)(7.85,0.425)(7.25,0.535)
\rput(8,-0.4){$x_2 = 8, k_2 \geq 1, k_2\minusminus, x_2' = 0$}
\pscurve{->}(7.25,0.2)(7.85,0.3)(8.45,0.0855)(7.85,-0.125)(7.25,-0.025)

\endpspicture
}
\end{picture}
}
\caption{Scheduler}
\label{fig:scheduler}
\end{figure}
This models a scheduler with two classes of
tasks $A_1$ and $A_2$, activated by interrupts $I_1$ and $I_2$.
Interrupt $I_1$ (resp.,
$I_2$) occurs at most once every 10 (resp., 20) seconds and activates
a task in class $A_1$ (resp., $A_2$), which takes 4 (resp., 8)
seconds to complete.
Tasks in $A_2$ have priority and preempt tasks in $A_1$.
It is required that tasks in $A_2$ never wait.

The \emph{Scheduler} automaton given in \textup{Figure~\ref{fig:scheduler}}
is the parallel composition of two component automata:
\emph{Interrupt} which models the
assumptions about the interrupt frequencies; and \emph{Task},
which models the execution of the tasks.
The \emph{Interrupt} automaton, which has a single location `Intpt',
has variables
$c_1$ and $c_2$; $c_i$ ($i = 1$,~$2$) measures the
time elapsed since interrupt $I_i$ occurred.
The \emph{Task} automaton has three locations:
`Idle' when no tasks are running;
and `Task1' and `Task2' when tasks in classes $A_1$
(resp., $A_2$) are active.
It has, for each $i = 1$,~$2$, variables
$x_i$, which measures the execution time of task $i$,
and $k_i$, which counts the number of pending tasks in class
task $i$.

The combined \emph{Scheduler} automaton has variables
$x_1$, $x_2$, $k_1$, $k_2$, $c_1$ and $c_2$ and
locations which are elements of the Cartesian product
of the sets of locations for \emph{Interrupt} and  \emph{Task}.
As \emph{Interrupt} has just one location,
each \emph{Task} location $\ell$ is used to denote the
corresponding \emph{Scheduler} location;
here, the initial $\Init(\ell)$, derivative $\Act(\ell)$ and invariant
$\Inv(\ell)$ polyhedra for the \emph{Scheduler}
are the concatenation of the corresponding component polyhedra
for the \emph{Task} and \emph{Interrupt} automata
(informally, a concatenation of polyhedra $\cP \in \Pset_m$ and
$\cQ \in \Pset_n$ can be obtained by first embedding $\cP$ into
a vector space of dimension $n+m$ and then add a suitably renamed-apart
version of the constraints defining~$\cQ$).
Each transition $(\ell, a, \cP, \ell')$ in the \emph{Task} automaton
not triggered by interrupts $I_1$ and $I_2$
has a transition $(\ell, a, \cQ, \ell')$ in the product automaton
where $\cQ \in \Pset_6$ is obtained by embedding $\cP$ into
a vector space of dimension $6$.
Letting $i =$~$1$, $2$,
for transitions $(\ell, I_i, \cP, \ell')$
and $(\mathrm{Intpt}, I_i, \cP', \mathrm{Intpt})$
in the \emph{Task} and \emph{Interrupt} automata, respectively,
there is a transition  $(\ell, I_i, \cQ, \ell')$
in the product automaton where $\cQ \in \Pset_6$
is obtained by concatenating $\cP$ and $\cP'$.
\end{example}

Given a linear hybrid automaton, the aim of an analyzer is to check,
or even find sufficient conditions that ensure, that a valid
\emph{run} of the system cannot \emph{reach} a location and vector of
values that violate some requirement of the system. For instance, in
Example~\ref{ex:water-monitor}, we need to show that the water level
always lies between 1\centimeter and 12\centimeter;
\ifthenelse{\boolean{TR}}{%%
in Example~\ref{ex:fischer-protocol}, we need to find conditions on $a$
and $b$ so that at most one processor can be in its critical section at
any one time;
}{}%% \ifthenelse{\boolean{TR}}
in Example~\ref{ex:scheduler}, we need to show
that no task in $A_2$ will ever wait. To show how
polyhedral computations can be used to prove such properties, we first
define more formally such a run and how reachable sets may be computed.
Note that these definitions follow,
with only minor changes, the approach in \cite{HalbwachsPR97}.

Letting $\cH = (\Loc, \Init, \Act, \Inv, \Lab, \Trans)$
be a linear hybrid automaton in $n$ dimensions,
a \emph{state} $s$ of $\cH$ consists of a pair $(\ell, \vect{v})$,
where $\ell \in \Loc$ and $\vect{v} \in \Inv(\ell)$.
Given states $s = (\ell, \vect{v})$ and $s' = (\ell', \vect{v}')$,
a time delay $t \in \nonnegRset$ and a
vector $\vect{w} \in \Act(\ell)$,
\ifthenelse{\boolean{TR}}{%%
\[
  s \timedstep{t}{\vect{w}} s'
\]
}{%%
\(
  s \timedstep{t}{\vect{w}} s'
\)
}%% \ifthenelse{\boolean{TR}}
is a \emph{step} of $\cH$ provided that,
for all $t' \in \left[0, t\right)$,
$\vect{v} + t' \vect{w} \in \Inv(\ell)$ and,
for some $(\ell, a, \cP, \ell') \in \Trans$,
\(
  (\vect{v} + t \vect{w}) \mathop{::} \vect{v}' \in \cP.
\)
A \emph{run} of $\cH$ is a  sequence (finite or infinite) of steps
\ifthenelse{\boolean{TR}}{%%
\begin{equation}
\label{eq:automaton-run}
s_0 \timedstep{t_0}{\vect{w}_0} s_1 \timedstep{t_1}{\vect{w}_1} s_2 \cdots
\end{equation}
}{%%
\(
s_0 \timedstep{t_0}{\vect{w}_0} s_1 \timedstep{t_1}{\vect{w}_1} s_2 \cdots
\),
}%%\ifthenelse{\boolean{TR}}{%%
where the initial state $s_0 = (\ell_0, \vect{v}_0)$ satisfies the condition
$\vect{v}_0 \in \Init(\ell_0)$.
An infinite run diverges if the sum $\sum_{i\geq 0} t_i$ diverges.
For each divergent run
\ifthenelse{\boolean{TR}}{%%
given by~\eqref{eq:automaton-run}
}{}%%\ifthenelse{\boolean{TR}}
where,
for $i \geq 0$, $s_i = (\ell_i, \vect{v}_i)$,
we associate a (state) \emph{behavior} $\beta$
which is a total function from time to states:
that is, $\beta(0) = s_0$ and, for each $t > 0$,
$\beta(t) \defeq (\ell_i, \vect{v})$, where
\ifthenelse{\boolean{TR}}{%%
\begin{equation*}
  i = \min \biggl\{\,
             k \in \Nset
           \biggm|
             \sum_{j=0}^k t_j > t
           \,\biggr\}
\qquad \text{and} \qquad
  \vect{v} = \vect{v}_i + \vect{w}_i \biggl(t - \sum_{j < i} t_j\biggr).
\end{equation*}
}{%%
\(
  i = \min \bigl\{\,
             k \in \Nset
           \bigm|
             \sum_{j=0}^k t_j > t
           \,\bigr\}
\)
and
\(
  \vect{v} = \vect{v}_i + \vect{w}_i \bigl(t - \sum_{j < i} t_j\bigr)
\).
}%%\ifthenelse{\boolean{TR}}{%%
A state $s$ is \emph{reachable} if there exists a divergent run
with behavior $\beta$ and time $t \in \nonnegRset$ such that
$\beta(t) = s$.
The set of all \emph{reachable values $R_\ell$} for a location $\ell$
is defined as:
\[
   R_\ell
     \defeq
       \bigl\{\,
         \vect{v} \in \Rset^n
       \bigm|
         \exists t \in \nonnegRset \st
           \beta(t) = (\ell, \vect{v})
       \,\bigr\}.
\]
The set of reachable values $R_\ell$
at a location $\ell$ can be characterized by a system of
fixpoint equations that are defined in terms of
sets of reachable values $R_{\ell'}$ at locations $\ell'$ where
$\bigl(\ell', a, \cP, \ell\bigr) \in \Trans$.
These equations use the following operations on sets of vectors in
$\Rset^n$.
Let $\cP, \cQ \in \Pset_{2n}$ and $S \sseq \Rset^n$. Then
\begin{align*}
  \psi_{\cP}(S)
    &\defeq
      \{\,
        \vect{v}' \in \Rset^n
      \mid
        \vect{v} \in S,
        \vect{v} \mathop{::} \vect{v}' \in \cP
      \,\};\\
  S \timeelapse \cQ
    &\defeq
      \{\,
        \vect{v} + t \vect{w} \in \Rset^n
      \mid
        \vect{v} \in S, \vect{w} \in \cQ, t \in \nonnegRset
      \,\}.
\end{align*}
Note that, if $S \in \Pset_n$, then also $\psi_{\cP}(S) \in \Pset_n$
and $S \timeelapse \cQ \in \Pset_n$.
The `$\mathop{\timeelapse}$' operator,
called the \emph{time elapse} operator, was first proposed
in~\cite{HalbwachsPR97}.
We can now provide the fixpoint equation for $R_\ell$:
\begin{equation}
\label{eq:reach-ell}
   R_\ell
     = \biggl(
         \Bigl(
           \Init(\ell)
             \union
               \bigunion_{(\ell', a, \cP, \ell) \in \Trans}
                 \psi_{\cP}(R_{\ell'}) \inters \Inv(\ell)
         \Bigr)
           \timeelapse \Act(\ell)
       \biggr)
         \inters \Inv(\ell).
\end{equation}
Informally, the fixpoint equation for $R_\ell$ says that the
reachable values at the location $\ell$ are obtained by letting the
time elapse either from an initial value for $\ell$ or from a value
obtained from an incoming transition.
However, the fixpoint Equation~\eqref{eq:reach-ell}
cannot handle strict constraints correctly and needs modifying;
this is illustrated in the following example.
\begin{example}
\label{ex:water-monitor-reach-all}
Consider again \textup{Example~\ref{ex:water-monitor}}.
Then, just applying \textup{Equation~\eqref{eq:reach-ell}}
\ifthenelse{\boolean{TR}}{%%
(as proposed in~\textup{\cite{HalbwachsPR94,HalbwachsPR97}}),
}{%%
(as proposed in~\textup{\cite{HalbwachsPR97}}),
}%%\ifthenelse{\boolean{TR}}{%%
the sets of reachable values at locations $\ell_1, \ell_2, \ell_3$ are
empty.
The reason for this is that, for example, at location $\ell_0$,
the strict constraint $w < 10$ must hold, while in the transition from
$\ell_0$ to $\ell_1$, the transition condition $w = 10$ has to hold.
On the other hand, it follows from the definition of a step,
that since one of the derivative constraints at $\ell_0$ is
$\dot{w} = 1$; the water level $w$ may continue to increase
up to the topological closure of $R_{\ell_0}$
which is consistent with $w = 10$.
\end{example}
To resolve this problem,
in Equation~\eqref{eq:reach-ell} defining the concrete computation,
 $R_{\ell'}$ needs to be replaced by
\ifthenelse{\boolean{TR}}{%%
\begin{equation}
\label{eq:reach-ell-strict}
   c(R_{\ell'})
     \inters
       \bigl(R_{\ell'} \timeelapse \Act(\ell')\bigr),
\end{equation}
}{%%
\(
   c(R_{\ell'})
     \inters
       \bigl(R_{\ell'} \timeelapse \Act(\ell')\bigr)
\),
}%%\ifthenelse{\boolean{TR}}{%%
where $c(R_\ell')$ denotes the topological closure of $R_\ell' \sseq \Rset^n$.

\ifthenelse{\boolean{TR}}{%%
Observe that, although the linear hybrid automata
are specified by means of polyhedra,
the reachable set $R_\ell$ for a linear hybrid automaton and location $\ell$
may not be in the form of a convex polyhedron.
Thus, to verify that some states of an automaton are unreachable
using the standard polyhedral computations,
approximations are needed.
In particular,
in the fixpoint Equation~\eqref{eq:reach-ell}
(or \eqref{eq:reach-ell-strict}),
the set operations have to be replaced
by the corresponding polyhedral operations.
In fact all the operations in \eqref{eq:reach-ell} except set union
can be used as they are, since they transform polyhedra to polyhedra.
Just the set union operation has to be replaced by the poly-hull operation
`$\polyhull$' described in Section~\ref{sec:preliminaries}.
}{%%\ifthenelse{\boolean{TR}}{%%
Observe that, although the linear hybrid automata
are specified by means of polyhedra,
the reachable set $R_\ell$ for a linear hybrid automaton and location $\ell$
may not be a convex polyhedron since Equation~\eqref{eq:reach-ell}
uses the set union operation.
Therefore, to verify that some states of an automaton are unreachable
using standard polyhedral computations,
set union has to be replaced by the poly-hull operation
$\polyhull$ described in Section~\ref{sec:preliminaries}.
}%%\ifthenelse{\boolean{TR}}{%%
Thus the following fixpoint equation computes an approximation
$R_\ell^\sharp$ to the reachability set~$R_\ell$.
\begin{equation}
\label{eq:reach-ell-sharp}
   R_\ell^\sharp =
     \biggl(
       \Bigl(
         \Init(\ell) \polyhull
           \biguplus_{(\ell', a, \cP, \ell) \in \Trans}
             \psi_{\cP}(R_{\ell'}^\sharp) \inters \Inv(\ell)
       \Bigr)
         \timeelapse \Act(\ell)
       \biggr)
       \inters \Inv(\ell).
\end{equation}
As for the concrete fixpoint equation,
to correctly handle the strict constraints
\ifthenelse{\boolean{TR}}{%%
Equation~\eqref{eq:reach-ell-sharp}
needs to be modified by replacing $R_{\ell'}^\sharp$ with
\[
   c(R_{\ell'}^\sharp)
     \inters
       \bigl(R_{\ell'}^\sharp \timeelapse \Act(\ell')\bigr).
\]
}{%%
in Equation~\eqref{eq:reach-ell-sharp}
we need to replace $R_{\ell'}^\sharp$ with
\(
   c(R_{\ell'}^\sharp)
     \inters
       \bigl(R_{\ell'}^\sharp \timeelapse \Act(\ell')\bigr)
\).
}%%\ifthenelse{\boolean{TR}}{%%

\ifthenelse{\boolean{TR}}{%%
If we let $\vect{R}^\sharp$ denote the tuple
$\{\, R_\ell^\sharp \mid \ell \in \Loc \,\}$
we can write Equation~\eqref{eq:reach-ell-sharp} as
\[
  R_\ell^\sharp = F_\ell(\vect{R}^\sharp).
\]
For all $\ell \in \Loc$, we write
$\vect{R}_\ell^{\sharp(0)} = \emptyset$ and, for all $k \geq 1$,
$\vect{R}_\ell^{\sharp(k+1)} = F_\ell(\vect{R}_\ell^{\sharp(k)})$.
Then $\vect{R}^\sharp$ can be computed iteratively provided the sequence
$\vect{R}^{\sharp(0)}, \vect{R}^{\sharp(1)}, \ldots$ does not diverge.
To handle diverging sequences, we apply a widening
(see Section~\ref{sec:widening}).
Note that we do not have to apply it at all locations.
Let $\cW$ be a set of locations that cut all cyclic paths
in the graph of the hybrid automaton (that is, each loop
of the directed graph contains at least one location in $\cW$).
Then the following set of fixpoint equations is guaranteed to converge:
\begin{equation}
\label{eq:reach-ell-sharp-with-widening}
  R_\ell^\sharp
    =
      \begin{cases}
        R_\ell^\sharp \widen F_\ell(\vect{R}^\sharp),
          &\text{ if $\ell \in \cW$;} \\
        F_\ell(\vect{R}^\sharp),
          &\text{ if $\ell \in \Loc \setdiff \cW$.}
      \end{cases}
\end{equation}
}{%%
If we let $\vect{R}^\sharp$ denote the tuple
$\{\, R_\ell^\sharp \mid \ell \in \Loc \,\}$
we can write Equation~\eqref{eq:reach-ell-sharp} as
\(
  R_\ell^\sharp = F_\ell(\vect{R}^\sharp)
\).
For all $\ell \in \Loc$, we write
$\vect{R}_\ell^{\sharp(0)} = \emptyset$ and, for all $k \geq 1$,
$\vect{R}_\ell^{\sharp(k+1)} = F_\ell(\vect{R}_\ell^{\sharp(k)})$.
Then $\vect{R}^\sharp$ can be computed iteratively provided the sequence
$\vect{R}^{\sharp(0)}, \vect{R}^{\sharp(1)}, \ldots$ does not diverge.
To handle diverging sequences, we apply a widening
(see Section~\ref{sec:widening});
note that this only needs to be applied at sufficient locations
so that each cyclic path
in the graph of the hybrid automaton has at least one widening point.
}%%\ifthenelse{\boolean{TR}}{%%

\begin{example}
\label{ex:water-monitor-abstract}
Consider again \textup{Example~\ref{ex:water-monitor}}.
As there is a single loop passing through $\ell_0$,
it is sufficient to define the set of widening locations
as
\ifthenelse{\boolean{TR}}{%%
$\cW = \{\ell_0\}$.
}%%\ifthenelse{\boolean{TR}}{%%
{%%
$\{\ell_0\}$.
}%%\ifthenelse{\boolean{TR}}{%%

With the modified form of \textup{Equation~\eqref{eq:reach-ell-sharp}} and
the polyhedra widening of~\textup{\cite{CousotH78}},
the computation requires three
iterations resulting in polyhedra defined by constraint systems
$\cC_i$ for $0 \leq i \leq 3$ where:
\begin{align*}
  \cC_0 &= \{ 1 \leq w < 10 \},
& \quad
  \cC_1 &= \{ w - x = 10,\, 10 \leq w < 12 \}, \\
  \cC_2 &= \{ w + 2x = 16,\, 5 < w \leq 12 \},
& \quad
  \cC_3 &= \{ w + 2x = 5,\, 1 < w \leq 5 \}.
\end{align*}
\end{example}
\ifthenelse{\boolean{TR}}{%%
\begin{example}
\label{ex:fischer-protocol-abstract}
Consider again \textup{Example~\ref{ex:fischer-protocol}}.
The analysis terminates without widening in
just two iterations with the resulting polyhedron at~$\ell_5$
defined by the constraint system:
\begin{multline*}
  \cC = \{
          k = 2,\,
          10a \geq 9b,\,
          0 \leq b \leq x_1,\,
          9x_1 \leq 10x_2 \leq 11x_1, \\
          11x_1 + 10a \geq 10x_2 + 11b
        \}.
\end{multline*}
It therefore follows that, to ensure that there can be no run
with a state at location $\ell_5$, it is sufficient that $10a < 9b$.
\end{example}
}{}%% \ifthenelse{\boolean{TR}}

\begin{example}
\label{ex:scheduler-abstract}
Consider again \textup{Example~\ref{ex:scheduler}}.
By applying the above mentioned polyhedra widening
at location `Task2' only,
the analysis for the product automaton terminates in four iterations.
\ifthenelse{\boolean{TR}}{%
After projecting away the variables $c_1$ and $c_2$,
the reachable values are given by polyhedra
defined by constraint systems $\cC_{t0}$, $\cC_{t1}$, and $\cC_{t2}$
for locations `Idle', `Task1' and `Task2', respectively, where:
\begin{align*}
\cC_{t0} &=
   \{ x_1 = x_2 = k_1 = k_2 = 0 \},\\
\cC_{t1} &=
   \{ 0 \leq x_1 \leq 4,\, x_2 = 0,\, k_1 = 1,\, k_2 = 0 \},\\
\cC_{t2} &=
   \{ x_2 \geq 0,\, x_2 \leq 8,\, 4k_1 \geq x_1,\, x_1 \geq 0,\, k_2 = 1 \}.
\end{align*}
So it can be concluded that, at each location of the automaton, $k_2 \leq 1$
and, hence, no task in class $A_2$ will ever have to wait.
}{%%
After projecting onto variables $k_1$ and $k_2$,
the reachable values are given by polyhedra
defined by constraint systems $\cC_{t0}$, $\cC_{t1}$, and $\cC_{t2}$
for locations `Idle', `Task1' and `Task2', respectively, where:
\begin{equation*}
  \cC_{t0}
    = \{ k_1 = k_2 = 0 \},
\quad
  \cC_{t1} = \{ k_2 = 0,\, k_1 = 1 \},
\quad
  \cC_{t2}
    = \{ k_2 = 1 \}.
\end{equation*}
Therefore, since at all locations $k_2 \leq 1$,
no task in class $A_2$ will ever have to wait.
}%\ifthenelse{\boolean{TR}}{%
However, as noted in~\textup{\cite{HalbwachsPR97}},
because of the convex hull approximation,
with the polyhedral domain the
analyzer fails to show that $k_1 \leq 2$.
\ifthenelse{\boolean{TR}}{%
We therefore redid the analysis using a domain of powersets of
polyhedra (see \textup{Section~\ref{sec:generalizations-of-polyhedra}})
and, after taking the poly-hull of the final
sets and projecting away the variables $c_1$ and $c_2$,
we obtained the polyhedra
defined by constraint systems $\cC'_{t0}$, $\cC'_{t1}$ and $\cC'_{t2}$
for locations `Idle', `Task1' and `Task2', respectively, where:
\begin{align*}
  \cC'_{t0}
    &= \{ x_1 = x_2 = k_1 = k_2 = 0 \}, \\
  \cC'_{t1}
    &= \{ 0 \leq x_1 \leq 4,\, x_2 = 0,\, k_1 = 1,\, k_2 = 0 \}, \\
  \cC'_{t2}
    &= \{ 0 \leq x_1 \leq 4,\, 0 \leq x_2 \leq 8,\,
           4k_1 \geq x_1 \geq 2k_1 - 2, \\
      & \qquad \qquad  \qquad \qquad \qquad
           x_1 + x_2 \geq 10k_1 - 10,\,
               k_1 \leq 2,\, k_2 = 1 \}.
\end{align*}
This verifies that $k_1 \leq 2$ and $k_0 \leq 1$
in every state of any run of the automata.
}{%%
We therefore redid the analysis using a domain of powersets of
polyhedra (see \textup{Section~\ref{sec:generalizations-of-polyhedra}})
and, after taking the poly-hull of the final
sets and projecting onto variables $k_1$ and $k_2$,
we obtained the polyhedra
defined by constraint systems $\cC'_{t0}$, $\cC'_{t1}$ and $\cC'_{t2}$
for locations `Idle', `Task1' and `Task2', respectively, where:
\begin{equation*}
  \cC'_{t0}
    = \{ k_1 = k_2 = 0 \},
\quad
  \cC'_{t1}
    = \{ k_2 = 0,\, k_1 = 1 \},
\quad
  \cC'_{t2}
    = \{ k_1 \leq 2,\, k_2 = 1 \}.
\end{equation*}
}%%\ifthenelse{\boolean{TR}}
\end{example}

Hybrid systems with affine or nonlinear dynamics do not fit the above
specification of a linear system so that the verification techniques
described here are not directly applicable.  Nonetheless, by
partitioning the continuous state space and over-approximating the
dynamics in each of the partitions, the same techniques used to verify
linear hybrid automata can be used in these more general cases
\ifthenelse{\boolean{TR}}{%%
\cite{DoyenHR05,Frehse05,HenzingerH95b,HenzingerHW97b,SankaranarayananSM06}.
}{%%
\cite{DoyenHR05,Frehse05,HenzingerHW97b}.
}%\ifthenelse{\boolean{TR}}
Such an approach has also been successfully applied in the
verification of analog circuits, as discussed in the following
section.

\section{Analysis and Verification of Analog Systems}
\label{sec:analysis-of-analog-systems}

The idea of applying formal methods, that originated in the digital
world, to analog systems was put forward in \cite{HartongHB02}.
This is an important step forward with
respect to more traditional methods for the validation of analog
circuit designs.  A formal verification tool can, for example, ensure
that a design satisfies certain properties for entire sets of initial
states and continuous ranges of circuit parameters, something that
cannot be done with simulation.

\ifthenelse{\boolean{TR}}{%%
In \cite{DangDM04} and \cite{GuptaKR04}, polyhedral approximations
were successfully used in the verification of analog circuits.
Here, we use a simple example, taken from \cite{FrehseKR06},
on the verification of an oscillator circuit to
illustrate the approach.%
\footnote{For a more general view, we refer the interested reader
to the cited literature and to \cite{Maler06}.}
}{%%
To illustrate the approach, we describe a simple example of
verification of an oscillator circuit, taken from \cite{FrehseKR06}.
}%%\ifthenelse{\boolean{TR}}{%%
To verify properties of the (cyclic) behavior of such circuits, cyclic
invariants have to be determined.  To establish a cyclic invariant for
a given set of initial states and ranges for the circuit parameters,
one has to show that the circuit returns to a subset of those initial
states, which implies the system will keep traversing the same states
indefinitely.
\ifthenelse{\boolean{TR}}{%%
From such an invariant, a number of properties of the
oscillator can be established \cite{FrehseKRM06}.
}{}%%\ifthenelse{\boolean{TR}}{%%

\ifthenelse{\boolean{TR}}{}{%%
\begin{figure}
 \begin{minipage}[b]{0.4\linewidth}
   \centering
\subfigure[Circuit schematic]%
{\label{fig:tunnel-diode-oscillator-schematic}%
\parbox[t]{\linewidth}%
{\centering
\setlength{\unitlength}{0.240900pt}
\begin{picture}(300,360)(-20,20)
\put(-300,0){
\psset{xunit=0.7cm,yunit=0.7cm,runit=0.7cm}
\psset{origin={0,0}}
\pspicture*[](10,4)
\psset{dipolestyle=elektor}
\pnode(2,3){Vin}
\pnode(1.5,3){S} \pnode(3.5,3){A} \pnode(6.5,3){B} \pnode(7.7,3){C}
\pnode(1.5,1){Sm}\pnode(3.5,1){Am}\pnode(6.5,1){Bm}\pnode(7.7,1){Cm}
\Ucc[dipoleconvention=generator,tension,tensionoffset=0.8,labeloffset=1.2](Sm)(S){$V_\mathrm{in}$}
\resistor[dipolestyle=rectangle](Vin)(A){$R$}
\coil[dipolestyle=elektorcurved,intensitylabel=$I_L$](A)(B){$L$}
\diode[dipoleconvention=receptor,intensitylabel=$I_d$,tensionlabel=$V_d$,tensionoffset=-0.5,tensionlabeloffset=-0.9,](B)(Bm){}
\capacitor(C)(Cm){$C$}
\psset{intensitylabel=}
\wire(Am)(Bm)
\wire(Bm)(Cm)
\wire(B)(C)
\wire(S)(Vin)
\wire(Sm)(Am)
\pscircle*(B){2\pslinewidth}
\pscircle*(Bm){2\pslinewidth}
\endpspicture
}
\end{picture}
}} \\
\subfigure[Tunnel diode characteristic]%
{\label{fig:tunnel-diode-characteristic}%
\parbox[t]{\linewidth}%
{\centering
\input diode.tex
}}
 \end{minipage}
 \begin{minipage}[b]{0.6\linewidth}
  \centering
\subfigure[Reachable states (dashed)]%
{\label{fig:tunnel-diode-oscillator-invariant}%
\parbox[b]{\linewidth}%
{\centering
\rotatebox{90}{$\qquad\qquad\qquad I_L [\milliampere]$}%
\includegraphics*[scale=0.45, bb = 105 205 488 576]{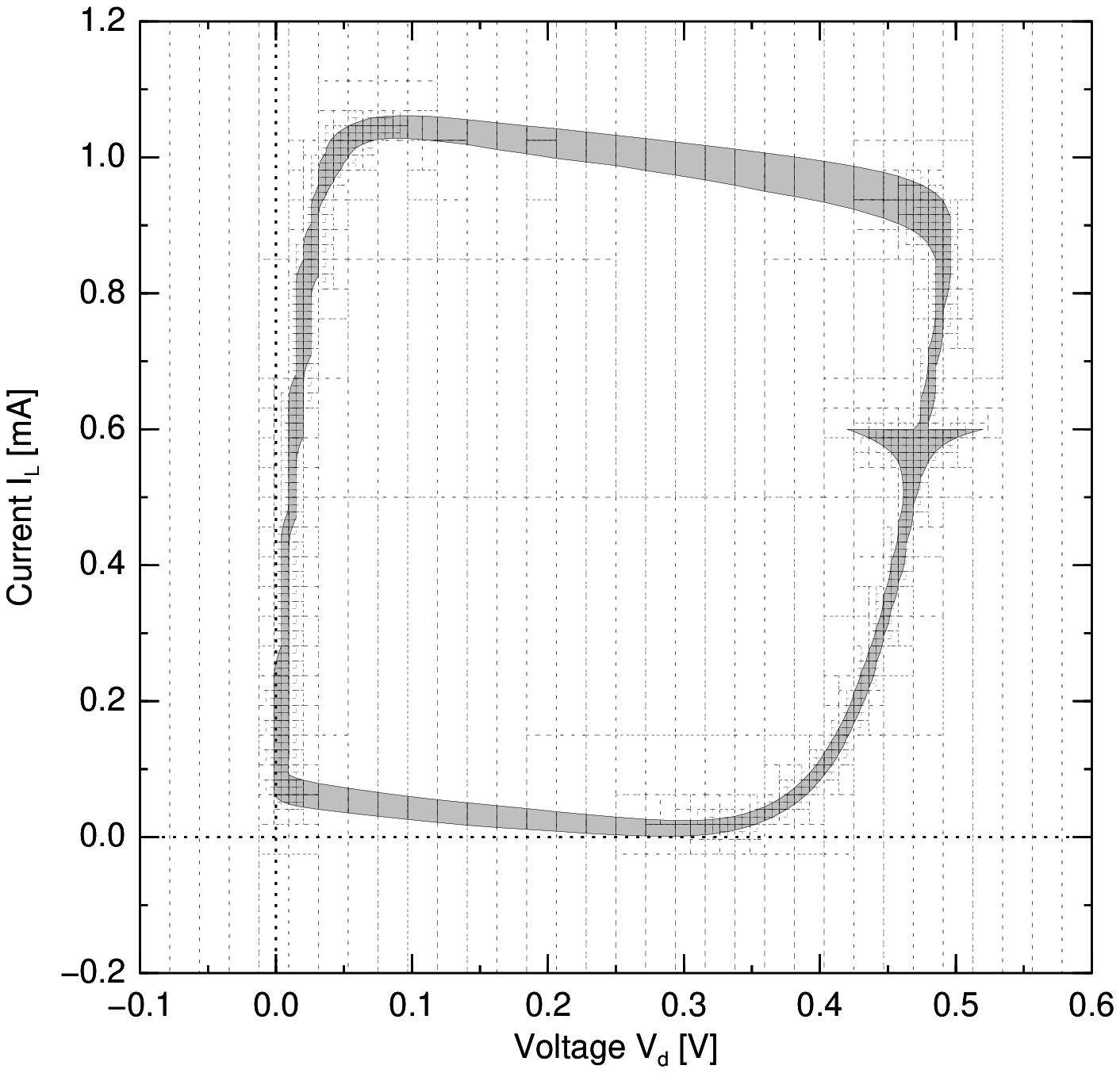} \\
$\quad\quad V_d [\volt]$
}
\rotatebox{90}{\quad\footnotesize{(Picture courtesy of Goran Frehse.)}}
}
 \end{minipage}
\caption{Tunnel-diode oscillator circuit}
\label{fig:tunnel-diode-oscillator}
\end{figure}
}%%\ifthenelse{\boolean{TR}}

Consider the tunnel-diode oscillator schematized
in Figure~\ref{fig:tunnel-diode-oscillator-schematic}.
\ifthenelse{\boolean{TR}}{%%
\begin{figure}
\subfigure[Circuit schematic]%
{\label{fig:tunnel-diode-oscillator-schematic}%
\parbox{0.5\linewidth}%
{\centering
\setlength{\unitlength}{0.240900pt}
\begin{picture}(300,360)(-20,20)
\put(-300,0){
\psset{xunit=0.7cm,yunit=0.7cm,runit=0.7cm}
\psset{origin={0,0}}
\pspicture*[](10,4)
\psset{dipolestyle=elektor}
\pnode(2,3){Vin}
\pnode(1.5,3){S} \pnode(3.5,3){A} \pnode(6.5,3){B} \pnode(8.5,3){C}
\pnode(1.5,1){Sm}\pnode(3.5,1){Am}\pnode(6.5,1){Bm}\pnode(8.5,1){Cm}
\Ucc[dipoleconvention=generator,tension,tensionoffset=0.8,labeloffset=1.2](Sm)(S){$V_\mathrm{in}$}
\resistor[dipolestyle=rectangle](Vin)(A){$R$}
\coil[dipolestyle=elektorcurved,intensitylabel=$I_L$](A)(B){$L$}
\diode[dipoleconvention=receptor,intensitylabel=$I_d$,tensionlabel=$V_d$,tensionoffset=-0.5,tensionlabeloffset=-0.9,](B)(Bm){}
\capacitor(C)(Cm){$C$}
\psset{intensitylabel=}
\wire(Am)(Bm)
\wire(Bm)(Cm)
\wire(B)(C)
\wire(S)(Vin)
\wire(Sm)(Am)
\pscircle*(B){2\pslinewidth}
\pscircle*(Bm){2\pslinewidth}
\endpspicture
}
\end{picture}
}}%
\subfigure[Tunnel diode characteristic]%
{\label{fig:tunnel-diode-characteristic}%
\parbox{0.5\linewidth}%
{\centering
\input diode.tex
}}
\caption{Tunnel-diode oscillator circuit}
\label{fig:tunnel-diode-oscillator}
\end{figure}
}{}%%\ifthenelse{\boolean{TR}}{%%
The state of the system at a given instant of time is completely characterized
by the values of the inductor current $I_L$ and the diode voltage drop $V_d$.
With these as the state variables, the system is described by the
second-order state equations
\ifthenelse{\boolean{TR}}{%%
\begin{align}
\label{eq:tdo-Vd/dt}
  \dot{V}_d &= 1/C\bigl(-I_d(V_d) + I_L\bigr), \\
\label{eq:tdo-IL/dt}
  \dot{I}_L &= 1/L(-V_d -R I_L + V_\mathrm{in}).
\end{align}
}{%%
$\dot{V}_d = 1/C\bigl(-I_d(V_d) + I_L\bigr)$
and
$\dot{I}_L = 1/L(-V_d -R I_L + V_\mathrm{in})$.
}%%\ifthenelse{\boolean{TR}}{%%
In \cite{FrehseKR06} it is shown how a cyclic invariant can be obtained
for this circuit using the PHAVer system.  First, a piecewise affine envelope
is constructed for the tunnel diode characteristic $I_d(V_d)$ depicted
in Figure~\ref{fig:tunnel-diode-characteristic}: for the particular
example analyzed in \cite{FrehseKR06}, sufficient precision is obtained
by dividing the range $V_d \in [-0.1\volt, 0.6\volt]$
\ifthenelse{\boolean{TR}}{%%
into $64$ intervals,
resulting in a piecewise affine model of~\eqref{eq:tdo-Vd/dt}.
}{%%
into $64$ intervals.
}%%\ifthenelse{\boolean{TR}}{%%
Forward reachability computation with PHAVer can obtain the
set of states depicted in Figure~\ref{fig:tunnel-diode-oscillator-invariant}.
\ifthenelse{\boolean{TR}}{%%
\begin{figure}
\hfill
\parbox[t]{0.5\linewidth}%
{\centering
\rotatebox{90}{$\qquad\qquad\qquad\qquad I_L [\milliampere]$}%
\includegraphics*[scale=0.5, bb = 105 205 488 576]{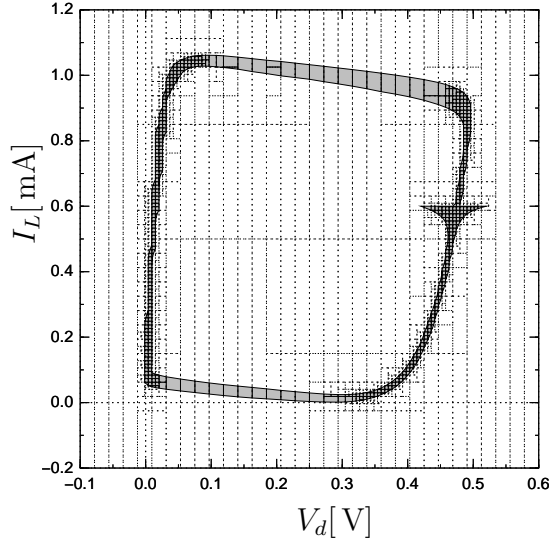} \\
$\qquad\qquad V_d [\volt]$
}
\hfill
\rotatebox{90}{\qquad\quad\footnotesize{(Picture courtesy of Goran Frehse.)}}
\caption{Reachable states of the tunnel-diode oscillator (dashed)}
\label{fig:tunnel-diode-oscillator-invariant}
\end{figure}
}{}%%\ifthenelse{\boolean{TR}}{%%
These are the states reachable from the set of initial states corresponding
to $V_d \in [0.42\volt, 0.52\volt]$ and $I_L = 0.6 \milliampere$
(the base of the downward-facing triangular shape in
Figure~\ref{fig:tunnel-diode-oscillator-invariant}).
\ifthenelse{\boolean{TR}}{%%
Taking into account that
}{%%
As
}%%\ifthenelse{\boolean{TR}}{%%
the loop shape constituted by the reachable states
is traversed clockwise, it can be seen that the inductor current $I_L$ returns
to the initial value of $0.6 \milliampere$ with a diode voltage drop
that is well within the initial range $[0.42\volt, 0.52\volt]$.
The set of reachable states so obtained is thus an invariant of
the circuit.

In \cite{FrehseKR06} it is shown that, due to over-approximation, forward
reachability can fail to determine invariants of more complex circuits.
A new technique combining forward and backward reachability with iterative
refinement of the partitions is thus proposed and shown to be more powerful
and efficient.

\ifthenelse{\boolean{TR}}{%%
\section{Families of Polyhedral Approximations for Analysis and Verification}
}{%%
\section{Families of Polyhedral Approximations}
}%%\ifthenelse{\boolean{TR}}{%%
\label{sec:families-of-polyhedral-approximations}

For several applications of static analysis and verification,
an approximation based on the domain of convex polyhedra can be
regarded as the most appropriate choice.
In this section we discuss alternative options (simplifications,
generalizations, and combinations with other numerical domains)
that might be considered when trying either to reduce the cost
of the analysis, or to increase the precision of the computed results.

\subsection{Simplifications of Polyhedra}
\label{sec:simplifications-of-polyhedra}

\ifthenelse{\boolean{TR}}{%%
There are contexts where approximations based on the domain of
}{%%
There are contexts where approximations based on general
}%% \ifthenelse{\boolean{TR}}
convex polyhedra, no matter which implementation is adopted,
incur an unacceptable computational cost.
In such cases, the static analysis may resort to further simplifications
so as to obtain useful results within reasonable time and space bounds.

A first, almost traditional approach is based on the identification
of suitable syntactic subclasses of polyhedra.
The abstract domain of \emph{bounding boxes} (or intervals~\cite{CousotC76})
is based on polyhedra that can be represented as finite conjunctions
of constraints of the form $\pm x_i \leq d$ or $\pm x_i < d$,
leading to the specification of operations whose worst-case complexity
is linear in the number of space dimensions.
As a more precise alternative, the class of \emph{potential constraints}
\ifthenelse{\boolean{TR}}{%%
\cite{AllenK85,Bagnara97th,Bellman57,Davis87,Dill89,LarsenLPY97},
also known as \emph{bounded differences},
}{%%
\cite{Bellman57},
also known as \emph{bounded differences} \cite{Bagnara97th,Davis87},
}%% \ifthenelse{\boolean{TR}}
allows for constraints
of the form $x_i - x_j \leq d$ or $\pm x_i \leq d$;
\ifthenelse{\boolean{TR}}{%%
the generalization proposed in~\cite{BalasundaramK89},
also admits constraints of the form $x_i + x_j \leq d$,
leading to the abstract domain of \emph{octagons}~\cite{Mine01b}.
}{%%
the abstract domain of \emph{octagons}~\cite{Mine05th}
also admits constraints of the form $x_i + x_j \leq d$.
}%%\ifthenelse{\boolean{TR}}{%%
In these last two cases, the operators are characterized by a worst-case
time complexity which is cubic in the number of space dimensions.
For all of the approximations mentioned above, improved efficiency also
follows from the fact that the corresponding computations are simple enough
to allow for the adoption of floating-point data types: in contrast,
the specification of safe and efficient floating-point operations
for general polyhedra is an open problem, so that polyhedra libraries
have to be based on unbounded precision data types.

Several alternative (syntactic and/or semantic) simplification schemes
have been put forward in the recent literature.
The \emph{Two Variables per Linear Inequality} abstract domain
is proposed in~\cite{SimonKH02},
where constraints take the syntactic form $ax_i + bx_j \leq d$.
In~\cite{SankaranarayananSM05}, an arbitrary family of polyhedra
is chosen before starting the analysis by fixing the slopes of a
finite number of linear inequalities,
which are called the \emph{template constraints};
linear programming techniques are then used to compute precise
approximations in the considered class of shapes.
In contrast, in~\cite{SankaranarayananCSM06}, general polyhedra are allowed,
but the corresponding operations (in particular, the poly-hull and the image
of affine relations) are approximated by less precise variants so as
to ensure a polynomial worst-case complexity in the size of the inputs.
An even more flexible approach is proposed in~\cite{Frehse05},
where arbitrary polyhedra are approximated, when they become too complex,
by limiting the number of constraints in their description and/or
the magnitude of the coefficients occurring in the constraints.
These more dynamic approximation schemes are promising,
in particular for those applications where nothing is known in advance
about the syntactic form of the constraints that will be computed
during the analysis.

An important observation to be made is that there is no actual need
to prefer \emph{a priori} (and therefore commit to) a specific
abstract domain: the analysis tool may be based on several abstractions,
safely switching from more precise, possibly costly domains to more
efficient, possibly imprecise ones, and vice versa, depending on the context.
When replacing a generic polyhedron by a simpler one,
the problem of the identification of a good over-approximation
has to be solved.
Depending on the context, the approaches may vary significantly.
At one extreme, when efficiency is really critical, the adoption
of syntactic techniques should be pursued:
for an interesting example, we refer the reader to one of the
simplification heuristics used in~\cite{Frehse05}, where the efficient
selection of a small number of linear inequalities out of
a constraint system is driven by a simple, yet effective reasoning
on the measure of the angles formed by the corresponding half-spaces.
At the other extreme, linear programming (LP) optimization techniques
may be used so as to obtain the best match in the considered
class of geometric shapes. For instance, the precise approximation
of a polyhedron by a bounding box (resp., a bounded difference
or octagon) can be implemented by a linear (resp., quadratic) number
of optimizations of a class of LP problems, where the objective function
varies while the feasible region is invariant and defined by the
constraints of the polyhedron.
Note that, if correctness has to be preserved, it is essential that no
rounding error is made on the wrong side, so that classical
floating-point implementations of LP solvers have to be considered unsafe,
unless the computed results can be certified by some other tool.
Alternatively, it is possible to consider LP implementations based on
unbounded precision data types.

When the number of space dimensions to be modeled is beyond a given
threshold, the whole analysis space can be split into a finite number
of smaller, more manageable components, thereby realizing a further
simplification scheme that can be combined with those described above.
The splitting strategy varies considerably.
\ifthenelse{\boolean{TR}}{%%
In~\cite{HalbwachsMG06,HalbwachsMP-V03}, Cartesian factoring techniques
}{%%
In~\cite{HalbwachsMG06}, Cartesian factoring techniques
}%%\ifthenelse{\boolean{TR}}{%%
are used so as to dynamically partition the space dimensions of
a polyhedron into independent subsets; the orthogonal factors are then
approximated by lower dimensional polyhedra with no precision penalty.
In an alternative approach described in~\cite{BlanchetCCFMMMR03}, many
(possibly overlapping) small subsets of space dimensions, called
\emph{variable packs}, are identified before the start of the
analysis by means of syntactic conditions; the relations holding
between the variables in each pack are then approximated by using an
octagonal abstraction.
A variation of this is described in~\cite{VenetB04},
where non-overlapping variable packs are dynamically computed
(and possibly merged) during the analysis, whereas the relations
between the variables in a pack are approximated by means of
potential constraints. In~\cite{VenetB04} it is also observed that,
since the average size of variables packs is small (5 variables),
more precise approximations based on general polyhedra should be feasible.

\subsection{Generalizations of Polyhedra}
\label{sec:generalizations-of-polyhedra}

There are applications where the restriction to the domain of convex
polyhedra is intrinsically inadequate. This may happen, not only when
the verification property of interest is itself non-convex,
but also when the adopted computation strategy requires that a
convex property is proved by passing through a non-convex
intermediate approximation.
This was the case in Example~\ref{ex:scheduler-abstract} of
Section~\ref{sec:analysis-of-hybrid-systems}, where the upper bound
($k_1 \leq 2$) on the number of waiting processes for class $A_1$
was obtained by switching from the domain of convex polyhedra
to the domain of finite sets of polyhedra.

\ifthenelse{\boolean{TR}}{%%
The finite powerset domain construction~\cite{Bagnara98SCP}
}{%%
The finite powerset domain construction~\cite{Bagnara98SCP}
}%%\ifthenelse{\boolean{TR}}{%%
is a special case of \emph{disjunctive completion}~\cite{CousotC79},
a systematic technique to derive an enhanced abstract domain
starting from an existing one. A finite powerset domain
implements disjunctions by maintaining an explicit (hence finite)
and \emph{non-redundant} collection of elements of the base-level domain:
non-redundancy means that a collection is made of maximal elements
with respect to the approximation ordering, so that no element
subsumes another element in the collection.

For a better understanding of the concepts, which are described
in completely general terms in~\cite{BagnaraHZ06STTT},
let us consider the application of the finite powerset construction
to the domain of convex polyhedra.
This instantiation (which is the one also adopted for the examples
developed in~\cite{BagnaraHZ06STTT}) can be used to model nonlinear
systems as described, e.g., in Section~\ref{sec:analysis-of-analog-systems}.
Then, an element of the abstract domain is a finite set of maximal
convex polyhedra, so that no polyhedron in the set is contained in
another polyhedron in the set.
The powerset domain is a lattice: the bottom and top elements
are $\emptyset$ and $\{ \Rset^n \}$, respectively;
the meet is obtained by removing redundancies from the set of all possible
binary intersections of an element in the first powerset with an element
in the second powerset; while the binary join is the non-redundant subset of
the union of the two arguments.
Most of the other abstract operations needed for a static analysis
using the finite powerset domain are easily obtained by ``lifting''
the corresponding operations defined on the base-level domain,
and then reinforcing non-redundancy. For instance, the computation
of the image of a finite powerset under an affine relation is obtained
by computing the image of each polyhedron in the collection.
However, the construction of a provably correct widening operator
has only recently been addressed in~\cite{BagnaraHZ06STTT}
(see Section~\ref{sec:widening}).
The generic specification of the abstract operators of
the finite powerset domain in terms of abstract operations
on the (arbitrary) base-level domain allows for the development
of a single implementation which is shared by all the possible
instances of the domain construction.

An alternative abstraction scheme has been proposed in~\cite{BagnaraR-CZ05}
for the case of finite conjunctions of polynomial inequalities.
Intuitively, a polynomial constraint can be approximated by means of
a linear constraint in a higher dimension vector space, so that the
different terms of the polynomial (e.g., $x_0$, $x_0x_1$, $x_0^2$) are mapped
to different and independent space dimensions; these linear constraints
are then used to perform an almost classical linear relation analysis
based on convex polyhedra.
Due to the linearization step, most of the precision of the polynomial
constraints is initially lost; however, some of the relations holding
between the different terms of the original polynomial can be recovered
by adding further constraints that are redundant when interpreted
in the polynomial world, but do contribute to precision in the
linearized space. In particular, in~\cite{BagnaraR-CZ05}
the polynomial constraints are mapped into finitely generated
\emph{polynomial cones} and a degree-bounded product closure operator
is systematically applied so as to improve accuracy.
As a trivial example, let the polynomial terms $x_0$, $x_1$ and $x_0x_1$
be mapped to the space dimensions $y_0$, $y_1$ and $y_2$, respectively.
Then, the linearization of the polynomial constraints $x_0 \geq 0$
and $x_1 \geq 0$ will produce a polyhedron that,
while satisfying $y_0 \geq 0$ and $y_1 \geq 0$,
leaves variable $y_2$ totally unconstrained.
By applying the product closure operator we also obtain
the linear constraint $y_2 \geq 0$,
thereby recovering the non-negativity of term $x_0x_1$.

\subsection{Combinations with other Numerical Abstractions}
\label{sec:other-abstractions}

\ifthenelse{\boolean{TR}}{%%
We observe that there
}{%%
There
}%%\ifthenelse{\boolean{TR}}{%%
are two basic kinds of numerical abstractions for
approximating the values of the program variables:
outer \emph{limits} (or bounds within which the values must lie)
and the pattern of \emph{distribution} of these values.
The first can be approximated by (constructions based on)
convex polyhedra, while the second can be approximated
by sets of congruences defining lattices of points
\ifthenelse{\boolean{TR}}{%%
we call \emph{grids}~\cite{BagnaraDHMZ07,Granger91,Granger97}.
}{%%
we call \emph{grids}~\cite{BagnaraDHMZ07,Granger97}.
}%%\ifthenelse{\boolean{TR}}{%%
Before considering how these and similar domains may be combined,
we give a brief overview of the domain of grids.

Any vector that satisfies
$\langle \vect{a}, \vect{v} \rangle = b + \mu f$,
for some $\mu \in \Zset$, is said to \emph{satisfy} the
congruence relation $\langle \vect{a}, \vect{v} \rangle \equiv_f b$.
A \emph{congruence system} $\cK$ is a finite set of congruence
relations in $\Rset^n$.  A \emph{grid} is the set of all vectors in
$\Rset^n$ that satisfy the congruences in $\cK$.  The domain of grids
$\Gset_{n}$ is the set of all grids in $\Rset^n$ ordered by
the set inclusion relation, so that the empty set and $\Rset^n$ are the
bottom and top elements of $\Gset_n$ respectively and the intersection
of two grids is itself a grid.  Thus, as for the domain of
polyhedra, the domain of grids forms a lattice
$(\Gset_n, \sseq, \emptyset, \Rset^n, \polyhull, \inters)$
where $\polyhull$ denotes the join
operation returning the least grid greater than or equal to the two
arguments.  For more details concerning all aspects of the domain of grids,
see~\cite{BagnaraDHMZ07}.

The distribution information captured by grids has a number
of applications in its own right: for instance, to ensure that
external memory accesses obey the alignment restriction imposed by the
host architecture, and to enable several transformations for efficient
parallel execution as well as optimizations that enhance cache
behavior.  However, here we are primarily concerned with applications
that can benefit from the combination of the domain of grids with that of
convex polyhedra.  For instance, knowing the frequency (and position)
of the points in a grid, we can shrink the polyhedra so that the
bounding hyperplanes pass through the grid values; if this leads to a
polyhedron with reduced dimension (such as a single point) or one that
is empty, it can lead, not only to improved precision, but also a more
efficient use of resources by
the analyzer~\cite{Ancourt91th,NookalaR00,QuintonRR96}.

Generic constructions, such as direct and reduced product, can be used
to provide a formal basis for the combination of the grid and
polyhedral domains \cite{CousotC79} although the exact choice of
product construction used to build the grid-polyhedral domain needs
further study.  Both the direct and reduced products have problems:
the direct product has no provision for communication between the
component domains, thereby losing precision; while the reduced
product, which is the most precise refinement of the direct product,
has exponential complexity. It is expected that, for grid-polyhedra,
the most useful product construction will lie between these
extremes. For instance, as equalities are common entities for both
constraint and congruence systems, if an equality is found to hold in
one component, it is safe to just add this to the other component.  In
addition, in an element of the grid-polyhedral domain, any hyperplane
that bounds the polyhedron component could be moved inwards until it
intersects with points of the grid with only linear cost on the number
of dimensions. Of course, this reduction on its own is not optimal
since the grid points in the intersection may not lie in the
polyhedron itself.  For optimality or, more generally, so as to gain
additional precision, we need to experiment with various forms of the
branch-and-bound and cutting-plane algorithms~\cite{KrishnanM06}
already well-researched for integer linear programming.  What is
needed is a range of options for the product construction allowing the
user to decide on the complexity/precision trade-off.  Further work on
this is needed, including an investigation of other proposals for generic
products that lie between the direct and reduced product, such as the
local decreasing iteration method~\cite{Granger92} and the open
product construction~\cite{CortesiLCVH00}.

\section{Polyhedral Computations Peculiar to Analysis and Verification}
\label{sec:peculiar-polyhedral-computations}

As observed in the previous sections, the analysis of the run-time
behavior of a system can be broken down into a set of
basic operations on the chosen abstract domains.
This means that each abstract domain should provide adequate computational
support for such a set and, where
appropriate, further operations that might be useful for tuning the
cost/precision ratio. In this section, we discuss several key issues
relevant to the design and implementation of an abstract domain of,
or based on, convex polyhedra.
Before going into further detail, it should be stressed that the
particular context of the application plays a significant and
non-trivial role here.
For instance, in many computational complexity studies,
it is assumed that a small number of operations (often, just a single one)
can have arbitrarily large operands;
also, it is typically required that exact results have to be computed.
These assumptions taken together may be inappropriate in the
context of static analysis:
it is quite often the case that a large number of operations will
have only small or medium sized operands; also, whenever
facing an efficiency issue, the exactness requirement can be dropped
(provided soundness is maintained).
As a consequence, the evaluation of alternative algorithmic strategies
should be largely based on practical experimentation.

\subsection{The Double Description Method}
\label{sec:double-description}

Convex polyhedra are typically specified by a finite system
of linear inequality constraints and for this representation
there are known algorithms
(e.g., based on Fourier-Motzkin elimination~\cite{LassezM92,Schrijver99})
for most of the operations already mentioned.

An alternative approach is based on the \emph{double description} method
due to Motzkin et al.~\cite{MotzkinRTT53}. This method was originally defined
on the set of topologically closed convex polyhedra, a sub-lattice
\(
  (\CPset_n, \sseq, \emptyset, \Rset^n, \polyhull, \union)
\)
of the lattice of (not necessarily closed, or NNC) polyhedra $\Pset_n$.
In the double description method, a closed polyhedron may be described by
using a system of non-strict linear inequalities or by using
a \emph{generator} system that records its key geometric features.
The following is the main theoretical result,
which is a simple consequence of well-known theorems
by Minkowski and Weyl~\cite{StoerW70}.

\begin{theorem}
\label{thm:minkowski-weyl}
The set $\cP \sseq \Rset^n$ is a topologically closed convex polyhedron
if and only if there exist finite sets $R, P \sseq \Rset^n$
of cardinality $r$ and $p$, respectively,
such that $\vect{0} \notin R$ and
$\cP$ can be generated from $(R, P)$ as follows:
\[
  \cP = \{\,
          R \vect{\rho} + P \vect{\pi} \in \Rset^n
        \mid
          \vect{\rho} \in \nonnegRset^r,
          \vect{\pi} \in \nonnegRset^p,
          \textstyle{\sum_{i=1}^p \pi_i = 1}
        \,\}.
\]
\end{theorem}

\noindent
Intuitively, a point of a polyhedron $\cP$ is obtained by adding
a convex combination of the vectors in $P$ (the generating points)
to a conic combination of the vectors in $R$ (the generating \emph{rays}).
%%The polyhedron is empty if and only if $P$ is empty.

It turns out that constraint and generator descriptions are duals:
each representation can be computed starting from the other one.
Clever implementations of this conversion procedure, improving on the
\ifthenelse{\boolean{TR}}{%%
Chernikova's algorithms~\cite{Chernikova64,Chernikova65,Chernikova68},
}{%%
Chernikova's algorithm~\cite{Chernikova68},
}%%\ifthenelse{\boolean{TR}}{%%
are the starting point for the development of software libraries that,
while being characterized by a worst case computational cost which is
exponential in the size of the input, turn out to be practically useful.
A common characteristic of these implementations is the exploitation
of \emph{incrementality}, whereby most of the computational work
done for an operation is reused to efficiently compute small variations
of the corresponding result.
Further computational enhancements are obtained by the adoption of
suitable heuristics, ranging from the efficient handling
of adjacency information~\cite{LeVerge92}, to a careful choice of
ordering strategies for the computation of intermediate
results~\cite{Avis00,AvisB95,FukudaP96};
the overall construction typically relies on a tight integration
of the basic algorithms with a carefully chosen set of data structures
\ifthenelse{\boolean{TR}}{%%
\cite{PPL-USER-0-9}.
}{%%
\cite{BagnaraHZ08SCP}.
}%%\ifthenelse{\boolean{TR}}{%%

An important motivation for the adoption of an implementation
based on the double description method is that the ability to switch
from a constraint description to a generator description, or vice versa,
can be usefully exploited to provide simple implementations for the
basic operations on polyhedra.
For instance, set intersection is easily implemented by taking the union
of the constraint systems representing the two arguments,
whereas the poly-hull is implemented by joining the generator systems
representing the two arguments;
and the test for emptiness can be implemented by checking that
the generator system has no points.
Moreover, a test for subset inclusion $\cP \sseq \cQ$ can be
implemented by checking if each point and each ray in a generator
system describing $\cP$ satisfies all linear inequalities in a
constraint system describing $\cQ$.
As a further example, the time elapse operation specified
in Section~\ref{sec:analysis-of-hybrid-systems},
can be implemented using the generator systems for the argument
polyhedra~\cite{HalbwachsPR97}.
That is a generator system for the polyhedron $\cP \timeelapse \cQ$
can be obtained by adopting the same set of generating points as $\cP$ and
by defining its set of rays as the union of
the set of generating rays for $\cP$ with the set of
all the generators (both points and rays) for $\cQ$.

As seen in Section~\ref{sec:analysis-of-computer-programs},
in the context of the analysis of imperative languages
one of the most frequent statements is variable assignment,
where the expression assigned is safely approximated by
an affine relation $\reld{\psi}{\Rset^n}{\Rset^n}$.
The (direct or inverse) image of an affine relation can be naively
computed by embedding the input polyhedron $\cP \sseq \Rset^n$ into
the space $\Rset^{2n}$, intersecting it with the constraints
defining $\psi$ and finally projecting the result back on $\Rset^n$.
However, due to the moves to/from a higher dimensional space,
this approach suffers from significant overheads.
Quite often, the expression assigned is a simple affine function of
the variables' values and can thus be exactly modeled by computing
the image of a single-update affine function.
%% : it is therefore important
%% that such a frequent special case is implemented as efficiently as possible.
With the double description method, the images of affine functions
are much more efficiently computed by applying them
directly to the generators of the argument
polyhedron.
A dual approach, using the constraint description of the polyhedron,
allows for the computation of the preimages of affine functions,
which can be of interest for a backward semantic construction,
where the initial values of program variables are approximated
starting from their final values.
Similar efficiency arguments motivate the study of specific implementations
for single-update bounded affine relations and other special subclasses
of affine relations.

\subsection{Widening and Narrowing}
\label{sec:widening}

The first widening operator for the domain of convex polyhedra,
the so-called \emph{standard widening}
\ifthenelse{\boolean{TR}}{%%
proposed in \cite{CousotH78} and refined in \cite{Halbwachs79th},
}{%%
proposed in \cite{CousotH78},
}%% \ifthenelse{\boolean{TR}}{%%
can be informally described as follows:
suppose that in the post-fixpoint iteration sequence we compute
as successive iterates the polyhedra $\cP_i$ and $\cP_{i+1}$;
then, the widening keeps all and only the constraints defining $\cP_i$
that are also satisfied by $\cP_{i+1}$.
This simple idea, which is basically borrowed from the widening operator
defined on the domain of intervals~\cite{CousotC76}, is quite effective
in ensuring the termination of the analysis (the number of constraints
decreases at each iteration);
by avoiding the application of the widening in the first few
\ifthenelse{\boolean{TR}}{%%
iterations of the analysis~\cite{Cousot81}
}{%%
iterations of the analysis
}%%\ifthenelse{\boolean{TR}}{%%
and/or by applying the ``widening up-to'' technique of~\cite{Halbwachs93},
it also provides, in the main, an adequate level of precision.

Some application fields, however, are particularly sensitive to
the precision of the deduced numerical information, to the point
that some authors propose to give up the termination guarantee
and use so-called \emph{extrapolation} operators: examples include
the operators defined in \cite{HenzingerH95} and \cite{HenzingerPW01},
%%for the \textsc{HyTech} system,
as well as the proposals in \cite{BultanGP99} and \cite{DelzannoP99}
for sets of polyhedra and
the heuristics sketched in \cite{BessonJT99}.

In~\cite{BagnaraHRZ05SCP} this precision problem is
reconsidered in a more general context and a framework is proposed
that is able to improve upon the precision of a given widening
while keeping the termination guarantee. The approach, which
builds on theoretical results put forward in work on
termination analysis, combines an existing widening
operator, whose termination guarantee should be \emph{formally certifiable},
with an arbitrary number of precision improving heuristics.
Its feasibility was demonstrated by instantiating the framework
so as to produce a new widening on polyhedra improving upon the
\ifthenelse{\boolean{TR}}{%%
precision of~\cite{Halbwachs79th}
}{%%
precision of the standard widening
}%% \ifthenelse{\boolean{TR}}{%%
in a significant percentage of benchmarks.

For the more challenging case of an abstract domain obtained
by the finite powerset domain construction, several generic
schemes of widenings have been proposed in~\cite{BagnaraHZ06STTT}
that are able to ``lift'' a widening defined on the base-level domain
without compromising its termination guarantee.
The instantiation of such a generic approach led to the definition
of the first non-trivial and provably correct widenings
on a domain of finite sets of convex polyhedra.
Being highly parametric, the widening schemes proposed
in~\cite{BagnaraHZ06STTT} can be instantiated according to
the needs of the specific application, as done in~\cite{GulavaniR06}.
One of the heuristic approaches adopted in~\cite{BagnaraHZ06STTT}
to control the precision/complexity trade-off of the widenings,
originally proposed in~\cite{BultanGP99}, attempts at reducing
the cardinality of a polyhedral collection by merging
two of its elements whenever their set union happens to be a convex
polyhedron. The implementation of such a heuristic could significantly
benefit from the results and algorithms presented
\ifthenelse{\boolean{TR}}{%%
in~\cite{BaranyF05,BemporadFT01}.
}{%%
in~\cite{BemporadFT01}.
}%%\ifthenelse{\boolean{TR}}{%%

It is also worth mentioning that, once a post-fixpoint approximation
has been obtained by means of an upward iteration sequence with widening,
its precision can be improved by means of a downward iteration,
possibly using a \emph{narrowing operator}
\ifthenelse{\boolean{TR}}{%%
\cite{CousotC76,CousotC77,CousotC92fr,CousotC92plilp}.
}{%%
\cite{CousotC76,CousotC92fr}.
}%%\ifthenelse{\boolean{TR}}{%%
To the best of our knowledge, no narrowing has ever been defined on
the domain of convex polyhedra: applications simply stop the downward
computation after a small number of iterations.

\subsection{Not Necessarily Closed Convex Polyhedra}
\label{sec:nnc-polyhedra}

Most static analysis applications computing linear
inequality relations between program variables
consider the domain $\CPset_n$ of topologically closed polyhedra.
One of the underlying motivations is that sometimes (e.g., when
working with integer valued variables only) strict inequalities
can be filtered away by suitable syntactic manipulations;
even when this is not the case, the topological closure approximation
may be interpreted as a quick and practical workaround to the fact that
some software libraries do not fully support computations on
NNC polyhedra. However, there are applications
\ifthenelse{\boolean{TR}}{%%
\cite{AlurCHH93,ColonS01,HalbwachsPR94,HalbwachsPR97}
}{%%
\cite{AlurCHH93,ColonS01,HalbwachsPR97}
}%%\ifthenelse{\boolean{TR}}{%%
where the ability of encoding and propagating strict inequalities
might be crucial for the usefulness
of the final results.

The first proposal for a systematic implementation of strict
inequalities in a software library based on the double description method
\ifthenelse{\boolean{TR}}{%%
was put forward in~\cite{HalbwachsPR94}:
}{%%
was put forward in~\cite{HalbwachsPR97}:
}%%\ifthenelse{\boolean{TR}}{%%
a syntactic translation embeds
an $n$-dimensional NNC polyhedron $\cP \in \Pset_n$
into an $(n+1)$-dimensional closed polyhedron $\cR \in \CPset_{n+1}$,
by adding a single \emph{slack variable} $\epsilon$, satisfying
the additional side constraints $0 \leq \epsilon \leq 1$.
Namely, any strict inequality constraint
$\langle \vect{a}, \vect{x} \rangle > b$
is translated into the non-strict inequality constraint
$\langle \vect{a}, \vect{x} \rangle - \epsilon \geq b$.
The computation is thus performed on the closed representation
$\cR \in \CPset_{n+1}$, with only minor adaptations to the basic
algorithms so as to
\ifthenelse{\boolean{TR}}{%%
also
}{}%%\ifthenelse{\boolean{TR}}{%%
take into account the \emph{implicit}
strict constraint $\epsilon > 0$.

While this idea is quite effective, the resulting software library
no longer enjoys all of the properties of the underlying double
description implementation: NNC polyhedra cannot be suitably described
using generator systems, and the geometric intuitions are
lost under the ``implementation details.''
These problems motivated the studies
\ifthenelse{\boolean{TR}}{%%
in~\cite{BagnaraHZ03a,BagnaraHZ05FAC,BagnaraRZH02},
}{%%
in~\cite{BagnaraHZ05FAC},
}%%\ifthenelse{\boolean{TR}}{%%
where a proper generalization of the double description method
to NNC polyhedra was proposed.
The main improvement was the identification of the \emph{closure point}
as a new kind of generator for NNC polyhedra, leading to the following
result generalizing Theorem~\ref{thm:minkowski-weyl}:

\begin{theorem}
\label{thm:NNC-minkowski-weyl}
The set $\cP \sseq \Rset^n$ is an NNC polyhedron if and only if
there exist finite sets $R, P, C \sseq \Rset^n$
of cardinality $r$, $p$ and $c$
%%, respectively,
such that $\vect{0} \notin R$ and
\ifthenelse{\boolean{TR}}{%%
\[
  \cP
    = \sset{
        R \vect{\rho} + P \vect{\pi} + C \vect{\gamma} \in \Rset^n
      }{
        \vect{\rho} \in \nonnegRset^r,
        \vect{\pi} \in \nonnegRset^p, \vect{\pi} \neq \vect{0},
        \vect{\gamma} \in \nonnegRset^c, \\
        \sum_{i=1}^p \pi_i + \sum_{i=1}^c \gamma_i = 1
      }.
\]
}{%%
\[
  \cP
    = \bigl\{\,
        R \vect{\rho} + P \vect{\pi} + C \vect{\gamma}
      \bigm|
        \vect{\rho} \in \nonnegRset^r,
        \vect{\pi} \in \nonnegRset^p \setminus \{ \vect{0} \},
        \vect{\gamma} \in \nonnegRset^c,
        \textstyle{\sum_{i=1}^p} \pi_i
          + \textstyle{\sum_{i=1}^c} \gamma_i = 1
      \,\bigr\}.
\]
}%\ifthenelse{\boolean{TR}}{%%
\end{theorem}

\noindent
The new condition $\vect{\pi} \neq \vect{0}$ ensures that
at least one of the points of $P$ plays an active role
in any convex combination of the vectors of $P$ and $C$.
As a consequence, the vectors of $C$ are closure points of $\cP$,
i.e., points that belong to the topological closure of $\cP$,
but may not belong to $\cP$ itself.

Thanks to the introduction of (strict inequalities and) closure
points, most of the pros of the double description method now also
\ifthenelse{\boolean{TR}}{%%
apply to the domain of NNC polyhedra:
}{%%
apply to the domain of NNC polyhedra~\cite{BagnaraHZ05FAC}:
}%%\ifthenelse{\boolean{TR}}{%%
simpler, higher-level
implementations of operations on NNC polyhedra can be specified,
reasoned about and justified in terms of any one of the two dual
descriptions; important implementation issues (such as the need to
identify and remove all kinds of redundancies in the
\ifthenelse{\boolean{TR}}{%%
descriptions~\cite{BagnaraHZ05FAC,BagnaraRZH02})
}{%%
descriptions)
}%%\ifthenelse{\boolean{TR}}{%%
can be provided with proper
solutions; different lower-level encodings
(e.g., an alternative management of the
\ifthenelse{\boolean{TR}}{%%
slack variable~\cite{BagnaraHZ03a,BagnaraHZ05FAC})
}{%%
slack variable)
}%%\ifthenelse{\boolean{TR}}{%%
can be investigated and experimented with, without affecting the user
of the software library.
It would be interesting, from both a theoretical and practical
point of view, to provide a more direct encoding of NNC polyhedra, i.e.,
one that is not based on the use of slack variables; this requires
the specification and the corresponding proof of correctness of a
direct NNC conversion algorithm, potentially achieving a major
efficiency improvement.

\section{Conclusion}
\label{sec:conclusion}

In the field of automatic analysis and verification of software and
hardware systems,
approximate reasoning on numerical quantities is crucial.
As first recognized in 1978 \cite{CousotH78},
polyhedral computation algorithms can be used for the automatic
inference of numerical assertions that correctly (though usually
not completely) characterize the behavior of a system at some
level of abstraction.

Until the end of the 1990's these techniques were not in
widespread use, mainly due to the unavailability of robust and
efficient implementations of convex polyhedra.
As far as we know, the first published libraries of polyhedral algorithms
suitable for analysis and verification purposes have been
\emph{Polylib},%
\TRfootnote{\url{http://www.ee.byu.edu/faculty/wilde/polyhedra.html}.}
released in 1995, written by Wilde at IRISA \cite{Wilde93th}
and based on earlier work by Le~Verge \cite{LeVerge92},
and the polyhedra library of \emph{POLINE} (POLyhedra INtegrated Environment)
written by Halbwachs and Proy at Verimag and also released in 1995.
Both libraries used machine integers to represent the coefficients of
linear equalities and inequalities, something that could easily result
into (undetected) overflows.  While Polylib provided only a fraction of the
functionalities offered by POLINE's library (which offered, among other
things, support for NNC polyhedra), it was available in source
format.  The POLINE's library, instead, was distributed only in binary
form for the Sun-4 platform (freely, until about the year 1996; under rather
restrictive conditions afterward).  POLINE included also a system called
POLKA (POLyhedra desK cAlculator) and an analyzer for linear hybrid automata.
A variation of a subset of POLINE's library was incorporated into the
\emph{HyTech} tool~\cite{HenzingerHW97b}.%
\TRfootnote{\url{http://embedded.eecs.berkeley.edu/research/hytech/}.}

The work of Wilde and Le~Verge,
which was extended by Loechner \cite{Loechner99},
led to the creation of \emph{PolyLib}.%
\TRfootnote{\url{http://icps.u-strasbg.fr/polylib/}.}
The \emph{New Polka} library by Jeannet,%
\TRfootnote{\url{http://pop-art.inrialpes.fr/people/bjeannet/newpolka/index.html}.}
first released in 2000 and originally based on both IRISA's Polylib
and POLINE's library, incorporates the idea
---suggested by Fukuda and Prodon \cite{FukudaP96}---
of lexicographically sorting the matrices representing constraints and
generators.
New Polka, which supports both closed and NNC polyhedra,
together with Min\'e's \emph{Octagon Abstract Domain Library}
\ifthenelse{\boolean{TR}}{%%
\cite{Mine01b,Mine05th}%
}{%%
\cite{Mine05th}%
}%%\ifthenelse{\boolean{TR}}{%%
\TRfootnote{\url{http://www.di.ens.fr/~mine/oct/}}
and an interval library called \emph{ITV}, is now included in
the \emph{APRON} library.%
\TRfootnote{\url{http://apron.cri.ensmp.fr/library/}.}
Finally, the \emph{Parma Polyhedra Library} (PPL),
initially inspired by New Polka and first released in 2001,
is developed and maintained by the authors of this paper.%
\TRfootnote{\url{http://www.cs.unipr.it/ppl}.}
The PPL supports both closed and NNC polyhedra, bounding boxes,
bounded difference and octagonal shapes, grids and combinations
of the above including the finite powerset construction
\ifthenelse{\boolean{TR}}{%%
\cite{BagnaraHZ06TR,BagnaraHZ08SCP}.
}{%%
\cite{BagnaraHZ08SCP}.
}%%\ifthenelse{\boolean{TR}}{%%

The above libraries have all been designed specifically for
applications of analysis and verification such as those described
in this paper.  However, two libraries that were designed for solving vertex
enumeration/convex hull problems have successfully been used in
static analysis and computer-aided verification tools:
Fukuda's \emph{cddlib},%
\TRfootnote{\url{http://www.ifor.math.ethz.ch/~fukuda/cdd_home/}.}
an implementation of the double description method \cite{MotzkinRTT53};
and \emph{lrslib},%
\TRfootnote{\url{http://cgm.cs.mcgill.ca/~avis/C/lrs.html}.}
the implementation by Avis of the reverse search algorithm \cite{Avis00}.

All the libraries mentioned in the last two paragraphs are distributed under
free software licenses and support the use of unbounded numeric coefficients.
This, together with the ever increasing available computing power and
the growing interest in ensuring the correctness of critical systems,
has caused, in the 2000's, the continuous emergence of new tools and
applications of polyhedral computations in the area of formal methods.
As a consequence, this is much more of a new beginning than an end
to research in this area.
As explained in
Sections~\ref{sec:families-of-polyhedral-approximations}
and~\ref{sec:peculiar-polyhedral-computations}, several open issues remain.
Most of them have to do with the need for effectively managing the
complexity-precision trade-off: the encouraging results obtained with
today's tools are pushing us to apply them to more complex systems
for a possibly
more precise analysis and/or verification of more complex properties.

\ifthenelse{\boolean{TR}}{%%
\section*{Acknowledgments}
We thank Goran Frehse for the discussion we had on the subject of polyhedra
simplifications and for contributing the PostScript code we used to produce
Figure~\ref{fig:tunnel-diode-oscillator-invariant}.
}{%%
\textbf{Acknowledgments\ }
We thank Goran Frehse for the discussion we had on polyhedra
simplifications and for the PostScript code for
Figure~\ref{fig:tunnel-diode-oscillator-invariant}.
}%%\ifthenelse{\boolean{TR}}{%%

%\bibliographystyle{elsart-num-sort}
%\bibliography{ppl,ppl_citations,mybib}

\begin{thebibliography}{100}
\expandafter\ifx\csname url\endcsname\relax
  \def\url#1{\texttt{#1}}\fi
\expandafter\ifx\csname urlprefix\endcsname\relax\def\urlprefix{URL }\fi

\bibitem{AlefeldH83}
G.~Alefeld, J.~Herzberger, Introduction to Interval Computation, Academic
  Press, New York, 1983.

\bibitem{AllenK85}
J.~F. Allen, H.~A. Kautz, A model of naive temporal reasoning, in: J.~R. Hobbs,
  R.~Moore (eds.), Formal Theories of the Commonsense World, Ablex, Norwood,
  NJ, 1985, pp. 251--268.

\bibitem{AlurCHHHNOSY95}
R.~Alur, C.~Courcoubetis, N.~Halbwachs, T.~A. Henzinger, P.-H. Ho, X.~Nicollin,
  A.~Olivero, J.~Sifakis, S.~Yovine, The algorithmic analysis of hybrid
  systems, Theoretical Computer Science 138 (1995) 3--34.

\bibitem{AlurCHH93}
R.~Alur, C.~Courcoubetis, T.~A. Henzinger, P.-H. Ho, Hybrid automata: An
  algorithmic approach to the specification and verification of hybrid systems,
  in: Hybrid Systems I, vol. 736 of Lecture Notes in Computer Science, 1993.

\bibitem{Ancourt91th}
C.~Ancourt, G\'en\'eration automatique de codes de transfert pour
  multiprocesseurs \`a m\'emoires locales, Ph.D. thesis, Universit\'e de Paris
  VI, Paris, France (Mar. 1991).

\bibitem{Avis00}
D.~Avis, {lrs}: A revised implementation of the reverse search vertex
  enumeration algorithm, in: G.~Kalai, G.~M. Ziegler (eds.), Polytopes ---
  Combinatorics and Computation, vol.~29 of Oberwolfach Seminars,
  {Birkh\"auser-Verlag}, 2000, pp. 177--198.

\bibitem{AvisB95}
D.~Avis, D.~Bremner, How good are convex hull algorithms?, in: Proceedings of
  the Eleventh Annual Symposium on Computational Geometry, ACM Press,
  Vancouver, B.C., Canada, 1995.

\bibitem{Bagnara97th}
R.~Bagnara, Data-flow analysis for constraint logic-based languages, Ph.D.
  thesis, Dipartimento di Informatica, Universit\`a di Pisa, Pisa, Italy,
  printed as Report TD-1/97 (Mar. 1997).

\bibitem{Bagnara98SCP}
R.~Bagnara, A hierarchy of constraint systems for data-flow analysis of
  constraint logic-based languages, Science of Computer Programming 30~(1--2)
  (1998) 119--155.

\bibitem{BagnaraDHMZ07}
R.~Bagnara, K.~Dobson, P.~M. Hill, M.~Mundell, E.~Zaffanella, Grids: A domain
  for analyzing the distribution of numerical values, in: G.~Puebla (ed.),
  Logic-based Program Synthesis and Transformation, 16th International
  Symposium, vol. 4407 of Lecture Notes in Computer Science, Springer-Verlag,
  Berlin, Venice, Italy, 2007.

\bibitem{BagnaraHPZ07TR}
R.~Bagnara, P.~M. Hill, A.~Pescetti, E.~Zaffanella, On the design of generic
  static analyzers for modern imperative languages, Tech. Rep.~{\tt
  arXiv:cs.PL/0703116}, Dipartimento di Matematica, Universit\`a di Parma,
  Italy, available from \url{http://arxiv.org/} (2007).

\bibitem{BagnaraHRZ05SCP}
R.~Bagnara, P.~M. Hill, E.~Ricci, E.~Zaffanella, Precise widening operators for
  convex polyhedra, Science of Computer Programming 58~(1--2) (2005) 28--56.

\bibitem{BagnaraHZ03a}
R.~Bagnara, P.~M. Hill, E.~Zaffanella, A new encoding and implementation of not
  necessarily closed convex polyhedra, in: M.~Leuschel, S.~Gruner, S.~{Lo
  Presti} (eds.), Proceedings of the 3rd Workshop on Automated Verification of
  Critical Systems, Southampton, UK, 2003, published as TR Number
  DSSE-TR-2003-2, University of Southampton.

\bibitem{BagnaraHZ05FAC}
R.~Bagnara, P.~M. Hill, E.~Zaffanella, Not necessarily closed convex polyhedra
  and the double description method, Formal Aspects of Computing 17~(2) (2005)
  222--257.

\bibitem{BagnaraHZ06TR}
R.~Bagnara, P.~M. Hill, E.~Zaffanella, The {Parma Polyhedra Library}: Toward a
  complete set of numerical abstractions for the analysis and verification of
  hardware and software systems, Quaderno 457, Dipartimento di Matematica,
  Universit\`a di Parma, Italy, available at
  \url{http://www.cs.unipr.it/Publications/}. Also published as {\tt
  arXiv:cs.MS/0612085}, available from \url{http://arxiv.org/}. (2006).

\bibitem{PPL-USER-0-9}
R.~Bagnara, P.~M. Hill, E.~Zaffanella, The {Parma Polyhedra Library} User's
  Manual, Department of Mathematics, University of Parma, Parma, Italy, release
  0.9 ed., available at \url{http://www.cs.unipr.it/ppl/} (Mar. 2006).

\bibitem{BagnaraHZ06STTT}
R.~Bagnara, P.~M. Hill, E.~Zaffanella, Widening operators for powerset domains,
  Software Tools for Technology Transfer 8~(4/5) (2006) 449--466. (As the
  figures in the journal version of this paper have been improperly printed
  ---rendering them useless---, we recommend that interested readers download
  an electronic copy from the PPL's web site at
  \url{http://www.cs.unipr.it/ppl/}.)

\bibitem{BagnaraHZ08SCP}
R.~Bagnara, P.~M. Hill, E.~Zaffanella, The {Parma Polyhedra Library}: Toward a
  complete set of numerical abstractions for the analysis and verification of
  hardware and software systems, Science of Computer ProgrammingTo appear.
  Journal version of \cite{BagnaraHZ06TR}.

\bibitem{BagnaraRZH02}
R.~Bagnara, E.~Ricci, E.~Zaffanella, P.~M. Hill, Possibly not closed convex
  polyhedra and the {Parma Polyhedra Library}, in: M.~V. Hermenegildo,
  G.~Puebla (eds.), Static Analysis: Proceedings of the 9th International
  Symposium, vol. 2477 of Lecture Notes in Computer Science, Springer-Verlag,
  Berlin, Madrid, Spain, 2002.

\bibitem{BagnaraR-CZ05}
R.~Bagnara, E.~Rodr{\'\i}guez-Carbonell, E.~Zaffanella, Generation of basic
  semi-algebraic invariants using convex polyhedra, in: C.~Hankin, I.~Siveroni
  (eds.), Static Analysis: Proceedings of the 12th International Symposium,
  vol. 3672 of Lecture Notes in Computer Science, Springer-Verlag, Berlin,
  London, UK, 2005.

\bibitem{BalasundaramK89}
V.~Balasundaram, K.~Kennedy, A technique for summarizing data access and its
  use in parallelism enhancing transformations, in: B.~Knobe (ed.), Proceedings
  of the ACM SIGPLAN'89 Conference on Programming Language Design and
  Implementation (PLDI), vol. 24(7) of ACM SIGPLAN Notices, ACM Press,
  Portland, Oregon, USA, 1989.

\bibitem{BaranyF05}
I.~B\'ar\'any, K.~Fukuda, A case when the union of polytopes is convex, Linear
  Algebra and its Applications 397 (2005) 381--388.

\bibitem{Bellman57}
R.~Bellman, Dynamic Programming, Princeton University Press, 1957.

\bibitem{BemporadFT01}
A.~Bemporad, K.~Fukuda, F.~D. Torrisi, Convexity recognition of the union of
  polyhedra, Computational Geometry: Theory and Applications 18~(3) (2001)
  141--154.

\bibitem{BenoyK97}
F.~Benoy, A.~King, Inferring argument size relationships with
  {CLP($\mathcal{R}$)}, in: J.~P. Gallagher (ed.), Logic Program Synthesis and
  Transformation: Proceedings of the 6th International Workshop, vol. 1207 of
  Lecture Notes in Computer Science, Springer-Verlag, Berlin, Stockholm,
  Sweden, 1997.

\bibitem{BessonJT99}
F.~Besson, T.~P. Jensen, J.-P. Talpin, Polyhedral analysis for synchronous
  languages, in: A.~Cortesi, G.~Fil\'e (eds.), Static Analysis: Proceedings of
  the 6th International Symposium, vol. 1694 of Lecture Notes in Computer
  Science, Springer-Verlag, Berlin, Venice, Italy, 1999.

\bibitem{Birkhoff67}
G.~Birkhoff, Lattice Theory, vol. XXV of Colloquium Publications, 3rd ed.,
  American Mathematical Society, Providence, Rhode Island, USA, 1967.

\bibitem{BlanchetCCFMMMR03}
B.~Blanchet, P.~Cousot, R.~Cousot, J.~Feret, L.~Mauborgne, A.~Min\'e,
  D.~Monniaux, X.~Rival, A static analyzer for large safety-critical software,
  in: Proceedings of the ACM SIGPLAN 2003 Conference on Programming Language
  Design and Implementation (PLDI'03), ACM Press, San Diego, California, USA,
  2003.

\bibitem{BultanGP99}
T.~Bultan, R.~Gerber, W.~Pugh, Model-checking concurrent systems with unbounded
  integer variables: Symbolic representations, approximations, and experimental
  results, ACM Transactions on Programming Languages and Systems 21~(4) (1999)
  747--789.

\bibitem{Chernikova64}
N.~V. Chernikova, Algorithm for finding a general formula for the non-negative
  solutions of system of linear equations, U.S.S.R. Computational Mathematics
  and Mathematical Physics 4~(4) (1964) 151--158.

\bibitem{Chernikova65}
N.~V. Chernikova, Algorithm for finding a general formula for the non-negative
  solutions of system of linear inequalities, U.S.S.R. Computational
  Mathematics and Mathematical Physics 5~(2) (1965) 228--233.

\bibitem{Chernikova68}
N.~V. Chernikova, Algorithm for discovering the set of all solutions of a
  linear programming problem, U.S.S.R. Computational Mathematics and
  Mathematical Physics 8~(6) (1968) 282--293.

\bibitem{ColonS01}
M.~A. Col\'on, H.~B. Sipma, Synthesis of linear ranking functions, in:
  T.~Margaria, W.~Yi (eds.), Tools and Algorithms for Construction and Analysis
  of Systems, 7th International Conference, TACAS 2001, vol. 2031 of Lecture
  Notes in Computer Science, Springer-Verlag, Berlin, Genova, Italy, 2001.

\bibitem{CortesiLCVH00}
A.~Cortesi, B.~{Le Charlier}, P.~{Van Hentenryck}, Combinations of abstract
  domains for logic programming: Open product and generic pattern construction,
  Science of Computer Programming 38~(1--3) (2000) 27--71.

\bibitem{Cousot81}
P.~Cousot, Semantic foundations of program analysis, in: S.~S. Muchnick, N.~D.
  Jones (eds.), Program Flow Analysis: Theory and Applications, chap.~10,
  Prentice Hall, Englewood Cliffs, NJ, USA, 1981, pp. 303--342.

\bibitem{Cousot05}
P.~Cousot, Proving program invariance and termination by parametric
  abstraction, lagrangian relaxation and semidefinite programming, in:
  R.~Cousot (ed.), Verification, Model Checking and Abstract Interpretation:
  Proceedings of the 6th International Conference (VMCAI 2005), vol. 3385 of
  Lecture Notes in Computer Science, Springer-Verlag, Berlin, Paris, France,
  2005.

\bibitem{CousotC76}
P.~Cousot, R.~Cousot, Static determination of dynamic properties of programs,
  in: B.~Robinet (ed.), Proceedings of the Second International Symposium on
  Programming, Dunod, Paris, France, Paris, France, 1976.

\bibitem{CousotC77}
P.~Cousot, R.~Cousot, Abstract interpretation: A unified lattice model for
  static analysis of programs by construction or approximation of fixpoints,
  in: Proceedings of the Fourth Annual ACM Symposium on Principles of
  Programming Languages, ACM Press, New York, 1977.

\bibitem{CousotC79}
P.~Cousot, R.~Cousot, Systematic design of program analysis frameworks, in:
  Proceedings of the Sixth Annual ACM Symposium on Principles of Programming
  Languages, ACM Press, New York, 1979.

\bibitem{CousotC92fr}
P.~Cousot, R.~Cousot, Abstract interpretation frameworks, Journal of Logic and
  Computation 2~(4) (1992) 511--547.

\bibitem{CousotC92plilp}
P.~Cousot, R.~Cousot, Comparing the {Galois} connection and widening/narrowing
  approaches to abstract interpretation, in: M.~Bruynooghe, M.~Wirsing (eds.),
  Proceedings of the 4th International Symposium on Programming Language
  Implementation and Logic Programming, vol. 631 of Lecture Notes in Computer
  Science, Springer-Verlag, Berlin, Leuven, Belgium, 1992.

\bibitem{CousotC92}
P.~Cousot, R.~Cousot, Inductive definitions, semantics and abstract
  interpretation, in: Proceedings of the Nineteenth Annual ACM Symposium on
  Principles of Programming Languages, ACM Press, Albuquerque, New Mexico, USA,
  1992.

\bibitem{CousotH78}
P.~Cousot, N.~Halbwachs, Automatic discovery of linear restraints among
  variables of a program, in: Conference Record of the Fifth Annual ACM
  Symposium on Principles of Programming Languages, ACM Press, Tucson, Arizona,
  1978.

\bibitem{DangDM04}
T.~Dang, A.~Donz{\'e}, O.~Maler, Verification of analog and mixed-signal
  circuits using hybrid system techniques, in: A.~J. Hu, A.~K. Martin (eds.),
  Proceedings of the 5th International Conference on Formal Methods in
  Computer-Aided Design, vol. 3312 of Lecture Notes in Computer Science,
  Springer-Verlag, Berlin, Austin, Texas, USA, 2004.

\bibitem{Davis87}
E.~Davis, Constraint propagation with interval labels, Artificial Intelligence
  32~(3) (1987) 281--331.

\bibitem{DelzannoP99}
G.~Delzanno, A.~Podelski, Model checking in {CLP}, in: R.~Cleaveland (ed.),
  Tools and Algorithms for Construction and Analysis of Systems, 5th
  International Conference, TACAS '99, vol. 1579 of Lecture Notes in Computer
  Science, Springer-Verlag, Berlin, Amsterdam, The Netherlands, 1999.

\bibitem{Dill89}
D.~L. Dill, Timing assumptions and verification of finite-state concurrent
  systems, in: J.~Sifakis (ed.), Proceedings of the International Workshop on
  Automatic Verification Methods for Finite State Systems, vol. 407 of Lecture
  Notes in Computer Science, Springer-Verlag, Berlin, Grenoble, France, 1989.

\bibitem{DooseM05}
D.~Doose, Z.~Mammeri, Polyhedra-based approach for incremental validation of
  real-time systems, in: L.~T. Yang, M.~Amamiya, Z.~Liu, M.~Guo, F.~J. Rammig
  (eds.), Proceedings of the International Conference on Embedded and
  Ubiquitous Computing (EUC 2005), vol. 3824 of Lecture Notes in Computer
  Science, Springer-Verlag, Berlin, Nagasaki, Japan, 2005.

\bibitem{DorRS01}
N.~Dor, M.~Rodeh, S.~Sagiv, Cleanness checking of string manipulations in {C}
  programs via integer analysis, in: P.~Cousot (ed.), Static Analysis: 8th
  International Symposium, SAS 2001, vol. 2126 of Lecture Notes in Computer
  Science, Springer-Verlag, Berlin, Paris, France, 2001.

\bibitem{DoyenHR05}
L.~Doyen, T.~A. Henzinger, J.-F. Raskin, Automatic rectangular refinement of
  affine hybrid systems, in: P.~Pettersson, W.~Yi (eds.), Proceedings of the
  3rd International Conference on Formal Modeling and Analysis of Timed Systems
  (FORMATS 2005), vol. 3829 of Lecture Notes in Computer Science,
  Springer-Verlag, Berlin, Uppsala, Sweden, 2005.

\bibitem{Ellenbogen04th}
R.~Ellenbogen, Fully automatic verification of absence of errors via
  interprocedural integer analysis, Master's thesis, School of Computer
  Science, Tel-Aviv University, Tel-Aviv, Israel (Dec. 2004).

\bibitem{Frehse04}
G.~Frehse, Compositional verification of hybrid systems with discrete
  interaction using simulation relations, in: Proceedings of the IEEE
  Conference on Computer Aided Control Systems Design (CACSD 2004), Taipei,
  Taiwan, 2004.

\bibitem{Frehse05}
G.~Frehse, {PHAVer}: Algorithmic verification of hybrid systems past {HyTech},
  in: M.~Morari, L.~Thiele (eds.), Hybrid Systems: Computation and Control:
  Proceedings of the 8th International Workshop (HSCC 2005), vol. 3414 of
  Lecture Notes in Computer Science, Springer-Verlag, Berlin, Z{\"u}rich,
  Switzerland, 2005.

\bibitem{FrehseHK04}
G.~Frehse, Z.~Han, B.~Krogh, Assume-guarantee reasoning for hybrid
  {I/O}-automata by over-approximation of continuous interaction, in:
  Proceedings of the 43rd IEEE Conference on Decision and Control (CDC 2004),
  Atlantis, Paradise Island, Bahamas, 2004.

\bibitem{FrehseKR06}
G.~Frehse, B.~H. Krogh, R.~A. Rutenbar, Verifying analog oscillator circuits
  using forward/backward refinement, in: Proceedings of the 9th Conference on
  Design, Automation and Test in Europe (DATE 06), ACM SIGDA, Munich, Germany,
  2006, {CD-ROM} publication.

\bibitem{FrehseKRM06}
G.~Frehse, B.~H. Krogh, R.~A. Rutenbar, O.~Maler, Time domain verification of
  oscillator circuit properties, in: Proceedings of the First Workshop on
  Formal Verification of Analog Circuits (FAC 2005), vol. 153 of Electronic
  Notes in Theoretical Computer Science, Elsevier Science B.V., Edinburgh,
  Scotland, 2006.

\bibitem{FukudaP96}
K.~Fukuda, A.~Prodon, Double description method revisited, in: M.~Deza,
  R.~Euler, Y.~Manoussakis (eds.), Combinatorics and Computer Science, 8th
  Franco-Japanese and 4th Franco-Chinese Conference, Brest, France, July 3-5,
  1995, Selected Papers, vol. 1120 of Lecture Notes in Computer Science,
  Springer-Verlag, Berlin, 1996.

\bibitem{GobertLC07}
F.~Gobert, B.~{Le Charlier}, A system to check operational properties of logic
  programs, in: M.-L. Potet, P.-Y. Schobbens, H.~Toussaint, G.~Saval (eds.),
  Approches Formelles dans l'Assistance au D\'eveloppement de Logiciels: Actes
  de la 8e conf\'erence, Universit\'e de Namur, Belgium, 2007.

\bibitem{GopanRS05}
D.~Gopan, T.~W. Reps, M.~Sagiv, A framework for numeric analysis of array
  operations, in: Proceedings of the 32nd ACM SIGPLAN-SIGACT Symposium on
  Principles of Programming Languages, Long Beach, California, USA, 2005.

\bibitem{Granger91}
P.~Granger, Static analysis of linear congruence equalities among variables of
  a program, in: S.~Abramsky, T.~S.~E. Maibaum (eds.), TAPSOFT'91: Proceedings
  of the International Joint Conference on Theory and Practice of Software
  Development, Volume 1: Colloquium on Trees in Algebra and Programming
  (CAAP'91), vol. 493 of Lecture Notes in Computer Science, Springer-Verlag,
  Berlin, Brighton, UK, 1991.

\bibitem{Granger92}
P.~Granger, Improving the results of static analyses programs by local
  decreasing iteration, in: R.~K. Shyamasundar (ed.), Proceedings of the 12th
  Conference on Foundations of Software Technology and Theoretical Computer
  Science, vol. 652 of Lecture Notes in Computer Science, Springer-Verlag,
  Berlin, New Delhi, India, 1992.

\bibitem{Granger97}
P.~Granger, Static analyses of congruence properties on rational numbers
  (extended abstract), in: P.~{Van Hentenryck} (ed.), Static Analysis:
  Proceedings of the 4th International Symposium, vol. 1302 of Lecture Notes in
  Computer Science, Springer-Verlag, Berlin, Paris, France, 1997.

\bibitem{GulavaniR06}
B.~S. Gulavani, S.~K. Rajamani, Counterexample driven refinement for abstract
  interpretation, in: H.~Hermanns, J.~Palsberg (eds.), Proceedings of the 12th
  International Conference on Tools and Algorithms for the Construction and
  Analysis of Systems (TACAS 2006), vol. 3920 of Lecture Notes in Computer
  Science, Springer-Verlag, Berlin, Vienna, Austria, 2006.

\bibitem{GuptaKR04}
S.~Gupta, B.~H. Krogh, R.~A. Rutenbar, Towards formal verification of analog
  designs, in: Proceedings of the 2004 International Conference on
  Computer-Aided Design, IEEE Computer Society / ACM, San Jose, CA, USA, 2004.

\bibitem{Halbwachs79th}
N.~Halbwachs, D\'etermination automatique de relations lin\'eaires
  v\'erifi\'ees par les variables d'un programme, {Th\`ese de
  3\textsuperscript{\`eme} cycle d'informatique}, Universit\'e scientifique et
  m\'edicale de Grenoble, Grenoble, France (Mar. 1979).

\bibitem{Halbwachs93}
N.~Halbwachs, Delay analysis in synchronous programs, in: C.~Courcoubetis
  (ed.), Computer Aided Verification: Proceedings of the 5th International
  Conference, vol. 697 of Lecture Notes in Computer Science, Springer-Verlag,
  Berlin, Elounda, Greece, 1993.

\bibitem{HalbwachsMG06}
N.~Halbwachs, D.~Merchat, L.~Gonnord, Some ways to reduce the space dimension
  in polyhedra computations, Formal Methods in System Design 29~(1) (2006)
  79--95.

\bibitem{HalbwachsMP-V03}
N.~Halbwachs, D.~Merchat, C.~Parent-Vigouroux, Cartesian factoring of polyhedra
  in linear relation analysis, in: R.~Cousot (ed.), Static Analysis:
  Proceedings of the 10th International Symposium, vol. 2694 of Lecture Notes
  in Computer Science, Springer-Verlag, Berlin, San Diego, California, USA,
  2003.

\bibitem{HalbwachsPR94}
N.~Halbwachs, Y.-E. Proy, P.~Raymond, Verification of linear hybrid systems by
  means of convex approximations, in: B.~{Le Charlier} (ed.), Static Analysis:
  Proceedings of the 1st International Symposium, vol. 864 of Lecture Notes in
  Computer Science, Springer-Verlag, Berlin, Namur, Belgium, 1994.

\bibitem{HalbwachsPR97}
N.~Halbwachs, Y.-E. Proy, P.~Roumanoff, Verification of real-time systems using
  linear relation analysis, Formal Methods in System Design 11~(2) (1997)
  157--185.

\bibitem{HartongHB02}
W.~Hartong, L.~Hedrich, E.~Barke, On discrete modeling and model checking for
  nonlinear analog systems, in: E.~Brinksma, K.~G. Larsen (eds.), Computer
  Aided Verification: Proceedings of the 14th International Conference, vol.
  2404 of Lecture Notes in Computer Science, Springer-Verlag, Berlin,
  Copenhagen, Denmark, 2002.

\bibitem{HenriksenG06}
K.~S. Henriksen, J.~P. Gallagher, Abstract interpretation of {PIC} programs
  through logic programming, in: Proceedings of the 6th IEEE International
  Workshop on Source Code Analysis and Manipulation, IEEE Computer Society
  Press, Sheraton Society Hill, Philadelphia, PA, USA, 2006.

\bibitem{Henzinger96}
T.~A. Henzinger, The theory of hybrid automata, in: Proceedings of the 11th
  Annual Symposium on Logic in Computer Science (LICS), IEEE Computer Society
  Press, 1996.

\bibitem{HenzingerH95b}
T.~A. Henzinger, P.-H. Ho, Algorithmic analysis of nonlinear hybrid systems,
  in: P.~Wolper (ed.), Computer Aided Verification: Proceedings of the 7th
  International Conference, vol. 939 of Lecture Notes in Computer Science,
  Springer-Verlag, Berlin, Li\`ege, Belgium, 1995.

\bibitem{HenzingerH95}
T.~A. Henzinger, P.-H. Ho, A note on abstract interpretation strategies for
  hybrid automata, in: P.~J. Antsaklis, W.~Kohn, A.~Nerode, S.~Sastry (eds.),
  Hybrid Systems II, vol. 999 of Lecture Notes in Computer Science,
  Springer-Verlag, Berlin, 1995.

\bibitem{HenzingerHW97b}
T.~A. Henzinger, P.-H. Ho, H.~Wong-Toi, {\sc HyTech}: A model checker for
  hybrid systems, Software Tools for Technology Transfer 1~(1+2) (1997)
  110--122.

\bibitem{HenzingerPW01}
T.~A. Henzinger, J.~Preussig, H.~Wong-Toi, Some lessons from the {\sc hytech}
  experience, in: Proceedings of the 40th Annual Conference on Decision and
  Control, IEEE Computer Society Press, 2001.

\bibitem{HymansU04}
C.~Hymans, E.~Upton, Static analysis of gated data dependence graphs, in:
  R.~Giacobazzi (ed.), Static Analysis: Proceedings of the 11th International
  Symposium, vol. 3148 of Lecture Notes in Computer Science, Springer-Verlag,
  Berlin, Verona, Italy, 2004.

\bibitem{Kahn87}
G.~Kahn, Natural semantics, in: F.-J. Brandenburg, G.~Vidal-Naquet, M.~Wirsing
  (eds.), Proceedings of the 4th Annual Symposium on Theoretical Aspects of
  Computer Science, vol. 247 of Lecture Notes in Computer Science,
  Springer-Verlag, Berlin, Passau, Germany, 1987.

\bibitem{Karr76}
M.~Karr, Affine relationships among variables of a program, Acta Informatica 6
  (1976) 133--151.

\bibitem{KrishnanM06}
K.~Krishnan, J.~Mitchell, A unifying framework for several cutting plane
  methods for semidefinite programming, Optimization Methods and Software
  21~(1) (2006) 57--74.

\bibitem{KruegelKMRV05}
C.~Kruegel, E.~Kirda, D.~Mutz, W.~Robertson, G.~Vigna, Automating mimicry
  attacks using static binary analysis, in: Proceedings of Security~'05, the
  14th USENIX Security Symposium, Baltimore, MD, USA, 2005.

\bibitem{LarsenLPY97}
K.~Larsen, F.~Larsson, P.~Pettersson, W.~Yi, Efficient verification of
  real-time systems: Compact data structure and state-space reduction, in:
  Proceedings of the 18th IEEE Real-Time Systems Symposium (RTSS'97), IEEE
  Computer Society Press, San Francisco, CA, 1997.

\bibitem{LassezM92}
J.-L. Lassez, M.~J. Maher, On {F}ourier's algorithm for linear arithmetic
  constraints, J. Autom. Reasoning 9~(3) (1992) 373--379.

\bibitem{LeVerge92}
H.~{Le Verge}, A note on {Chernikova's} algorithm, \emph{Publication interne}
  635, IRISA, Campus de Beaulieu, Rennes, France (1992).

\bibitem{Leroy06}
X.~Leroy, Coinductive big-step operational semantics, in: P.~Sestoft (ed.),
  Programming Languages and Systems, Proceedings of the 14th European Symposium
  on Programming, vol. 3924 of Lecture Notes in Computer Science,
  Springer-Verlag, Berlin, Vienna, Austria, 2006.

\bibitem{Loechner99}
V.~Loechner, {\it PolyLib\/}: A library for manipulating parameterized
  polyhedra, Available at \url{http://icps.u-strasbg.fr/~loechner/polylib/},
  declares itself to be a continuation of \cite{Wilde93th} (Mar. 1999).

\bibitem{Maler06}
O.~Maler, Analog circuit verification: A state of an art, in: Proceedings of
  the First Workshop on Formal Verification of Analog Circuits (FAC 2005), vol.
  153 of Electronic Notes in Theoretical Computer Science, Elsevier Science
  B.V., Edinburgh, Scotland, 2006.

\bibitem{MesnardB05TPLP}
F.~Mesnard, R.~Bagnara, {cTI}: A constraint-based termination inference tool
  for {ISO-Prolog}, Theory and Practice of Logic Programming 5~(1{\&}2) (2005)
  243--257.

\bibitem{Mine01b}
A.~Min\'e, The octagon abstract domain, in: Proceedings of the Eighth Working
  Conference on Reverse Engineering (WCRE'01), IEEE Computer Society Press,
  Stuttgart, Germany, 2001.

\bibitem{Mine05th}
A.~Min\'e, Weakly relational numerical abstract domains, Ph.D. thesis, \'Ecole
  Polytechnique, Paris, France (Mar. 2005).

\bibitem{MotzkinRTT53}
T.~S. Motzkin, H.~Raiffa, G.~L. Thompson, R.~M. Thrall, The double description
  method, in: H.~W. Kuhn, A.~W. Tucker (eds.), Contributions to the Theory of
  Games -- Volume II, No.~28 in Annals of Mathematics Studies, Princeton
  University Press, Princeton, New Jersey, 1953, pp. 51--73.

\bibitem{MullerS00}
O.~M\"{u}ller, T.~Stauner, Modelling and verification using linear hybrid
  systems, Mathematical and Computer Modelling of Dynamical Systems 6~(1)
  (2000) 71--89.

\bibitem{Muller-OlmS04ICALP}
M.~{M\"uller-Olm}, H.~Seidl, A note on {Karr's} algorithm, in: J.~Diaz,
  J.~{Karhum\"aki}, A.~et~al. (eds.), Automata, Languages and Programming:
  Proceedings of the 31st International Colloquium (ICALP 2004), vol. 3142 of
  Lecture Notes in Computer Science, Springer-Verlag, Berlin, Turku, Finland,
  2004.

\bibitem{NakanishiJPF99}
T.~Nakanishi, K.~Joe, C.~D. Polychronopoulos, A.~Fukuda, The modulo interval: A
  simple and practical representation for program analysis, in: Proceedings of
  the 1999 International Conference on Parallel Architectures and Compilation
  Techniques, IEEE Computer Society, Newport Beach, California, USA, 1999.

\bibitem{NookalaR00}
S.~P.~K. Nookala, T.~Risset, A library for {Z}-polyhedral operations,
  \emph{Publication interne} 1330, IRISA, Campus de Beaulieu, Rennes, France
  (2000).

\bibitem{Plotkin81}
G.~Plotkin, A structural approach to operational semantics, Tech. Rep. DAIMI
  FN-19, Computer Science Department, University of Aarhus, Denmark (1981).

\bibitem{QuintonRR96}
P.~Quinton, S.~Rajopadhye, T.~Risset, On manipulating {Z}-polyhedra, Tech. Rep.
  1016, IRISA, Campus Universitaire de Bealieu, Rennes, France (Jul. 1996).

\bibitem{SankaranarayananCSM06}
S.~Sankaranarayanan, M.~Col{\'o}n, H.~B. Sipma, Z.~Manna, Efficient strongly
  relational polyhedral analysis, in: E.~A. Emerson, K.~S. Namjoshi (eds.),
  Verification, Model Checking and Abstract Interpretation: Proceedings of the
  7th International Conference (VMCAI 2006), vol. 3855 of Lecture Notes in
  Computer Science, Springer-Verlag, Berlin, Charleston, SC, USA, 2006.

\bibitem{SankaranarayananSM05}
S.~Sankaranarayanan, H.~B. Sipma, Z.~Manna, Scalable analysis of linear systems
  using mathematical programming, in: R.~Cousot (ed.), Verification, Model
  Checking and Abstract Interpretation: Proceedings of the 6th International
  Conference (VMCAI 2005), vol. 3385 of Lecture Notes in Computer Science,
  Springer-Verlag, Berlin, Paris, France, 2005.

\bibitem{SankaranarayananSM06}
S.~Sankaranarayanan, H.~B. Sipma, Z.~Manna, Fixed point iteration for computing
  the time elapse operator, in: J.~Hespanha, A.~Tiwari (eds.), Hybrid Systems:
  Computation and Control: Proceedings of the 9th International Workshop (HSCC
  2006), vol. 3927 of Lecture Notes in Computer Science, Springer-Verlag,
  Berlin, Santa Barbara, CA, USA, 2006.

\bibitem{Schmidt95}
D.~A. Schmidt, Natural-semantics-based abstract interpretation (preliminary
  version), in: A.~Mycroft (ed.), Static Analysis: Proceedings of the 2nd
  International Symposium, vol. 983 of Lecture Notes in Computer Science,
  Springer-Verlag, Berlin, Glasgow, UK, 1995.

\bibitem{Schmidt97}
D.~A. Schmidt, Abstract interpretation of small-step semantics, in: M.~Dam
  (ed.), Analysis and Verification of Multiple-Agent Languages, vol. 1192 of
  Lecture Notes in Computer Science, Springer-Verlag, Berlin, 1997, pp. 76--99,
  5th LOMAPS Workshop Stockholm, Sweden, June 24--26, 1996, Selected Papers.

\bibitem{Schmidt98}
D.~A. Schmidt, Trace-based abstract interpretation of operational semantics,
  {LISP} and Symbolic Computation 10~(3) (1998) 237--271.

\bibitem{Schrijver99}
A.~Schrijver, Theory of Linear and Integer Programming, Wiley Interscience
  Series in Discrete Mathematics and Optimization, John Wiley \& Sons, 1999.

\bibitem{ShahamKS00}
R.~Shaham, E.~K. Kolodner, S.~Sagiv, Automatic removal of array memory leaks in
  {J}ava, in: D.~A. Watt (ed.), Proceedings of the 9th International Conference
  on Compiler Construction (CC 2000), vol. 1781 of Lecture Notes in Computer
  Science, Springer-Verlag, Berlin, Berlin, Germany, 2000.

\bibitem{SimonKH02}
A.~Simon, A.~King, J.~M. Howe, Two variables per linear inequality as an
  abstract domain, in: M.~Leuschel (ed.), Logic Based Program Synthesis and
  Tranformation, 12th International Workshop, vol. 2664 of Lecture Notes in
  Computer Science, Springer-Verlag, Berlin, Madrid, Spain, 2002.

\bibitem{SohnVG91}
K.~Sohn, A.~{Van Gelder}, Termination detection in logic programs using
  argument sizes (extended abstract), in: Proceedings of the Tenth {ACM}
  {SIGACT-SIGMOD-SIGART} Symposium on Principles of Database Systems, ACM,
  Association for Computing Machinery, Denver, Colorado, United States, 1991.

\bibitem{SongCR06}
H.~Song, K.~J. Compton, W.~C. Rounds, {SPHIN:} a model checker for
  reconfigurable hybrid systems based on {SPIN}, in: R.~Lazic, R.~Nagarajan
  (eds.), Proceedings of the 5th International Workshop on Automated
  Verification of Critical Systems, vol. 145 of Electronic Notes in Theoretical
  Computer Science, University of Warwick, UK, 2006.

\bibitem{StoerW70}
J.~Stoer, C.~Witzgall, Convexity and Optimization in Finite Dimensions {I},
  Springer-Verlag, Berlin, 1970.

\bibitem{vanHeeOSV06}
K.~{van Hee}, O.~Oanea, N.~Sidorova, M.~Voorhoeve, Verifying generalized
  soundness for workflow nets, in: I.~Virbitskaite, A.~Voronkov (eds.),
  Perspectives of System Informatics: Proceedings of the Sixth International
  Andrei Ershov Memorial Conference, vol. 4378 of Lecture Notes in Computer
  Science, Springer-Verlag, Berlin, Akademgorodok, Novosibirsk, Russia, 2006.

\bibitem{VenetB04}
A.~Venet, G.~Brat, Precise and efficient static array bound checking for large
  embedded {C} programs, in: Proceedings of the ACM SIGPLAN 2004 Conference on
  Programming Language Design and Implementation (PLDI'04), ACM Press,
  Washington, DC, USA, 2004.

\bibitem{Wilde93th}
D.~K. Wilde, A library for doing polyhedral operations, {Master's thesis},
  Oregon State University, Corvallis, Oregon, also published as IRISA
  \emph{Publication interne} 785, Rennes, France, 1993 (Dec. 1993).

\end{thebibliography}

\begin{thebibliography}{10}
\expandafter\ifx\csname url\endcsname\relax
  \def\url#1{\texttt{#1}}\fi
\expandafter\ifx\csname urlprefix\endcsname\relax\def\urlprefix{URL }\fi

\bibitem{AlefeldH83}
G.~Alefeld, J.~Herzberger, Introduction to Interval Computation, Academic
  Press, New York, 1983.

\bibitem{AlurCHH93}
R.~Alur, C.~Courcoubetis, T.~A. Henzinger, P.-H. Ho, Hybrid automata: An
  algorithmic approach to the specification and verification of hybrid systems,
  in: Hybrid Systems I, vol. 736 of Lecture Notes in Computer Science, 1993.

\bibitem{Ancourt91th}
C.~Ancourt, G\'en\'eration automatique de codes de transfert pour
  multiprocesseurs \`a m\'emoires locales, Ph.D. thesis, Universit\'e de Paris
  VI, Paris, France (Mar. 1991).

\bibitem{Avis00}
D.~Avis, {lrs}: A revised implementation of the reverse search vertex
  enumeration algorithm, in: G.~Kalai, G.~M. Ziegler (eds.), Polytopes ---
  Combinatorics and Computation, vol.~29 of Oberwolfach Seminars,
  {Birkh\"auser-Verlag}, 2000, pp. 177--198.

\bibitem{AvisB95}
D.~Avis, D.~Bremner, How good are convex hull algorithms?, in: Proceedings of
  the Eleventh Annual Symposium on Computational Geometry, ACM Press,
  Vancouver, B.C., Canada, 1995.

\bibitem{Bagnara97th}
R.~Bagnara, Data-flow analysis for constraint logic-based languages, Ph.D.
  thesis, Dipartimento di Informatica, Universit\`a di Pisa, Pisa, Italy,
  printed as Report TD-1/97 (Mar. 1997).

\bibitem{Bagnara98SCP}
R.~Bagnara, A hierarchy of constraint systems for data-flow analysis of
  constraint logic-based languages, Science of Computer Programming 30~(1--2)
  (1998) 119--155.

\bibitem{BagnaraDHMZ07}
R.~Bagnara, K.~Dobson, P.~M. Hill, M.~Mundell, E.~Zaffanella, Grids: A domain
  for analyzing the distribution of numerical values, in: G.~Puebla (ed.),
  Logic-based Program Synthesis and Transformation, 16th International
  Symposium, vol. 4407 of Lecture Notes in Computer Science, Springer-Verlag,
  Berlin, Venice, Italy, 2007.

\bibitem{BagnaraHPZ07TR}
R.~Bagnara, P.~M. Hill, A.~Pescetti, E.~Zaffanella, On the design of generic
  static analyzers for modern imperative languages, Tech. Rep.~{\tt
  arXiv:cs.PL/0703116}, Dipartimento di Matematica, Universit\`a di Parma,
  Italy, available from \url{http://arxiv.org/} (2007).

\bibitem{BagnaraHRZ05SCP}
R.~Bagnara, P.~M. Hill, E.~Ricci, E.~Zaffanella, Precise widening operators for
  convex polyhedra, Science of Computer Programming 58~(1--2) (2005) 28--56.

\bibitem{BagnaraHZ05FAC}
R.~Bagnara, P.~M. Hill, E.~Zaffanella, Not necessarily closed convex polyhedra
  and the double description method, Formal Aspects of Computing 17~(2) (2005)
  222--257.

\bibitem{BagnaraHZ06STTT}
R.~Bagnara, P.~M. Hill, E.~Zaffanella, Widening operators for powerset domains,
  Software Tools for Technology Transfer 8~(4/5) (2006) 449--466. (As the
  figures in the journal version of this paper have been improperly printed
  ---rendering them useless---, we recommend that interested readers download
  an electronic copy from the PPL's web site at
  \url{http://www.cs.unipr.it/ppl/}.)

\bibitem{BagnaraHZ07TRa}
R.~Bagnara, P.~M. Hill, E.~Zaffanella, Applications of polyhedral computations
  to the analysis and verification of hardware and software systems,
  {\tt arXiv:cs.CG/0701122}, available from \url{http://arxiv.org/}. (2007).

\bibitem{BagnaraHZ08SCP}
R.~Bagnara, P.~M. Hill, E.~Zaffanella, The {Parma Polyhedra Library}: Toward a
  complete set of numerical abstractions for the analysis and verification of
  hardware and software systems, Science of Computer ProgrammingTo appear.

\bibitem{BagnaraR-CZ05}
R.~Bagnara, E.~Rodr{\'\i}guez-Carbonell, E.~Zaffanella, Generation of basic
  semi-algebraic invariants using convex polyhedra, in: C.~Hankin, I.~Siveroni
  (eds.), Static Analysis: Proceedings of the 12th International Symposium,
  vol. 3672 of Lecture Notes in Computer Science, Springer-Verlag, Berlin,
  London, UK, 2005.

\bibitem{Bellman57}
R.~Bellman, Dynamic Programming, Princeton University Press, 1957.

\bibitem{BemporadFT01}
A.~Bemporad, K.~Fukuda, F.~D. Torrisi, Convexity recognition of the union of
  polyhedra, Computational Geometry: Theory and Applications 18~(3) (2001)
  141--154.

\bibitem{BenoyK97}
F.~Benoy, A.~King, Inferring argument size relationships with
  {CLP($\mathcal{R}$)}, in: J.~P. Gallagher (ed.), Logic Program Synthesis and
  Transformation: Proceedings of the 6th International Workshop, vol. 1207 of
  Lecture Notes in Computer Science, Springer-Verlag, Berlin, Stockholm,
  Sweden, 1997.

\bibitem{BessonJT99}
F.~Besson, T.~P. Jensen, J.-P. Talpin, Polyhedral analysis for synchronous
  languages, in: A.~Cortesi, G.~Fil\'e (eds.), Static Analysis: Proceedings of
  the 6th International Symposium, vol. 1694 of Lecture Notes in Computer
  Science, Springer-Verlag, Berlin, Venice, Italy, 1999.

\bibitem{Birkhoff67}
G.~Birkhoff, Lattice Theory, vol. XXV of Colloquium Publications, 3rd ed.,
  American Mathematical Society, Providence, Rhode Island, USA, 1967.

\bibitem{BlanchetCCFMMMR03}
B.~Blanchet, P.~Cousot, R.~Cousot, J.~Feret, L.~Mauborgne, A.~Min\'e,
  D.~Monniaux, X.~Rival, A static analyzer for large safety-critical software,
  in: Proceedings of the ACM SIGPLAN 2003 Conference on Programming Language
  Design and Implementation (PLDI'03), ACM Press, San Diego, California, USA,
  2003.

\bibitem{BultanGP99}
T.~Bultan, R.~Gerber, W.~Pugh, Model-checking concurrent systems with unbounded
  integer variables: Symbolic representations, approximations, and experimental
  results, ACM Transactions on Programming Languages and Systems 21~(4) (1999)
  747--789.

\bibitem{Chernikova68}
N.~V. Chernikova, Algorithm for discovering the set of all solutions of a
  linear programming problem, U.S.S.R. Computational Mathematics and
  Mathematical Physics 8~(6) (1968) 282--293.

\bibitem{ColonS01}
M.~A. Col\'on, H.~B. Sipma, Synthesis of linear ranking functions, in:
  T.~Margaria, W.~Yi (eds.), Tools and Algorithms for Construction and Analysis
  of Systems, 7th International Conference, TACAS 2001, vol. 2031 of Lecture
  Notes in Computer Science, Springer-Verlag, Berlin, Genova, Italy, 2001.

\bibitem{CortesiLCVH00}
A.~Cortesi, B.~{Le Charlier}, P.~{Van Hentenryck}, Combinations of abstract
  domains for logic programming: Open product and generic pattern construction,
  Science of Computer Programming 38~(1--3) (2000) 27--71.

\bibitem{Cousot05}
P.~Cousot, Proving program invariance and termination by parametric
  abstraction, lagrangian relaxation and semidefinite programming, in:
  R.~Cousot (ed.), Verification, Model Checking and Abstract Interpretation:
  Proceedings of the 6th International Conference (VMCAI 2005), vol. 3385 of
  Lecture Notes in Computer Science, Springer-Verlag, Berlin, Paris, France,
  2005.

\bibitem{CousotC76}
P.~Cousot, R.~Cousot, Static determination of dynamic properties of programs,
  in: B.~Robinet (ed.), Proceedings of the Second International Symposium on
  Programming, Dunod, Paris, France, Paris, France, 1976.

\bibitem{CousotC79}
P.~Cousot, R.~Cousot, Systematic design of program analysis frameworks, in:
  Proceedings of the Sixth Annual ACM Symposium on Principles of Programming
  Languages, ACM Press, New York, 1979.

\bibitem{CousotC92fr}
P.~Cousot, R.~Cousot, Abstract interpretation frameworks, Journal of Logic and
  Computation 2~(4) (1992) 511--547.

\bibitem{CousotC92}
P.~Cousot, R.~Cousot, Inductive definitions, semantics and abstract
  interpretation, in: Proceedings of the Nineteenth Annual ACM Symposium on
  Principles of Programming Languages, ACM Press, Albuquerque, New Mexico, USA,
  1992.

\bibitem{CousotH78}
P.~Cousot, N.~Halbwachs, Automatic discovery of linear restraints among
  variables of a program, in: Conference Record of the Fifth Annual ACM
  Symposium on Principles of Programming Languages, ACM Press, Tucson, Arizona,
  1978.

\bibitem{Davis87}
E.~Davis, Constraint propagation with interval labels, Artificial Intelligence
  32~(3) (1987) 281--331.

\bibitem{DelzannoP99}
G.~Delzanno, A.~Podelski, Model checking in {CLP}, in: R.~Cleaveland (ed.),
  Tools and Algorithms for Construction and Analysis of Systems, 5th
  International Conference, TACAS '99, vol. 1579 of Lecture Notes in Computer
  Science, Springer-Verlag, Berlin, Amsterdam, The Netherlands, 1999.

\bibitem{DooseM05}
D.~Doose, Z.~Mammeri, Polyhedra-based approach for incremental validation of
  real-time systems, in: L.~T. Yang, M.~Amamiya, Z.~Liu, M.~Guo, F.~J. Rammig
  (eds.), Proceedings of the International Conference on Embedded and
  Ubiquitous Computing (EUC 2005), vol. 3824 of Lecture Notes in Computer
  Science, Springer-Verlag, Berlin, Nagasaki, Japan, 2005.

\bibitem{DorRS01}
N.~Dor, M.~Rodeh, S.~Sagiv, Cleanness checking of string manipulations in {C}
  programs via integer analysis, in: P.~Cousot (ed.), Static Analysis: 8th
  International Symposium, SAS 2001, vol. 2126 of Lecture Notes in Computer
  Science, Springer-Verlag, Berlin, Paris, France, 2001.

\bibitem{DoyenHR05}
L.~Doyen, T.~A. Henzinger, J.-F. Raskin, Automatic rectangular refinement of
  affine hybrid systems, in: P.~Pettersson, W.~Yi (eds.), Proceedings of the
  3rd International Conference on Formal Modeling and Analysis of Timed Systems
  (FORMATS 2005), vol. 3829 of Lecture Notes in Computer Science,
  Springer-Verlag, Berlin, Uppsala, Sweden, 2005.

\bibitem{Ellenbogen04th}
R.~Ellenbogen, Fully automatic verification of absence of errors via
  interprocedural integer analysis, Master's thesis, School of Computer
  Science, Tel-Aviv University, Tel-Aviv, Israel (Dec. 2004).

\bibitem{Frehse05}
G.~Frehse, {PHAVer}: Algorithmic verification of hybrid systems past {HyTech},
  in: M.~Morari, L.~Thiele (eds.), Hybrid Systems: Computation and Control:
  Proceedings of the 8th International Workshop (HSCC 2005), vol. 3414 of
  Lecture Notes in Computer Science, Springer-Verlag, Berlin, Z{\"u}rich,
  Switzerland, 2005.

\bibitem{FrehseKR06}
G.~Frehse, B.~H. Krogh, R.~A. Rutenbar, Verifying analog oscillator circuits
  using forward/backward refinement, in: Proceedings of the 9th Conference on
  Design, Automation and Test in Europe (DATE 06), ACM SIGDA, Munich, Germany,
  2006, {CD-ROM} publication.

\bibitem{FukudaP96}
K.~Fukuda, A.~Prodon, Double description method revisited, in: M.~Deza,
  R.~Euler, Y.~Manoussakis (eds.), Combinatorics and Computer Science, 8th
  Franco-Japanese and 4th Franco-Chinese Conference, Brest, France, July 3-5,
  1995, Selected Papers, vol. 1120 of Lecture Notes in Computer Science,
  Springer-Verlag, Berlin, 1996.

\bibitem{GopanRS05}
D.~Gopan, T.~W. Reps, M.~Sagiv, A framework for numeric analysis of array
  operations, in: Proceedings of the 32nd ACM SIGPLAN-SIGACT Symposium on
  Principles of Programming Languages, Long Beach, California, USA, 2005.

\bibitem{Granger92}
P.~Granger, Improving the results of static analyses programs by local
  decreasing iteration, in: R.~K. Shyamasundar (ed.), Proceedings of the 12th
  Conference on Foundations of Software Technology and Theoretical Computer
  Science, vol. 652 of Lecture Notes in Computer Science, Springer-Verlag,
  Berlin, New Delhi, India, 1992.

\bibitem{Granger97}
P.~Granger, Static analyses of congruence properties on rational numbers
  (extended abstract), in: P.~{Van Hentenryck} (ed.), Static Analysis:
  Proceedings of the 4th International Symposium, vol. 1302 of Lecture Notes in
  Computer Science, Springer-Verlag, Berlin, Paris, France, 1997.

\bibitem{GulavaniR06}
B.~S. Gulavani, S.~K. Rajamani, Counterexample driven refinement for abstract
  interpretation, in: H.~Hermanns, J.~Palsberg (eds.), Proceedings of the 12th
  International Conference on Tools and Algorithms for the Construction and
  Analysis of Systems (TACAS 2006), vol. 3920 of Lecture Notes in Computer
  Science, Springer-Verlag, Berlin, Vienna, Austria, 2006.

\bibitem{Halbwachs93}
N.~Halbwachs, Delay analysis in synchronous programs, in: C.~Courcoubetis
  (ed.), Computer Aided Verification: Proceedings of the 5th International
  Conference, vol. 697 of Lecture Notes in Computer Science, Springer-Verlag,
  Berlin, Elounda, Greece, 1993.

\bibitem{HalbwachsMG06}
N.~Halbwachs, D.~Merchat, L.~Gonnord, Some ways to reduce the space dimension
  in polyhedra computations, Formal Methods in System Design 29~(1) (2006)
  79--95.

\bibitem{HalbwachsPR97}
N.~Halbwachs, Y.-E. Proy, P.~Roumanoff, Verification of real-time systems using
  linear relation analysis, Formal Methods in System Design 11~(2) (1997)
  157--185.

\bibitem{HartongHB02}
W.~Hartong, L.~Hedrich, E.~Barke, On discrete modeling and model checking for
  nonlinear analog systems, in: E.~Brinksma, K.~G. Larsen (eds.), Computer
  Aided Verification: Proceedings of the 14th International Conference, vol.
  2404 of Lecture Notes in Computer Science, Springer-Verlag, Berlin,
  Copenhagen, Denmark, 2002.

\bibitem{Henzinger96}
T.~A. Henzinger, The theory of hybrid automata, in: Proceedings of the 11th
  Annual Symposium on Logic in Computer Science (LICS), IEEE Computer Society
  Press, 1996.

\bibitem{HenzingerH95}
T.~A. Henzinger, P.-H. Ho, A note on abstract interpretation strategies for
  hybrid automata, in: P.~J. Antsaklis, W.~Kohn, A.~Nerode, S.~Sastry (eds.),
  Hybrid Systems II, vol. 999 of Lecture Notes in Computer Science,
  Springer-Verlag, Berlin, 1995.

\bibitem{HenzingerHW97b}
T.~A. Henzinger, P.-H. Ho, H.~Wong-Toi, {\sc HyTech}: A model checker for
  hybrid systems, Software Tools for Technology Transfer 1~(1+2) (1997)
  110--122.

\bibitem{HenzingerPW01}
T.~A. Henzinger, J.~Preussig, H.~Wong-Toi, Some lessons from the {\sc hytech}
  experience, in: Proceedings of the 40th Annual Conference on Decision and
  Control, IEEE Computer Society Press, 2001.

\bibitem{HymansU04}
C.~Hymans, E.~Upton, Static analysis of gated data dependence graphs, in:
  R.~Giacobazzi (ed.), Static Analysis: Proceedings of the 11th International
  Symposium, vol. 3148 of Lecture Notes in Computer Science, Springer-Verlag,
  Berlin, Verona, Italy, 2004.

\bibitem{Kahn87}
G.~Kahn, Natural semantics, in: F.-J. Brandenburg, G.~Vidal-Naquet, M.~Wirsing
  (eds.), Proceedings of the 4th Annual Symposium on Theoretical Aspects of
  Computer Science, vol. 247 of Lecture Notes in Computer Science,
  Springer-Verlag, Berlin, Passau, Germany, 1987.

\bibitem{Karr76}
M.~Karr, Affine relationships among variables of a program, Acta Informatica 6
  (1976) 133--151.

\bibitem{KrishnanM06}
K.~Krishnan, J.~Mitchell, A unifying framework for several cutting plane
  methods for semidefinite programming, Optimization Methods and Software
  21~(1) (2006) 57--74.

\bibitem{KruegelKMRV05}
C.~Kruegel, E.~Kirda, D.~Mutz, W.~Robertson, G.~Vigna, Automating mimicry
  attacks using static binary analysis, in: Proceedings of Security~'05, the
  14th USENIX Security Symposium, Baltimore, MD, USA, 2005.

\bibitem{LassezM92}
J.-L. Lassez, M.~J. Maher, On {F}ourier's algorithm for linear arithmetic
  constraints, J. Autom. Reasoning 9~(3) (1992) 373--379.

\bibitem{LeVerge92}
H.~{Le Verge}, A note on {Chernikova's} algorithm, \emph{Publication interne}
  635, IRISA, Campus de Beaulieu, Rennes, France (1992).

\bibitem{Loechner99}
V.~Loechner, {\it PolyLib\/}: A library for manipulating parameterized
  polyhedra, Available at \url{http://icps.u-strasbg.fr/~loechner/polylib/},
  declares itself to be a continuation of \cite{Wilde93th} (Mar. 1999).

\bibitem{MesnardB05TPLP}
F.~Mesnard, R.~Bagnara, {cTI}: A constraint-based termination inference tool
  for {ISO-Prolog}, Theory and Practice of Logic Programming 5~(1{\&}2) (2005)
  243--257.

\bibitem{Mine05th}
A.~Min\'e, Weakly relational numerical abstract domains, Ph.D. thesis, \'Ecole
  Polytechnique, Paris, France (Mar. 2005).

\bibitem{MotzkinRTT53}
T.~S. Motzkin, H.~Raiffa, G.~L. Thompson, R.~M. Thrall, The double description
  method, in: H.~W. Kuhn, A.~W. Tucker (eds.), Contributions to the Theory of
  Games -- Volume II, No.~28 in Annals of Mathematics Studies, Princeton
  University Press, Princeton, New Jersey, 1953, pp. 51--73.

\bibitem{NakanishiJPF99}
T.~Nakanishi, K.~Joe, C.~D. Polychronopoulos, A.~Fukuda, The modulo interval: A
  simple and practical representation for program analysis, in: Proceedings of
  the 1999 International Conference on Parallel Architectures and Compilation
  Techniques, IEEE Computer Society, Newport Beach, California, USA, 1999.

\bibitem{NookalaR00}
S.~P.~K. Nookala, T.~Risset, A library for {Z}-polyhedral operations,
  \emph{Publication interne} 1330, IRISA, Campus de Beaulieu, Rennes, France
  (2000).

\bibitem{Plotkin81}
G.~Plotkin, A structural approach to operational semantics, Tech. Rep. DAIMI
  FN-19, Computer Science Department, University of Aarhus, Denmark (1981).

\bibitem{QuintonRR96}
P.~Quinton, S.~Rajopadhye, T.~Risset, On manipulating {Z}-polyhedra, Tech. Rep.
  1016, IRISA, Campus Universitaire de Bealieu, Rennes, France (Jul. 1996).

\bibitem{SankaranarayananCSM06}
S.~Sankaranarayanan, M.~Col{\'o}n, H.~B. Sipma, Z.~Manna, Efficient strongly
  relational polyhedral analysis, in: E.~A. Emerson, K.~S. Namjoshi (eds.),
  Verification, Model Checking and Abstract Interpretation: Proceedings of the
  7th International Conference (VMCAI 2006), vol. 3855 of Lecture Notes in
  Computer Science, Springer-Verlag, Berlin, Charleston, SC, USA, 2006.

\bibitem{SankaranarayananSM05}
S.~Sankaranarayanan, H.~B. Sipma, Z.~Manna, Scalable analysis of linear systems
  using mathematical programming, in: R.~Cousot (ed.), Verification, Model
  Checking and Abstract Interpretation: Proceedings of the 6th International
  Conference (VMCAI 2005), vol. 3385 of Lecture Notes in Computer Science,
  Springer-Verlag, Berlin, Paris, France, 2005.

\bibitem{Schmidt95}
D.~A. Schmidt, Natural-semantics-based abstract interpretation (preliminary
  version), in: A.~Mycroft (ed.), Static Analysis: Proceedings of the 2nd
  International Symposium, vol. 983 of Lecture Notes in Computer Science,
  Springer-Verlag, Berlin, Glasgow, UK, 1995.

\bibitem{Schrijver99}
A.~Schrijver, Theory of Linear and Integer Programming, Wiley Interscience
  Series in Discrete Mathematics and Optimization, John Wiley \& Sons, 1999.

\bibitem{ShahamKS00}
R.~Shaham, E.~K. Kolodner, S.~Sagiv, Automatic removal of array memory leaks in
  {J}ava, in: D.~A. Watt (ed.), Proceedings of the 9th International Conference
  on Compiler Construction (CC 2000), vol. 1781 of Lecture Notes in Computer
  Science, Springer-Verlag, Berlin, Berlin, Germany, 2000.

\bibitem{SimonKH02}
A.~Simon, A.~King, J.~M. Howe, Two variables per linear inequality as an
  abstract domain, in: M.~Leuschel (ed.), Logic Based Program Synthesis and
  Tranformation, 12th International Workshop, vol. 2664 of Lecture Notes in
  Computer Science, Springer-Verlag, Berlin, Madrid, Spain, 2002.

\bibitem{SohnVG91}
K.~Sohn, A.~{Van Gelder}, Termination detection in logic programs using
  argument sizes (extended abstract), in: Proceedings of the Tenth {ACM}
  {SIGACT-SIGMOD-SIGART} Symposium on Principles of Database Systems, ACM,
  Association for Computing Machinery, Denver, Colorado, United States, 1991.

\bibitem{SongCR06}
H.~Song, K.~J. Compton, W.~C. Rounds, {SPHIN:} a model checker for
  reconfigurable hybrid systems based on {SPIN}, in: R.~Lazic, R.~Nagarajan
  (eds.), Proceedings of the 5th International Workshop on Automated
  Verification of Critical Systems, vol. 145 of Electronic Notes in Theoretical
  Computer Science, University of Warwick, UK, 2006.

\bibitem{StoerW70}
J.~Stoer, C.~Witzgall, Convexity and Optimization in Finite Dimensions {I},
  Springer-Verlag, Berlin, 1970.

\bibitem{vanHeeOSV06}
K.~{van Hee}, O.~Oanea, N.~Sidorova, M.~Voorhoeve, Verifying generalized
  soundness for workflow nets, in: I.~Virbitskaite, A.~Voronkov (eds.),
  Perspectives of System Informatics: Proceedings of the Sixth International
  Andrei Ershov Memorial Conference, vol. 4378 of Lecture Notes in Computer
  Science, Springer-Verlag, Berlin, Akademgorodok, Novosibirsk, Russia, 2006.

\bibitem{VenetB04}
A.~Venet, G.~Brat, Precise and efficient static array bound checking for large
  embedded {C} programs, in: Proceedings of the ACM SIGPLAN 2004 Conference on
  Programming Language Design and Implementation (PLDI'04), ACM Press,
  Washington, DC, USA, 2004.

\bibitem{Wilde93th}
D.~K. Wilde, A library for doing polyhedral operations, {Master's thesis},
  Oregon State University, Corvallis, Oregon, also published as IRISA
  \emph{Publication interne} 785, Rennes, France, 1993 (Dec. 1993).

\end{thebibliography}

\ifthenelse{\boolean{TR}}{%%
\newcommand{\noopsort}[1]{}\hyphenation{ Ba-gna-ra Bie-li-ko-va Bruy-noo-ghe
  Common-Loops DeMich-iel Dober-kat Di-par-ti-men-to Er-vier Fa-la-schi
  Fell-eisen Gam-ma Gem-Stone Glan-ville Gold-in Goos-sens Graph-Trace
  Grim-shaw Her-men-e-gil-do Hoeks-ma Hor-o-witz Kam-i-ko Kenn-e-dy Kess-ler
  Lisp-edit Lu-ba-chev-sky Ma-te-ma-ti-ca Nich-o-las Obern-dorf Ohsen-doth
  Par-log Para-sight Pega-Sys Pren-tice Pu-ru-sho-tha-man Ra-guid-eau Rich-ard
  Roe-ver Ros-en-krantz Ru-dolph SIG-OA SIG-PLAN SIG-SOFT SMALL-TALK Schee-vel
  Schlotz-hauer Schwartz-bach Sieg-fried Small-talk Spring-er Stroh-meier
  Thing-Lab Zhong-xiu Zac-ca-gni-ni Zaf-fa-nel-la Zo-lo }

}{%%
\newcommand{\noopsort}[1]{}\hyphenation{ Ba-gna-ra Bie-li-ko-va Bruy-noo-ghe
  Common-Loops DeMich-iel Dober-kat Di-par-ti-men-to Er-vier Fa-la-schi
  Fell-eisen Gam-ma Gem-Stone Glan-ville Gold-in Goos-sens Graph-Trace
  Grim-shaw Her-men-e-gil-do Hoeks-ma Hor-o-witz Kam-i-ko Kenn-e-dy Kess-ler
  Lisp-edit Lu-ba-chev-sky Ma-te-ma-ti-ca Nich-o-las Obern-dorf Ohsen-doth
  Par-log Para-sight Pega-Sys Pren-tice Pu-ru-sho-tha-man Ra-guid-eau Rich-ard
  Roe-ver Ros-en-krantz Ru-dolph SIG-OA SIG-PLAN SIG-SOFT SMALL-TALK Schee-vel
  Schlotz-hauer Schwartz-bach Sieg-fried Small-talk Spring-er Stroh-meier
  Thing-Lab Zhong-xiu Zac-ca-gni-ni Zaf-fa-nel-la Zo-lo }

}%%\ifthenelse{\boolean{TR}}{%%

\end{document}